\documentclass[prd,nofootinbib,preprintnumbers]{revtex4-2}
\usepackage{amsfonts,amsmath}
\usepackage[colorlinks=true,linkcolor=red,citecolor=blue]{hyperref}
\usepackage[parfill]{parskip}
\usepackage{orcidlink}
\usepackage{subcaption}
\usepackage{booktabs}
\usepackage{graphicx}
\usepackage{float}

\usepackage{amssymb}

\newcommand{\be}{\begin{eqnarray}}
\newcommand{\ee}{\end{eqnarray}}
\newcommand{\ba}{\begin{array}}
\newcommand{\ea}{\end{array}}

\newcommand{\nn}{\nonumber}

\newcommand{\bi}{\begin{itemize}}
\newcommand{\ei}{\end{itemize}}

\newcommand{\half}{{\textstyle{\frac{1}{2}}}}

\begin{document}

\title{Toward an advanced phenomenology of $\pi N$ transition distribution amplitudes}

\author{
B.~Pire$^{1}$\orcidlink{0000-0003-4882-7800},
K.~Semenov-Tian-Shansky$^{2}$\orcidlink{0000-0001-8159-0900},
P.~Sznajder$^{3}$\orcidlink{0000-0002-2684-803X},
L.~Szymanowski$^{3}$\orcidlink{0000-0003-1623-8331}
}
\affiliation{
$^1$ Centre de Physique Th\'eorique, CNRS, École Polytechnique, I.P. Paris, 91128 Palaiseau, France \\
$^2$ Kyungpook National University, Daegu 41566, Korea\\
$^3$ National Centre for Nuclear Research, NCBJ, 02-093 Warsaw, Poland
}

\begin{abstract}
We introduce a new approach to modeling transition distribution amplitudes (TDAs) for the processes $e p \to e n \pi^+$ and $e p \to e p \pi^0$. The modeling is flexible, constrained by sparsely available experimental data, and satisfies theoretical requirements, including reduction to nucleon distribution amplitudes in the appropriate limit. We study the sensitivity of observable predictions to various modeling assumptions. We discuss unpolarized cross-sections, as well as the three non-vanishing polarization observables at leading twist, namely, the single transverse target spin asymmetry and the two double spin asymmetries that occur with a polarized lepton beam on either a longitudinally or transversely polarized target. The analysis is complemented by a simple  Monte Carlo study to provide guidance for exploring exclusive processes in the so-called backward kinematics. Our work aims to highlight the importance of future measurements to better constrain TDAs and to support upcoming experimental proposals.
\end{abstract}

\date{\today}

\maketitle

\section{Introduction}
\label{sec:intro}

The collinear factorization framework, see {\it e.g.} Ref.~\cite{Collins:2011zzd} for an overview, allows a systematic separation of the short-distance strong interaction dynamics, calculable within perturbative QCD, from the long-distance dynamics encoded in universal distributions, such as parton distribution functions (PDFs), generalized parton distributions (GPDs), (generalized) distribution amplitudes (DAs and GDAs).
These non-perturbative functions, which are directly related to the hadronic matrix elements of light-cone operators, serve as a bridge between the underlying partonic degrees of freedom and the observable properties of hadrons. They provide detailed insights into the spatial and momentum distributions of quarks and gluons inside hadrons. Consequently, the precise determination of these functions has become a principal objective of modern hadron-structure research, guiding an extensive experimental programme (see, {\it e.g.}, Ref.\cite{Diehl:2023nmm} for a recent review).

Transition distribution amplitudes (TDAs) have been introduced in 
Refs.~\cite{Frankfurt:1999fp,Pire:2004ie,Pire:2005ax} 
to extend the collinear factorization framework for the case of hard exclusive reactions in the so-called near-backward kinematical regime \cite{Gayoso:2021rzj}. 
Nucleon-to-photon and nucleon-to-meson TDAs are defined as Fourier transforms of the matrix elements of the tri-local quark light cone ($n^2=0$) operator%
\footnote{Throughout this paper, we imply the use of the light-cone gauge $A \cdot n=0$ and omit the Wilson lines necessary to ensure the QCD gauge invariance of the tri-local operator (\ref{Def_3q_operator}).}
\begin{equation}
\hspace{2em}\widehat{O}_{\rho \tau \chi} 
(\lambda_1n,\lambda_2n,\lambda_3n) =\varepsilon_{c_1 c_2 c_3}
  \Psi^{c_1}_{\rho}(\lambda_1 n)
  \Psi^{c_2}_{\tau}(\lambda_2 n)  \Psi^{c_3}_{\chi}(\lambda_3 n)\,,
  \label{Def_3q_operator}
\end{equation}
where $c_{1,2,3}$ stand for the color group indices  and $\rho$, $\tau$, $\chi$
denote the Dirac indices of the quark field operators; and quark  flavor indices are omitted for brevity.

A prominent example of reactions admitting a description in terms of TDAs
are hard exclusive processes with a baryon number transfer in the $u$-channel  \cite{Pire:2021hbl}, 
such as the backward deeply virtual Compton (DVCS) scattering
\begin{equation}
    \gamma^* (q)+  N(p_N) \to N' (p'_N) + \gamma(q'),
    \label{Def_DVCS}
\end{equation}
and backward  hard exclusive electroproduction of mesons 
${\cal M}=\{ \pi, \eta,  \rho, \omega, \phi, \, \ldots \}$
\begin{equation}
    \gamma^* (q) + N(p_N) \to N'(p'_N) +{\cal M} (p_{\cal M}).
    \label{Def_DVMP}
\end{equation}
The near-backward kinematical regime for the reactions 
(\ref{Def_DVCS}) 
and 
(\ref{Def_DVMP}) is defined by large values of the invariant energy 
$s = (p_N + q)^2 $ 
and photon virtuality $Q^2 = -q^2$, 
with their ratio given by the Bjorken variable, 
$x_B = Q^2/(2 p_N \cdot q)$,
held fixed, while absolute value of the $u$-channel invariant momentum transfer, $|u| = |(q - p'_N)^2|$, remains small.
This kinematical regime complements the more familiar near-forward scattering regime, in which 
$|t|$ is kept small, and a collinear factorized description in terms of GPDs (and meson DAs) is provided 
for the DVCS and hard exclusive meson electroproduction processes.
Nucleon-to-meson and nucleon-to-photon TDAs share common features with both GPDs and nucleon DAs, reflecting a closely related underlying physical picture. They characterize partonic correlations inside hadrons and provide access to the baryon-number momentum distribution and, through an impact-parameter representation, to its transverse spatial localization.

First data accessing this new kinematical region for exclusive reactions have been obtained at JLab \cite{CLAS:2017rgp, JeffersonLabFp:2019gpp} and their phenomenological study in the TDA framework are indeed promising. Feasibility studies of backward DVCS for the kinematical conditions of the future EIC were performed in \cite{Sweger:2023bmx}. Other processes such as 
$N \bar N \to \gamma^* {\cal M}$ 
\cite{PANDA:2014qiz}, 
$N \bar N \to J/\psi {\cal M}$
\cite{PANDA:2016scz}, 
$\pi  N \to \gamma^* N'$, 
$\pi  N \to  J/\psi N'$, 
$\gamma  N \to \gamma^* N'$, 
$\gamma  N \to J/\psi N'$ 
may also be studied \cite{Pire:2022kwu}. On the theory side, although no factorization proof exists yet and no next-to-leading order calculation has shown that corresponding ultraviolet divergences may be reabsorbed in the QCD evolution of DAs and of TDAs, recent progresses in the calculation of baryonic electromagnetic form factors at next-to-leading order (NLO) in the strong coupling \cite{Chen:2024fhj, Huang:2024ugd} 
demonstrate that the goal is under reach, at least at the level of a NLO leading twist calculation.

Similarly to the GPDs, extracting TDAs from experimental data is not straightforward. It requires the development of sufficiently flexible parametrizations that can be employed to estimate relevant observables. With dedicated experiments currently in preparation~\cite{Li:2020nsk}, the development of an advanced phenomenological framework for nucleon-to-meson TDAs is both timely and essential. This is the primary objective of the present work.

This paper is organized as follows. Section \ref{sec:theory} sums up the main features of the theory of TDAs. We first recall  the known results for amplitudes, unpolarized cross-section and single transverse target spin asymmetry. We  also propose new polarization observables in the form of double spin asymmetries, which can serve to discriminate between different TDA models and present an interesting challenge for both phenomenological analysis and experiment.
In section \ref{sec:modeling}, we first develop a general modeling strategy based on the spectral representation of TDAs. We show how the few existing data already help us to normalize $N\to \pi$ TDAs and discuss then the sensitivity of various observables to modeling assumptions. 
In section \ref{sec:mc} we then perform some Monte Carlo studies of the $e N\to e' N \pi^0$ process for both forward and backward kinematics. In section\ref{sec:summary}, we summarize our progresses in modeling in a flexible way the TDAs and plead for further measurements of backward cross-sections and polarization observables. Three appendices follow to fix our notations for nucleon DAs, to recall the cross-channel nucleon exchange contribution to $\pi N$ TDAs, and to recapitulate the known constraints coming from the soft pion theorem.
\section{Overview of TDAs and backward meson electroproduction}
\label{sec:theory}

In this section we briefly review the basic properties of nucleon-to-meson TDAs and
their application to description of hard exclusive meson electroproduction in the near-backward kinematical regime. 
In the current paper we primarily discuss the case of nucleon-to-pion TDAs, however,
a generalization for nucleon-to-vector meson and nucleon-to-photon TDAs, as well as
their cross-conjugate counterparts, meson- (or photon-) to-nucleon TDAs is straightforward and can be found {\it e.g.} in Ref.~\cite{Pire:2021hbl}.

\subsection{Kinematics of backward pion electroproduction}

\label{Sec_Kinematics}
Let us consider the near-backward kinematics regime for the hard process
\begin{equation}
\gamma^*(q,\lambda_\gamma)
+ N(p_N,s_N) \to N(p'_N,s'_N)+ \pi(p_\pi)\,,
\label{Reaction_cases}
\end{equation}
with the hard scale set by the virtuality of the initial photon $Q^2=-q^2$, and where $\lambda_\gamma$ is  the virtual photon
polarization while $s_N$, $s'_N$ 
stand for the nucleon  polarization variables. We employ the standard Mandelstam variables for the
hard reaction (\ref{Reaction_cases})
\begin{equation}
s= (q +p_N)^2 \equiv W^2; \ \ t=(p'_N-p_N)^2; \ \  u=(p'_N-q)^2 =(p_\pi-p_N)^2;
\end{equation}
with $s+t+u=2m_N^2 +m_\pi^2-Q^2$, where $m_N$ ($m_\pi$) denotes nucleon (pion) mass.  

\begin{figure}[!ht]
\begin{center}
\includegraphics[width=0.35\textwidth]{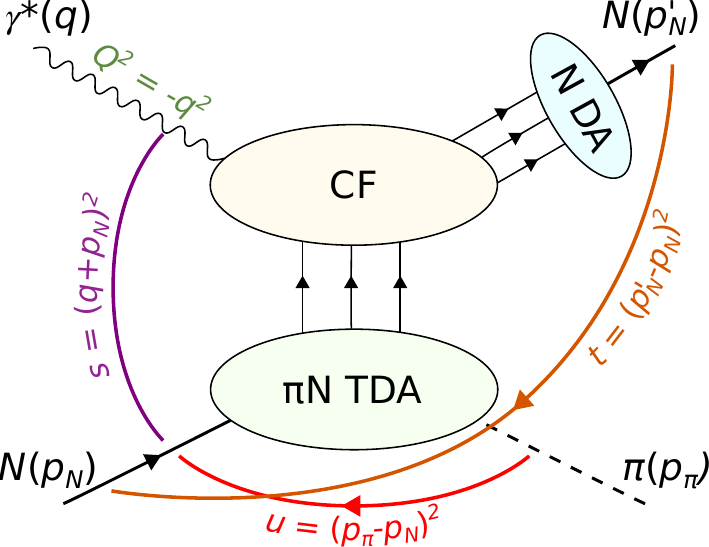}
\end{center}
\caption{Kinematical quantities and the collinear factorization mechanism for $\gamma^* N \to \pi N$ in the near-backward kinematical regime (large $Q^2$, $s$;
fixed $x_B$; $|u| \sim 0$). The lower blob, denoted $\pi N$ TDA, depicts the nucleon-to-pion  transition distribution amplitude; $N$DA blob depicts the
nucleon distribution amplitude; CF denotes the hard subprocess amplitude (coefficient function).}
\label{Fig_Factorization}
\end{figure}

The collinear factorization mechanism for the reaction (\ref{Reaction_cases}) in the near-backward kinematics is schematically presented in Fig.~\ref{Fig_Factorization}. The amplitude
is expressed as a convolution of a perturbatively calculable hard part, or coefficient function  (CF), with nucleon-to-pion TDAs and nucleon DA. 

We introduce the
pion-nucleon average momentum $P=\frac{p_N+p_\pi}{2}$ and the $u$-channel momentum transfer 
$\Delta=p_\pi-p_N$.
We introduce the
light-cone vectors
$p, n$ ($p^2=n^2=0$)
satisfying
$ 2p \cdot n =1$ and perform the Sudakov decomposition of momenta:
\be
v^\mu=v^+ p^\mu+v^-n^\mu+v^\mu_T; \ \ v^2=v^+v^-+ v_T^2,
\ee
where $v^+=2 (n \cdot v)$;  $v^-=2 (p \cdot v)$; and the transverse directions
are chosen perpendicular to the $z$-axis and satisfy $(p \cdot v_T)=(n \cdot v_T)=0$.
The explicit form of the Sudakov decomposition of the relevant momenta is specified {\it e.g.}
in Sec.~3.1 of Ref.~\cite{Pire:2021hbl}. 

We introduce the $u$-channel skewness variable which characterizes the longitudinal momentum transfer:
\begin{equation}
\xi \equiv  -\frac{(p_\pi-p_N) \cdot n}{(p_N+p_\pi) \cdot n}\,.
\label{Def_xi}
\end{equation}
Within the collinear factorization framework,  neglecting both the pion
and nucleon masses with respect to the hard scale  $Q$ 
and setting
$\Delta_T=0$
results in the approximate expression for the  skewness variable
(\ref{Def_xi}):
\begin{equation}
\xi \simeq \frac{Q^2}{2 W^2-Q^2} \simeq   \frac{x_B}{2 -x_B}.
\label{Xi_collinear}
\end{equation}

The transverse invariant momentum transfer $\Delta_T^2 \le 0$ can be expressed as
\begin{equation}
\Delta_T^2= \frac{1-\xi}{1+\xi} \left( u-2\xi \left[ \frac{m_N^2}{1+\xi}-
\frac{m_\pi^2}{1-\xi} \right] \right).
\label{Def_DT2}
\end{equation}
We introduce the threshold value of the cross-channel invariant momentum transfer $u_0$,
corresponding to the exactly backward scattering ($\Delta_T^2=0$)
\be
u_0=-\frac{2 \xi\left(m_{\pi}^2(1+\xi)-m_N^2(1-\xi)\right)}{1-\xi^2}.
\ee

In order to control the validity of the approximate expression for the skewness variable (\ref{Xi_collinear})
it is instructive to compare the invariant transverse momentum transfer from Eq.~(\ref{Def_DT2})
to its exact value 
\be
\Delta_T^2=-\frac{\Lambda^2\left(s, m_N^2, m_\pi^2\right)}{4 s}\left(1-\cos ^2 \theta_\pi^*\right),
\ee
where
\be
\Lambda(x, y, z)=\sqrt{x^2+y^2+z^2-2 x y-2 x z-2 y z}
\ee
is the Källen kinematical function; and the cosine of the $\gamma^* N$ CMS 
scattering angle $\theta_\pi^*$ is expressed as
\be
\cos \theta_\pi^*=-\frac{2 s\left(u-m_N^2-m_\pi^2\right)+\left(s+m_\pi^2-m_N^2\right)\left(s+m_N^2+Q^2\right)}{\Lambda\left(s, m_N^2, m_\pi^2\right) \Lambda\left(s,-Q^2, m_N^2\right)} .
\ee

\subsection{Definition and basic properties of nucleon-to-meson TDAs}

For definiteness, we present the definition of the leading twist-$3$ proton-to-$\pi^0$ TDA:
\be
\begin{aligned}
& 4 
(P \cdot n)^3 \int\left[\prod_{j=1}^3 \frac{d \lambda_j}{2 \pi}\right] e^{i \sum_{k=1}^3 x_k \lambda_k(P \cdot n)} 
\left\langle\pi^0\left(p_\pi\right)\right| \widehat{O}_{\rho \tau \chi}^{\,uud}\left(\lambda_1 n, \lambda_2 n, \lambda_3 n\right)\left|N^p\left(p_N, s_N\right)\right\rangle \\
& =\delta\left(x_1+x_2+x_3-2 \xi\right) i \frac{f_N}{f_\pi m_N} \sum_s\left(s^{\pi N}\right)_{\rho \tau, \chi} H_s^{\pi N}\left(x_1, x_2, x_3, \xi, u ; \mu^2\right).
\label{Def_piN_TDA}
\end{aligned}
\ee
Here $f_\pi=93$~MeV is the pion weak decay constant and the constant $f_N=5.0 \times 10^{-3} $~GeV$^2$ \cite{Chernyak:1987nv} determines the normalization of the nucleon light-cone wave function at the origin.  
The sum in 
(\ref{Def_piN_TDA}) 
stands over the set of leading-twist-$3$
Dirac structures
\be
(s^{\pi N})_{\rho \tau, \chi}=\left\{ (v_{1,2}^{\pi N} )_{\rho \tau, \chi}, (a_{1,2}^{\pi N})_{\rho \tau, \chi}, (t_{1,2,3,4}^{\pi N})_{\rho \tau, \chi}\right\}.
\ee
There exist several choices of the Dirac structures. In this
paper we employ the set of Eq.~(12) of Ref.~\cite{Pire:2011xv} constructed from the fully covariant components (vectors $P$, $\Delta$, as well as the Dirac matrices and the charge conjugation matrix $C$ and the nucleon Dirac spinor). The advantage of this set over the set of Dirac structures originally introduced in Ref.~\cite{Lansberg:2007ec}, involving vectors $p$, $n$, and $\Delta_T$, is the simple manifestation of the polynomiality property of the $x_i$-Mellin moments of TDAs. The relations between $\pi N$ TDAs corresponding to these two choices of the Dirac structures is specified in Eq. (4.59) of Ref.~\cite{Pire:2021hbl}.  
We employ the shortened notations for the set of $8$ invariant leading twist-$3$ $\pi N$ TDAs
\be
H^{\pi N}=\left\{V_{1,2}^{\pi N}, A_{1,2}^{\pi N}, T_{1,2,3,4}^{\pi N}\right\}.
\label{Set_TDA}
\ee

Nucleon-to-meson TDAs depend on three light-cone momentum fraction variables 
$x_i$
  with 
$i=\{1,\,2,\, 3\}$; 
the skewness variable 
$\xi \in [0;\,1]$; 
the invariant momentum transfer 
$\Delta^2 \equiv u$; 
and the factorization scale 
$\mu^2$, 
which will typically be omitted for brevity.  In what follows, we will also omit the dependence on $u$ wherever it is not relevant to the discussion.

\begin{figure}[!ht]
\begin{center}
\includegraphics[width=0.9\textwidth]{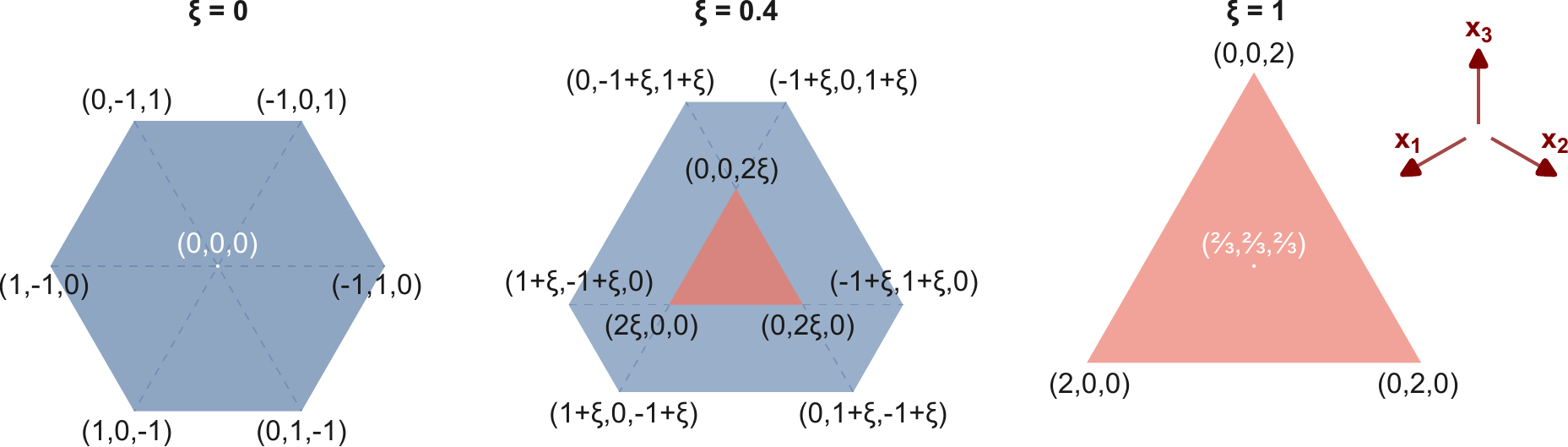}
\end{center}
\caption{Physical domains for TDAs in the barycentric coordinates for $\xi = 0$ (left);
$0 < \xi <1$ (centre)
and $\xi=1$ (right). Arrows show the
positive directions for momentum fractions $x_i$ on the barycentric plane. Note that in the case $\xi=0$ the equilateral triangle corresponding to the
ERBL-like domain shrinks to a point.
Fig.}
\label{Fig_Domains}
\end{figure}

The light-cone momentum  fraction variables $x_i$ specify
the distribution of the longitudinal momentum between the three active
partons, and satisfy the momentum conservation constraint $x_1+x_2+x_3=2 \xi$.
The support domain of TDAs in the variables $x_i$ is given by the intersection of the $3$ stripes 
$-1+\xi \le x_i \le 1+\xi$.
A convenient way to depict this domain  is to employ the barycentric coordinates, see Ref.~\cite{Pire:2010if}. The values of momentum fractions $x_i$
are specified by oriented distances from a point on the plane to three sides of the equilateral triangle marked in red in Fig.~\ref{Fig_Domains}.
The height of this equilateral triangle is defined by
the momentum conservation constraint $x_1+x_2+x_3=2 \xi$.

The shape of the support domain depends on the value of $\xi$. Particularly, for $\xi=0$ this  domain turns to be a regular hexagon, 
while for $\xi=1$ it reduces to an equilateral triangle.
For $0<\xi<1$ we identify the Efremov–Radyushkin–
Brodsky–Lepage (ERBL)-like domain corresponding to the equilateral triangle with the height $2 \xi$ and two types of the Dokshitzer-Gribov-Lipatov-Altarelli-Parisi (DGLAP)-like domains bounded by the lines $x_i=-1+\xi$; $x_i=0$; $x_i=1+\xi$.
The scale dependence of TDAs within the ERBL-like and DGLAP-like support domains is
governed, respectively, by the ERBL-type and DGLAP-type evolution equations, see Ref.~\cite{Pire:2005ax}.

Another convenient way to deal with the arguments of the TDAs is to use the so-called quark-diquark coordinates%
\footnote{We do not attribute any dynamical meaning to the quark-diquark coordinates introduced here; they merely provide a convenient parametrization of the variables $x_1, x_2, x_3$.}
introducing two independent variables (instead of $3$ with one constraint). There exist $3$ equivalent choices
$i = 1,\, 2,\, 3$ of
quark-diquark coordinates, depending on which pair of longitudinal  momenta is selected to constitute the momentum of a diquark:
\begin{equation}
w_{i}=x_{i}-\xi; \ \ \  v_{i}=\frac{1}{2} \sum_{k, l=1}^{3}
\varepsilon_{i k l} x_{k}\,,
\label{Def_qDq_coord}
\end{equation}
where $\varepsilon_{ikl}$ is the totally antisymmetric tensor ($\varepsilon_{123}=1$).
The support domain of TDAs in terms of quark-diquark coordinates can be parametrized as
\begin{equation}
-1 \leq w_{i} \leq 1 ; \quad-1+| \xi-\xi_{i}^{\prime}|   \leq v_{i} \leq 1-| \xi-\xi_{i}^{\prime}|,
\label{Support_TDA_wv}
\end{equation}
where
\begin{equation}
\xi_{i}^{\prime} \equiv \frac{\xi-w_{i}}{2}\,
\label{Def_xip_i}
\end{equation}
is the skewness variable that specifies the fraction of the longitudinal momentum
carried by a diquark system, which complements the $i$-th quark,

Similarly to GPDs, TDAs satisfy a highly non-trivial polynomiality property, which is a consequence of the underlying Lorentz invariance of the framework.
The Mellin moments of TDAs in variables $x_i$ computed over the $\xi$-dependent support domain turn to be polynomials \footnote{This simple manifestation of polynomiality is ensured for the set of TDAs defined 
with the set of the Dirac structures of Ref.~\cite{Pire:2011xv}. Otherwise, extra non-polynomial kinematical factors may arise for the Mellin moments of TDAs, see discussion in Ref.~\cite{Pire:2011xv}.} of specific order of the skewness variable $\xi$:
\begin{eqnarray}
 &&
\int_{-1+\xi}^{1+\xi}  dx_1 \int_{-1+\xi}^{1+\xi}  dx_2 \int_{-1+\xi}^{1+\xi}  dx_3
\delta(x_1+x_2+x_3-2\xi)
 x_1^{n_1} x_2^{n_2} x_3^{n_3}
H(x_1,x_2,x_3,\xi)  
= P_{n_1+n_2+n_3(+1)}(\xi)\,.
\label{Def_Mellin_moments_TDAs}
\end{eqnarray}
This polynomiality property is known as the Cavalieri condition
\cite{Gelfand_Graev}
in the theory of the Radon transform, which plays a central role in tomographic imaging.

The coefficients of the powers of $\xi$ in the polynomials $P_{n_1+n_2+n_3(+1)}(\xi)$ depend
on $u$ (as well as on the factorization scale $\mu^2$) and can be expressed through nucleon-to-meson
transition form factors of the local twist-$3$ operators:
\be
\widehat{O}_{\rho \tau \chi}^{\mu_1 \ldots \mu_{n_1}, \nu_1 \ldots \nu_{n_2}, \sigma_1 \ldots \sigma_{n_3}}(0)=\left[i \vec{D}^{\mu_1} \ldots i \vec{D}^{\mu_{n_1}} \Psi_\rho\right]\left[i \vec{D}^{\nu_1} \ldots i \vec{D}^{\nu_{n_2}} \Psi_\tau\right]\left[i \vec{D}^{\sigma_1} \ldots i \vec{D}^{\sigma_{n_3}} \Psi_\chi\right],
\label{O_Tw3_local}
\ee
where the covariant derivative is $\vec{D}^\mu=\vec{\partial}^\mu-i g A^{a \mu} \frac{\lambda^a}{2}$,
with $\lambda^a$, $a=1,\ldots,8$, being the Gell-Mann matrices. The nucleon-to-meson transition form factors of the operators 
(\ref{O_Tw3_local}) 
can be studied with the methods of the lattice QCD extending the framework
already existing for PDFs \cite{Lin:2017snn}, GPDs \cite{Hagler:2009ni,Alexandrou:2019ali} and baryon DAs \cite{QCDSF:2008qtn}. Moreover, lattice studies of 
matrix elements of local $3$-quark operators were performed in Refs.~\cite{Aoki:2017puj,Yoo:2021gql} 
in the context of calculation of the proton decay matrix element in grand-unified theories.

\subsection{Amplitudes}
\label{Sec_Amp}
With the reaction mechanism of Fig.~\ref{Fig_Factorization}, 
to the leading twist-$3$, and leading order (LO) in strong coupling, $\alpha_s \equiv \frac{g^2}{4 \pi}{}$, amplitudes of hard exclusive backward meson production reactions $\gamma^* N \to N' {\cal M} $
require calculation of the same 21 diagrams as occurring in the
textbook pQCD formulation of the nucleon electromagnetic form factors \cite{Chernyak:1984bm}.
The explicit expressions for the diagrams contributing to the hard scattering amplitude of backward pion 
\cite{Lansberg:2007ec}
and vector meson 
\cite{Pire:2015kxa}
production  
are summarized  in Tables~2 and 3 of
Ref~\cite{Pire:2021hbl}, see also Erratum to Ref.~\cite{Pire:2015kxa}.

The leading twist-$3$, LO helicity amplitudes of the reaction (\ref{Reaction_cases}) admit the following parametrization:
\be
\mathcal{M}_{s_N s_N^{\prime}}^{\lambda_\gamma}=\mathcal{C}_\pi \frac{1}{Q^4} \sum_{k=1}^2 \mathcal{S}_{s_N s_N^{\prime}}^{(k) \lambda_\gamma} \mathcal{I}^{(k)}\left(\xi, u\right),
\label{LO_helicity_amplitudes}
\ee
where the spin structures are defined as
\be
\begin{aligned}
& \mathcal{S}_{s_N s_N^{\prime}}^{(1) \lambda_\gamma} \equiv \bar{U}\left(p_N^{\prime}, s_N^{\prime}\right) \hat{\mathcal{E}}\left(q, \lambda_\gamma\right) \gamma^5 U\left(p_N, s_N\right); \\
& \mathcal{S}_{s_N s_N^{\prime}}^{(2) \lambda_\gamma} \equiv \frac{1}{m_N} \bar{U}\left(p_N^{\prime}, s_N^{\prime}\right) \hat{\mathcal{E}}\left(q, \lambda_\gamma\right) \hat{\Delta}_T \gamma^5 U\left(p_N, s_N\right).
\end{aligned}
\ee
Here we employ the Dirac hat notation $\hat{v} \equiv v_\mu \gamma^\mu$; and 
$\mathcal{E}\left(q, \lambda_\gamma\right)$ is the polarization vector of the virtual photon with helicity $\lambda_\gamma$.
The overall normalization constant ${\mathcal{C}}_\pi$ is expressed as
\begin{equation}
{\mathcal{C}}_\pi = -i \frac{(4 \pi \alpha_s)^2 \sqrt{4 \pi \alpha_{\rm em}} f_N^2 }{54 f_\pi },
\label{eq:Cpi}
\end{equation}
with $\alpha_{\rm em} \simeq \frac{1}{137}$ being the fine structure constant. 

The elementary amplitudes are expressed as integral convolutions,  both in TDA arguments
$x_i$
and nucleon DA arguments
$y_i$,
of the combination of
$T_\alpha^{(k)}$, $k=1,\,2$: 
\be
&&
\mathcal{I}^{(k)}(\xi,u) =
 \int_{-1+\xi}^{1+\xi} d_3 x \, 
 \delta(
\sum_{j=1}^3 x_j 
 -2\xi)
 \int_{-1}^1 d_3y \, \delta( \sum_{l=1}^3 y_l-1) \left( 2 \sum_{\alpha=1}^7 T^{(k)}_\alpha  +  \sum_{\alpha=8}^{14} T^{(k)}_\alpha
 \right),
 \label{Amplitude_M}
\ee
where the index $ \alpha $ refers to the diagram number%
\footnote{The contribution of diagrams with $\alpha=15,\, \ldots, 21$ coincides with that of $7$ first diagrams, hence the factor 2 in parenthesis in the r.h.s. of (\ref{Amplitude_M}). }, with 
$\alpha = 1,\, \ldots,\, 14 $, as labeled in Ref.~\cite{Chernyak:1984bm}.

Each integrand
$T^{(k)}_\alpha$ 
has the following factorized structure
\be
T^{(k)}_\alpha  \equiv &&K_\alpha(x_1,x_2,x_3,\xi) \times
\left[ \text{Combination of} \; \pi N \; \text{TDAs} \; H^{\pi N} (x_1,x_2,x_3,\xi, u), \; \text{Eq.~(\ref{Set_TDA})}\right] \nn \\ &&
\times Q_\alpha (y_1,y_2,y_3) \times
\left[ \text{Combination of nucleon DAs} \; \Phi^N(y_1,y_2,y_3),\; \text{Eq.~(\ref{Set_of_DAs})} \right].
\label{Structure_of_Amplitude}
\ee

Here 
$K_\alpha(x_1,x_2,x_3,\xi)$ 
and 
$Q_\alpha (y_1,y_2,y_3)$
denote singular convolution kernels arising from the denominators of collinear parton propagators present in the corresponding diagram.

The convolution kernels $ Q_\alpha(y_1, y_2, y_3) $ are analogous to those encountered in the case of nucleon form factors. The evaluation of the corresponding convolution integrals poses no particular difficulty, as the singularities lie at the boundaries of the domain, where the nucleon distribution amplitudes vanish. As a result, the $ y $-convolution integrals are regular and do not contribute any imaginary part to the amplitude.

To facilitate the treatment of the hard kernels 
$K_\alpha(x_1, x_2, x_3, \xi)$, 
it is convenient to switch to the quark–diquark coordinates (\ref{Def_qDq_coord}). 
For each diagram, there exists a preferred choice 
$i = 1, 2, 3$
of the quark–diquark coordinates 
$ (w_i, v_i) $, 
in which the corresponding convolution kernel 
$ K_\alpha$ simplifies to one of the following canonical forms:
\be
&&
K_I^{( \pm, \pm)}\left(w_i, v_i, \xi \right)   =\frac{1}{\left(w_i \pm \xi \mp i 0\right)} \frac{1}{\left(v_i \pm \xi_i^{\prime} \mp i 0\right)}; \nn \\ &&
K_{I I}^{(-, \pm)}\left(w_i, v_i, \xi\right)   =\frac{1}{\left(w_i-\xi+i 0\right)^2} \frac{1}{\left(v_i \pm \xi_i^{\prime} \mp i 0\right)},
\label{Def_kernels_KI_II}
\ee
where $\xi_i^{\prime}$ are defined in (\ref{Def_xip_i}).

The singularities of the hard kernels  
(\ref{Def_kernels_KI_II}) 
are located along the cross-over trajectories 
$ w_i = -\xi$, 
$ v_i = \pm \xi'$, 
which separate the ERBL-like and DGLAP-like support regions of TDAs, as well as along the lines $ w_i = \xi$, which entirely lie within the DGLAP-like domains. As a result, the convolution integrals over 
$x_i $ involving TDAs in Eq.~(\ref{Amplitude_M})
may generate a non-zero imaginary part in the amplitude. This is due to the fact that TDAs do not necessarily vanish on the cross-over trajectories $x_i = 0$, which separate the ERBL-like and DGLAP-like regimes, nor on the lines $ x_i = 2\xi$.

To ensure the consistency of the principal value regularization prescriptions in~(\ref{Def_kernels_KI_II}), the TDAs must be continuous at the cross-over lines 
$x_i=0$ Additionally, to guarantee the convergence of integrals involving the $K_{II}^{(-, \pm)} $ kernels,   TDAs must be at least once differentiable at the lines $x_i = 2\xi $.

\subsection{Cross-sections and polarization observables}
\label{Sec_CS}

In this subsection we present our conventions for the cross-section of backward pion
electroproduction reaction $e N \to e'N'\pi$. Following Refs.~\cite{Arens:1996xw,Diehl:2005pc} we  specify the link between the
experimentally  measured $5$-fold 
electroproduction cross-section to the cross-sections of the hard subprocess 
$\gamma^*N \to N' \pi$.
We also introduce new polarization observables, Double Spin Asymmetries (DSAs), which turn to be
non-vanishing to the leading twist accuracy with the framework involving TDAs. We argue that accessing these DSAs  can be helpful in establishing the validity of the
collinear factorized description involving TDAs for backward reactions. Moreover, DSAs are highly sensitive to the dependence of TDAs on $\xi$ and $u$, potentially offering a means to discriminate between different phenomenological models of TDAs.

We consider the exclusive electroproduction of pions off nucleons within the one-photon exchange approximation:
\be
e\left(k, s_e\right)+N\left(p_N, s_N\right) \rightarrow\left(\gamma^*\left(q, \lambda_\gamma\right)+N\left(p_N, s_N\right)\right)+e\left(k^{\prime}, s_e^{\prime}\right) \rightarrow e\left(k^{\prime}, s_e^{\prime}\right)+\pi\left(p_\pi\right)+N^{\prime}\left(p_N^{\prime}, s_N^{\prime}\right).
\label{Reaction_electroproduction}
\ee
Here $s_N$, $s'_N$ ($s_e$, $s'_e)$ denote the polarization of the initial and final
nucleons (leptons); and $\lambda_\gamma$ is the polarization of the virtual photon.

We employ the standard variables
\be
x_B \equiv \frac{Q^2}{2 p_N \cdot q}=\frac{Q^2}{Q^2+s-m_N^2} ; \quad y \equiv \frac{p_N \cdot q}{p_N \cdot k}=\frac{s+Q^2-m_N^2}{
s_{eN}
-m_N^2},
\ee
and introduce the variable 
$s_{eN}=(p_N+k)^2$
to avoid confusion with the Mandelstam variable 
$s=(p+q)^2$
of the hard subprocess~(\ref{Reaction_cases}).

\begin{figure}[!ht]
\begin{center}
\includegraphics[width=0.6\textwidth]{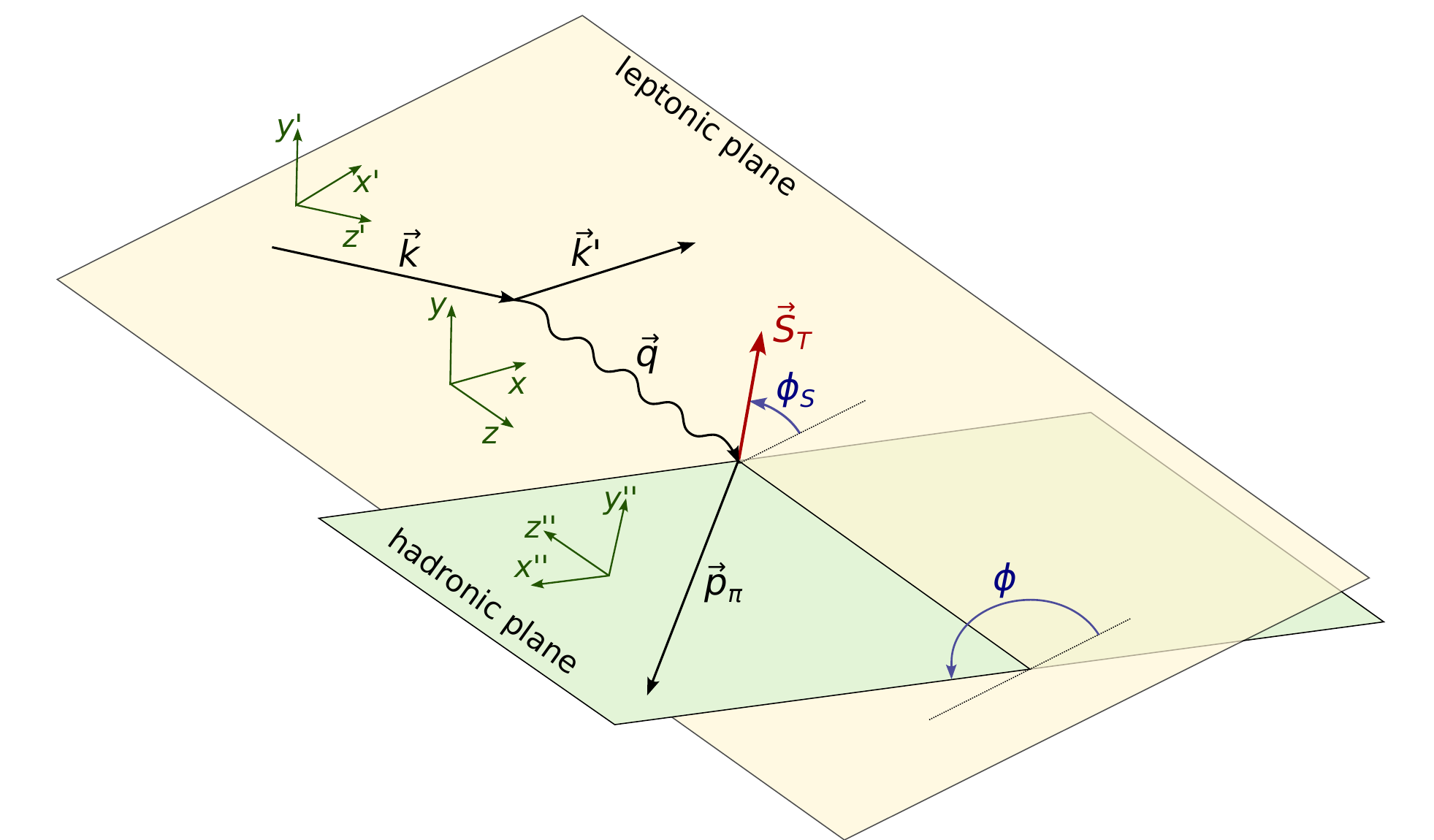}
\end{center}
\caption{ 
Kinematics of the process 
(\ref{Reaction_electroproduction}) in the target rest frame and
definition of the $C$, $C'$ 
and $C''$ 
reference systems. $\vec{p}_{\pi \,T}$ and $\vec{S}_T$ respectively are the components of $\vec{p_\pi}$ and $\vec{S}$ perpendicular to $\vec{q}$. (The target spin vector $\vec{S}$ is not shown.) $\phi$ and $\phi_S$ respectively are the azimuthal angles of $\vec{p}_\pi$ and $\vec{S}$ in the coordinate system with axes $x$, $y$, $z$, in accordance with the Trento conventions \cite{Bacchetta:2004jz}. 
}
\label{Fig_Planes}
\end{figure}

As it is well known, within the one-photon approximation, the cross-section of the electroproduction reaction can be written as 
a convolution of the leptonic tensor $L^{\nu \mu}$ and hadronic tensor $W_{\mu \nu}$:
\be
d \sigma(e N \rightarrow e' N' \pi) \propto L^{\nu \mu} W_{\mu \nu} \frac{d^3 k^{\prime}}{2 k^{\prime 0}} \frac{d^3 p_\pi}{2 p_\pi^0}.
\ee
The leptonic tensor reads
\be
L^{\nu \mu}=k^{\prime \nu} k^\mu+k^\nu k^{\prime \mu}-\left(k^{\prime} \cdot k\right) g^{\nu \mu}{-}i P_{\ell} \epsilon^{\nu \mu \alpha \beta} q_\alpha k_\beta
\label{Def_lept_tensor}
\ee
with the convention $\epsilon^{0123}=+1$ of Itzykson and Zuber
\cite{Itzykson}
for the Levi-Civita tensor. The lepton beam polarizations $P_\ell=\pm 1$, respectively, corresponds to a purely right-handed (left-handed) beam. As pointed out in Ref.~\cite{Arens:1996xw}, the leptonic tensor (\ref{Def_lept_tensor}) can be expressed as a linear combination of terms ${\cal E}^\nu_n {\cal E}^{\mu *}_m$ built of the virtual photon polarization vectors with definite helicities $m,n= \pm 1,\,0$ with the coefficients depending on $Q^2$, azimuthal angle $\phi$, $P_\ell$ and the polarization parameter $\varepsilon$, 
\be
\varepsilon=\frac{2(1-y)-2 x_B y\, m_N^2\left( 
s_{eN}
-m_N^2\right)^{-1}}{1+(1-y)^2+2 x_B y\, m_N^2
\left(
s_{eN}
-m_N^2
\right)^{-1}}.
\ee

The hadronic tensor
\be
W_{\mu \nu}=\sum_{\text {spins }} \langle N (p_N^{\prime},   s^\prime_N)\, \pi\left(p_\pi\right) | J_\mu(0 |N\left(p_N,s_N\right) \rangle^*
 \langle N (p_N^{\prime},   s^\prime_N)\, \pi\left(p_\pi\right) | J_\nu(0 |N\left(p_N,s_N\right) \rangle,
\ee
with $J_\mu$ denoting the electromagnetic current, encodes the complete nonperturbative QCD information about the hard subprocess~(\ref{Reaction_cases}).

The analysis of target polarization effects in the reaction~(\ref{Reaction_electroproduction}) presents certain subtleties. In the experimental setup, the target polarization is typically defined with respect to the lepton beam direction: either longitudinal or transverse. However, in the theoretical description of the underlying hard $\gamma^* N$ subprocess~(\ref{Reaction_cases}), it is more natural to define polarization states with respect to the direction of the virtual photon momentum. Here we follow this latter convention and define the target nucleon spin in the 
coordinate system $C''$ with axes $x'',\,y'',\,z''$, see Fig.~\ref{Fig_Planes},
\be
\vec{S} 
=\left(\begin{array}{c}
S_T \cos \left(\phi-\phi_S\right) \\
S_T \sin \left(\phi-\phi_S\right) \\
S_L
\end{array}\right),
\label{Def_target_spin}
\ee
with $\phi$ being the angle between the leptonic and hadronic planes and $\phi_S$ denoting the azimuthal angle of the transverse
component of the target nucleon spin within the coordinate system $C$.
The relation between this set  of polarization states with that defined with respect to the coordinate system $C^\prime$, with $z'$ axis aligned with the direction of the lepton beam, has been systematically discussed in Ref.~\cite{Diehl:2005pc}.

The contraction of the leptonic and hadronic tensors $L^{\nu \mu} W_{\mu \nu}$ can be written in terms of
quantities 
\be
\frac{d \sigma_{m n}}{dt}=\sum_{i j} \rho_{j i} \frac{d \sigma_{m n}^{i j}}{dt},  
\ee
where $\rho_{j i}$ is the spin density matrix of the target nucleon,
\be
\rho_{j i}=\frac{1}{2}\left[\delta_{j i}+\vec{S} \cdot \vec{\sigma}_{j i}\right] \stackrel{C^{\prime \prime}}{=} \frac{1}{2}\left(\begin{array}{cc}
1+S_L & S_T \exp \left[-i\left(\phi-\phi_S\right)\right] \\
S_T \exp \left[i\left(\phi-\phi_S\right)\right] & 1-S_L
\end{array}\right), 
\label{Def_rho}
\ee
with $\vec{\sigma} \equiv (\sigma_1,\, \sigma_2, \, \sigma_3)$ denoting the Pauli matrices. 
The spin density matrix (\ref{Def_rho}) is defined with respect to states described by the two-component spinors
\be
\chi_{+\frac{1}{2}}=\binom{1}{0}, \quad \chi_{-\frac{1}{2}}=\binom{0}{1};
\ee
corresponding to definite spin projection 
$+\frac{1}{2}$ and $-\frac{1}{2}$ along the $z''$ axis, and to right- and left-handed nucleon helicity in the $\gamma^* N$ 
center-of-mass frame.

The general structure of the $t$-integrated cross-section 
$ \frac{d^4 \sigma}{d x_B d Q^2 d \phi d \psi} \equiv \int dt \frac{d^5 \sigma}{d x_B d Q^2 dt d \phi d \psi}$
of
the pion electroproduction reaction (\ref{Reaction_electroproduction}) in terms 
of polarized photoabsorption cross-sections and interference 
terms $\sigma_{m n}^{i j}$,
with lower indices $m,\,n=0,\,+1,\,-1$ referring to the virtual photon polarization and upper indices $i,\,j=+\frac{1}{2}, \, -\frac{1}{2}$
referring to the target proton polarization, has been worked out in 
Refs.~\cite{Arens:1996xw,Diehl:2005pc}. Here, following the notations of Ref.~\cite{Diehl:2005pc}, we present
the general structure of the fully unintegrated $5$-fold cross-section of
the reaction (\ref{Reaction_electroproduction})%
\footnote{Ref.~\cite{Diehl:2005pc} employs a convention for the Levi-Civita tensor that differs from the one adopted in this work. Consequently, when comparing with Eq.~(29) of Ref.~\cite{Diehl:2005pc}, the signs of the single-polarization dependent contributions to the cross-section are reversed. In contrast, the signs of the unpolarized and double-polarization dependent terms remain unaffected.}:
\be
\begin{aligned}
& {\left[\frac{\alpha_{\mathrm{em}}}{8 \pi^3}\right.}\left.\frac{y^2}{1-\varepsilon} \frac{1-x_B}{x_B} \frac{1}{Q^2}\right]^{-1} \frac{d^5 \sigma}{d x_B d Q^2 dt d \phi d \psi} \\
&=  \underbrace{
\frac{1}{2}\left( \frac{d \sigma_{++}^{++}}{dt}+\frac{d \sigma_{++}^{--}}{dt}\right)
}_{d\sigma_T/dt}
+\varepsilon 
\frac{d \sigma_{00}^{++}}{dt}
-\varepsilon \cos (2 \phi) \operatorname{Re} \frac{ d  \sigma_{+-}^{++}}{dt}-\sqrt{\varepsilon(1+\varepsilon)} \cos \phi \operatorname{Re}\left( \frac{d \sigma_{+0}^{++}}{dt}+\frac{d \sigma_{+0}^{--}}{dt}\right) \\
&
{+}P_{\ell} 
\sqrt{\varepsilon(1-\varepsilon)} \sin \phi \operatorname{Im}
\left( \frac{d \sigma_{+0}^{++}}{dt}+\frac{d \sigma_{+0}^{--}}{dt}\right)
\\
&
{+}S_L\left[\varepsilon \sin (2 \phi) \operatorname{Im} \frac{d \sigma_{+-}^{++}}{dt}+\sqrt{\varepsilon(1+\varepsilon)} \sin \phi \operatorname{Im}\left( \frac{d \sigma_{+0}^{++}}{dt}-\frac{d \sigma_{+0}^{--}}{dt}\right)\right] \\
&+S_L P_{\ell}\Big[ 
\underbrace{ 
{\sqrt{1-\varepsilon^2} \frac{1}{2}\left( \frac{d \sigma_{++}^{++}}{dt}-\frac{d \sigma_{++}^{--}}{dt}\right)}
}_{A_{LL}^\mathrm{const.}}
-\sqrt{\varepsilon(1-\varepsilon)} \cos \phi \operatorname{Re}\left( \frac{d \sigma_{+0}^{++}}{dt}-\frac{d \sigma_{+0}^{--}}{dt}\right)\Big] \\
&
{+}S_T\Big[
\underbrace{ 
\operatorname{Im}\left( \frac{d \sigma_{++}^{+-}}{dt}
+
\varepsilon  
\frac{d \sigma_{00}^{+-}}{dt}
\right)
}_{A_{UT}^{\sin(\phi-\phi_s)}}\sin \left(\phi-\phi_S\right)
+\frac{\varepsilon}{2} \sin \left(\phi+\phi_S\right) \operatorname{Im} \frac{d \sigma_{+-}^{+-}}{dt}
\\
&
\left.
+\frac{\varepsilon}{2} \sin \left(3 \phi-\phi_S\right) \operatorname{Im} \frac{d \sigma_{+-}^{-+}}{dt} 
+\sqrt{\varepsilon(1+\varepsilon)} \sin \phi_S \operatorname{Im} \frac{ d \sigma_{+0}^{+-}}{dt}+\sqrt{\varepsilon(1+\varepsilon)} \sin \left(2 \phi-\phi_S\right) \operatorname{Im} \frac{d\sigma_{+0}^{-+}}{dt} \right] \\
& +S_T P_{\ell}\Big[ 
\underbrace{ 
{\sqrt{1-\varepsilon^2}  \operatorname{Re} \frac{d \sigma_{++}^{+-}}{dt} }
}_{A_{LT}^{\cos(\phi-\phi_s)}}\cos \left(\phi-\phi_S\right)  
\\
&\left.\quad-\sqrt{\varepsilon(1-\varepsilon)} \cos \phi_S \operatorname{Re} \frac{ d\sigma_{+0}^{+-}}{dt}-\sqrt{\varepsilon(1-\varepsilon)} \cos \left(2 \phi-\phi_S\right) \operatorname{Re} \frac{d \sigma_{+0}^{-+}}{dt}\right].
\label{CS_Diehl}
\end{aligned}
\ee
Here $P_{\ell}=\pm 1$ corresponds to a purely right-handed (left-handed) lepton beam, {\it cf.} Eq.~(\ref{Def_lept_tensor}); the components of the target nucleon spin (\ref{Def_target_spin}) are defined in the usual $C''$ frame, while the angle $\psi$ is the azimuthal angle of the transverse component of target nucleon spin within the reference frame $C'$ aligned with the lepton beam direction, see Fig.~\ref{Fig_Planes}. To the leading order in $1/Q$-expansion, $\psi =\phi_S$, see discussion in Ref.~\cite{Diehl:2005pc}. 
Also, for clarity of the subsequent discussion, we mark with underbraces the specific terms in the cross-section~(\ref{CS_Diehl}) that receive contributions at leading twist-$3$ accuracy from the collinear factorization mechanism involving $\pi N$ TDAs.

Polarized photoabsorption cross-sections and interference 
terms $ \frac{d \sigma_{m n}^{i j}}{dt}$ occurring in the cross-section 
formula (\ref{CS_Diehl})
can be put in correspondence with the helicity amplitudes $\mathcal{M}^{\lambda_\gamma=m}_{s_N=i \,;\, s'_N }$
of the hard subprocess $\gamma^* N \to \pi N'$:
\be
\frac{d \sigma_{m n}^{i j}\left(x_B, Q^2,t\right)}{dt} \propto  
\sum_{s'_N}\left(\mathcal{M}^{\lambda_\gamma=m}_{s_N=i \,;\, s'_N }\right)^* \mathcal{M}^{\lambda_\gamma=n}_{s_N=j \,;\, s'_N },
\label{Def_sigma_mn_ij}
\ee
see Refs.~\cite{Burkert:2004sk,Kim:2013gqa,MurdohThesis}.
Due to hermiticity and parity invariance they satisfy 
\be
\frac{ d \sigma_{n m}^{j i}}{dt}=\left( \frac{d\sigma_{m n}^{i j}}{dt}\right)^*, \quad \frac{ d \sigma_{-m-n}^{-i-j}}{dt}=(-1)^{m-n-i+j} \frac{\sigma_{m n}^{i j}}{dt}.
\ee
Note that $\frac{d\sigma_{n m}^{j i}}{dt}$ depend on the kinematical variables $x_B$, $Q^2$ and $t$ (or $u$), whereas the dependence on $\varepsilon$ and $\phi$ is contained in 
the leptonic tensor (\ref{Def_lept_tensor}), and the dependence on target spin is encoded in the spin-density matrix $\rho_{ij}$ (\ref{Def_rho}).

The terms in the second line of Eq.~(\ref{CS_Diehl}), describing an unpolarized beam and target, are usually presented as the spin averaged
cross-section:
\be
&&
\frac{d \sigma_0}{dt}=
\frac{1}{2}\left( \frac{d \sigma_{++}^{++}}{dt}+\frac{d \sigma_{++}^{--}}{dt}\right)
+\varepsilon 
\frac{d \sigma_{00}^{++}}{dt}
-\varepsilon \cos (2 \phi) \operatorname{Re} \frac{ d  \sigma_{+-}^{++}}{dt}-\sqrt{\varepsilon(1+\varepsilon)} \cos \phi \operatorname{Re}\left( \frac{d \sigma_{+0}^{++}}{dt}+\frac{d \sigma_{+0}^{--}}{dt}\right) \nn \\ &&
\equiv 
\frac{d \sigma_T}{dt}+\varepsilon \frac{d\sigma_L}{dt}
{+}\varepsilon \cos (2 \phi) \frac{d  \sigma_{T T}^{\cos (2 \phi)}}{dt} {+} \sqrt{ 
{2}
\varepsilon(1+\varepsilon)} \cos \phi \frac{d \sigma_{L T}^{\cos \phi}}{dt}.
\label{Def_sigma0}
\ee
The first two terms in r.h.s. of (\ref{Def_sigma0}) are often denoted as
$
 \frac{d \sigma_T}{dt}+\varepsilon \frac{d\sigma_L}{dt} \equiv \frac{d \sigma_U}{dt}$,
where the subscript $U$ refers to the ``unseparated'' cross-section.

Within the collinear factorization framework involving $\pi N$ TDAs, with the reaction mechanism 
depicted in Fig.~\ref{Fig_Factorization}, only few 
 photoabsorption cross-sections and interference 
terms $\sigma_{m n}^{i j}$ 
(\ref{Def_sigma_mn_ij})
in (\ref{CS_Diehl})
obtain contributions to the leading twist-$3$ accuracy. These turn to be those involving the transverse polarization
of the virtual photon
$\frac{d \sigma_{++}^{++}}{d t}$, $\frac{d \sigma_{++}^{--}}{d t}$  and
$\frac{d \sigma_{++}^{+-}}{d t}$.
For the unpolarized cross-section 
(\ref{Def_sigma0})  
this leads to dominance of the transverse cross-section  
$\frac{d \sigma_T}{d t}$.
To the leading order in $\alpha_s$, the leading twist-$3$ unpolarized cross-section
is expressed in terms of the convolution integrals ${\cal I}^{(1,2)}$ 
(\ref{Amplitude_M}) 
occurring in the
parametrization of helicity amplitudes (\ref{LO_helicity_amplitudes}):
\begin{equation}
\frac{d \sigma_T}{d t}=\left|\mathcal{C}_\pi\right|^2 \frac{1}{Q^6} \frac{1}{32 \pi \Lambda\left(s, m_N^2,-Q^2\right)\left(s-m_N^2\right)} \frac{1+\xi}{\xi}\left(\big|\mathcal{I}^{(1)}(\xi,u)\big|^2-\frac{\Delta_T^2}{m_N^2}\big|\mathcal{I}^{(2)}(\xi,u)\big|^2\right),
\label{eq:dsigmadt}
\end{equation}
where $\mathcal{C}_\pi$ is defined in (\ref{eq:Cpi}). We also provide the expression for the LO leading twist-$3$ cross-section
of  $\gamma^*_TN \to \pi N'$  differential in the solid angle of the produced pion in $\gamma^* N$ CMS  ($d\Omega_\pi \equiv d \cos \theta_\pi d \phi_\pi$)
\be
\frac{d^2 \sigma_T}{d \Omega_\pi}=\left|\mathcal{C}_\pi\right|^2 \frac{1}{Q^6} \frac{\Lambda\left(s, m_N^2, m_\pi^2\right)}{128 \pi^2 s\left(s-m_N^2\right)} \frac{1+\xi}{\xi}\left(\left|\mathcal{I}^{(1)}(\xi,u)\right|^2-\frac{\Delta_T^2}{m_N^2}\left|\mathcal{I}^{(2)}(\xi,u)\right|^2\right),
\label{eq:dsigmadOmega_pi}
\ee
where the connection to (\ref{eq:dsigmadt}) is established with help of the familiar change of variables 
\be
d t=\frac{\Lambda\left(s, m_N^2,-Q^2\right) \Lambda\left(s, m_N^2, m_\pi^2\right)}{2 s} d \cos \theta_\pi,
\ee
and the integration over the pion azimuthal angle $\phi_\pi$ is trivial. 

Apart from the unpolarized cross-section (\ref{eq:dsigmadt}), one may construct several polarization observables which 
obtain leading twist contributions within the reaction mechanism involving $\pi N$ TDAs. We introduce shortened notations
$\rightleftarrows$ 
for the longitudinal beam and 
$\mathrel{\substack{\Rightarrow \\[-0.5ex] \Leftarrow}}$
for longitudinal 
target polarizations; and $\Uparrow \Downarrow$
for the transverse target polarizations.

These include the transverse target spin single asymmetry \cite{Lansberg:2010mf,Pire:2012nz}
\be
A_{UT}
=
 \frac{d\sigma^{\Uparrow}-d\sigma^{\Downarrow}}{d\sigma^{ \Uparrow}+d\sigma^{ \Downarrow}}\,, 
\label{Def_ASTSA} 
\ee
sensitive to 
the $\sin \left(\phi-\phi_S\right)$-modulated term in the 6th line of the master cross-section 
equation~(\ref{CS_Diehl}).
There are also two double spin asymmetries, that were previously not considered in the literature.
The longitudinal-beam-longitudinal-target double spin asymmetry,
\be
A_{LL}
= \frac{\left(d\sigma^{\rightarrow \Rightarrow}-d\sigma^{\leftarrow \Rightarrow}\right)-\left(d\sigma^{\rightarrow \Leftarrow}-d\sigma^{\leftarrow \Leftarrow}\right)}{
 d\sigma^{\rightarrow \Rightarrow}+d\sigma^{\rightarrow \Leftarrow}+d\sigma^{\leftarrow \Leftarrow}+d\sigma^{\leftarrow \Rightarrow}
}\,, 
\label{Def_ALLDSA}
\ee
originates from the $\phi$-independent 
term in the 5th line of (\ref{CS_Diehl}).
Finally, the longitudinal-beam-transverse-target  double spin asymmetry, 
\be
A_{LT} =
\frac{\left(d\sigma^{\rightarrow \Uparrow}-d\sigma^{\rightarrow \Downarrow}\right)+\left(d\sigma^{\leftarrow \Downarrow}-d\sigma^{\leftarrow \Uparrow}\right)}{d\sigma^{\rightarrow \Uparrow}+d\sigma^{\leftarrow \Downarrow}+d\sigma^{\rightarrow \Downarrow}+d\sigma^{\leftarrow \Uparrow}}\,,
 \label{Def_ALTDSA} 
\ee
arises from the $\cos \left(\phi-\phi_S\right)$-modulated term in the 8th  line of (\ref{CS_Diehl}).

In order to single out the polarization observables 
(\ref{Def_ASTSA}), (\ref{Def_ALLDSA}), (\ref{Def_ALTDSA}) 
we employ the following set of projection operators
constructed with help of techniques of Ref.~\cite{Kleiss:1985yh}:
\bi
\item  Longitudinal target spin asymmetry:
\be
U(p_N, h_N) \bar{U}(p_N, h_N)-U(p_N, -h_N) \bar{U}(p_N, -h_N)= h_N (\hat{p}_N+m_N) \gamma_5,
\ee
where $h_N$ is the target nucleon helicity defined in $\gamma^*N$ center-of-mass system.

\item  Transverse target spin asymmetry:
\be
U(p_N, s_{N_T}) \bar{U}(p_N, s_{N_T})-U(p_N, -s_{N_T}) \bar{U}(p_N, -s_{N_T})=   (\hat{p}_N+m_N) \gamma_5 \hat{S}_{T};
\ee
\ei

To the leading twist-$3$ accuracy, this results in the following expressions for the single 
transverse target spin asymmetry 
(\ref{Def_ASTSA})
in terms of the convolution integrals ${\cal I}^{(1,2)}$ 
(\ref{Amplitude_M}):
\begin{equation}
A_{UT}^{\sin(\phi-\phi_s)}= 
K
\frac
{
 \mathrm{Im} (\mathcal{I}^{(1)}(\xi, u)\mathcal{I}^{(2)*}(\xi, u))
}
{
|\mathcal{I}^{(1)}(\xi, u)|^2
- \frac{\Delta_T^2}{m_N^2}| \mathcal{I}^{(2)}(\xi, u)|^2
},
\label{eq:STSA}
\end{equation}
where we introduce the notation
\be
K \equiv -2\frac{|\Delta_T|}{m_N}.
\label{Def_K}
\ee
For the double spin asymmetries (\ref{Def_ALLDSA}), (\ref{Def_ALTDSA}) we get
\begin{equation}
\frac{A_{LL}^\mathrm{const.}}{   \sqrt{1- \varepsilon^2} }  = \frac
{
|\mathcal{I}^{(1)}(\xi, u)|^2
+ \frac{\Delta_T^2}{m_N^2}| \mathcal{I}^{(2)}(\xi, u)|^2
}
{
|\mathcal{I}^{(1)}(\xi, u)|^2
- \frac{\Delta_T^2}{m_N^2}| \mathcal{I}^{(2)}(\xi, u)|^2
}\,,
\label{eq:DSA1}
\end{equation}
and 
\begin{equation}
\frac{A_{LT}^{\cos(\phi-\phi_S)}}{ \sqrt{1- \varepsilon^2}}
=
K
\frac{
\mathrm{Re} \left(\mathcal{I}^{(1)}(\xi, u)\,\mathcal{I}^{(2)*}(\xi, u)\right)
}
{
|\mathcal{I}^{(1)}(\xi, u)|^2
- \frac{\Delta_T^2}{m_N^2}| \mathcal{I}^{(2)}(\xi, u)|^2
}.
\label{eq:DSA2}
\end{equation}

Other spin asymmetries, like the electron beam spin asymmetry, vanish at leading twist and thus necessitate new theoretical developments which are outside the scope of the present work. Note however that the electron beam spin asymmetry has been measured in backward kinematics by CLAS12 \cite{CLAS:2020yqf} and found to be rather small at moderate values of $Q^2$ which is a positive signal for an early scaling property of the backward scattering amplitude.

\section{Modeling  transition distribution amplitudes}
\label{sec:modeling}

\subsection{A spectral representation of transition distribution amplitudes }

In the case of GPDs, an elegant approach to simultaneously implement the support property 
($  |x| \le 1$) and the polynomiality of Mellin moments in $x$ is through the spectral representation in terms of double distributions%
\footnote{In this subsection we suppress the dependence of GPDs/TDAs and corresponding spectral densities, double and quadruple distributions, on the invariant momentum transfer $\Delta^2 \equiv t$ or $u$ in cases it plays no role in the discussion.}
~\cite{Radyushkin:1997ki}:
\be
H (x, \xi)=\int_{\Omega} d \beta d \alpha \delta(\beta+\xi \alpha-x) F (\beta, \alpha). 
\label{DD_repesentation_GPD}
\ee
The spectral properties of double distributions,
$\Omega=\{|\beta| \le 1; \, |\alpha| \le 1-|\beta|\}$,
were established from the diagrammatic perspective with help of the $\alpha$-representation techniques \cite{Radyushkin:1983wh}. 
The formulation of GPDs in terms of double distributions 
$F(\beta, \alpha)$ 
provides a natural way to interpret GPDs as kinematic ``hybrids'' -- interpolating between forward parton distribution functions (PDFs)
$q(\beta)$ 
in the 
$\xi \to 0$ 
limit and distribution amplitudes 
$\phi(\alpha)$ in the $\xi \to 1$ limit.
This led to a successful phenomenological approach to modeling GPDs using the so-called factorized Ansatz for double distributions \cite{Musatov:1999xp}:
\be
F(\beta, \alpha)= h(\beta, \alpha) q(\beta).
\label{Rad_Ansatz}
\ee
This Ansatz combines the reduction of GPDs to known PDFs in the forward limit with a simple, asymptotic-type parametrization of meson DA, representing the double distribution as a product of a PDF and a universal profile function, 
$h(\beta, \alpha)$, 
which inherits the shape of the asymptotic meson DA.
The profile function is normalized in such a way that 
\be
\int_{-1+|\beta|}^{1-|\beta|} d \alpha h(\beta, \alpha)=1;
\ee
this ensures the forward limit, 
$H(x,\xi=0)=q(x)$,
for the GPD 
(\ref{DD_repesentation_GPD}).

It was then conjectured in Ref.~\cite{Polyakov:1999gs} that, in order to satisfy the polynomiality property
in its complete form, the representation~(\ref{DD_repesentation_GPD})
has to be modified. 
For singlet quark GPD the highest power of $\xi$ of its $N$-th ($N$-odd) Mellin moment in $x$ is $N+1$, while highest power from the representation (\ref{DD_repesentation_GPD}) is only $N-1$. Therefore, GPDs within the spectral representation are to be complemented with the 
so-called $D$-term, that has a pure ERBL support $|x| \le \xi$, and has a structure analogous to a contribution arising from an exchange with a meson-like state in the $t$-channel:
\be
H (x, \xi)=\int_{\Omega} d \beta d \alpha \delta(\beta+\xi \alpha-x) F (\beta, \alpha)+
\theta(|x| \le \xi) D\left( \frac{x}{\xi} \right).
\label{DD_repesentation_GPD_Dterm}
\ee
A method was later developed to incorporate the $D$-term within the double distribution representation by introducing two distinct spectral densities, see 
Ref.~\cite{Teryaev:2001qm} and a
discussion of Sec.~3.8.1 of Ref.~\cite{Belitsky:2005qn}:
\be
H(x, \xi)=\int_{\Omega} d \beta d \alpha \delta(\beta+\xi \alpha-x)\left\{F^{(0)} (\beta, \alpha)+\xi F^{(1)}(\beta, \alpha)\right\}.
\label{Two_component_GPD}
\ee

A generalization of the spectral representation techniques for the case of nucleon-to-meson
TDAs has been addressed in Ref~\cite{Pire:2010if}.
The spectral representation of TDAs in terms of the so-called quadruple distributions was found to ensure the support  and  polynomiality properties.  
The relation between a given TDA, $H(x_1, x_2, x_3, \xi)$,
and the corresponding spectral density, 
$F$ 
reads:
\begin{align}
H(x_1, x_2, x_3, \xi) & =
\left[
\prod_{i=1}^3
\int_{\Omega_i} d \beta_i d \alpha_i
\right]
\delta(x_1-\xi-\beta_1-\alpha_1 \xi) \,
\delta(x_2-\xi-\beta_2-\alpha_2 \xi) \nonumber \\
& \times
\delta(\beta_1+ \beta_2+ \beta_3)
\delta(\alpha_1+\alpha_2+\alpha_3+1)
F(\beta_1, \beta_2, \beta_3, \alpha_1, \alpha_2, \alpha_3)\,,
\label{eq:spectral_represention_x1x2}
\end{align}
where $\Omega_{i} =\{|\beta_i| \le 1 \cap |\alpha_i| \le 1-|\beta_i|\}$, $i=1,2,3$, denote three copies of the usual
spectral regions. 
The spectral density $F
(\beta_1, \, \beta_2, \, \beta_3, \, \alpha_1, \, \alpha_2, \alpha_3)$ is a function of six spectral variables, which are subject to two constraints, imposed by the $\delta$-functions in the second line of
Eq.~(\ref{eq:spectral_represention_x1x2}), so efficiently it is
a quadruple distribution. 

In contrast to the GPD case, a combination of $T$-invariance and hermiticity does not impose extra symmetry requirements for the quadruple distributions like parity in $\alpha_i$.
However, the intrinsic $3$-body nature of the problem results in specific symmetry properties of TDAs under the operation of the cyclic permutation of the variables $x_i$, see discussion of the symmetric properties of TDAs in Ref.~\cite{Pire:2011xv}.

These symmetry properties make an impact when switching from 
the three momentum fraction variables to the quark-diquark coordinates
(\ref{Def_qDq_coord}). 
For each of the three choices of the quark-diquark coordinates 
(\ref{Def_qDq_coord}), 
there exist an associated choices of combinations of the spectral variables.  Furthermore, the spectral representation within the quark-diquark coordinates can be organized into two distinct forms:
\begin{enumerate}
    \item the form originally suggested in Ref.~\cite{Pire:2010if},
that, in complete analogy to (\ref{DD_repesentation_GPD}), allows imposing constraints on TDAs in the forward limit, $\xi=0$.  For a specific choice of quark-diquark coordinates, $w \equiv w_3$ and $v \equiv v_3$, (\ref{eq:spectral_represention_x1x2}) is rewritten as: 
\begin{align}
H(w, v, \xi) & = 
\int_{-1}^1 d \sigma
\int_{-1+\frac{| \sigma| }{2}}^{1-\frac{| \sigma| }{2}} d \rho
\int_{-1+| \sigma| }^{1-|  \rho- \frac{\sigma}{2}| -| \rho+ \frac{\sigma}{2}| } d \omega
\int_{-\frac{1}{2}+| \rho- \frac{\sigma}{2}| +\frac{\omega}{2}}^{\frac{1}{2}-| \rho+ \frac{\sigma}{2}| -\frac{ \omega}{2}} d \nu
\delta(w-\sigma-\omega \xi) \nonumber \\
& \times  \delta(v-\rho-\nu \xi) \,
 F_3(\sigma,\, \rho,\, \omega,\, \nu),
\label{eq:spectral_represention_wv}
\end{align}
where
\begin{align}
\sigma = \beta_3 \,, \quad 
2\rho = \beta_{1} -\beta_{2} \,, \quad
\omega = \alpha_3 \,, \quad  
2\nu = \alpha_{1} - \alpha_{2} \,;
\label{Spectral_par_3}
\end{align}
and
\be
F_3\left(\sigma , \rho , \omega, \nu \right) \equiv F\left(\rho -\frac{\sigma }{2},-\rho -\frac{\sigma }{2}, \sigma , \nu-\frac{1+\omega}{2},-\nu-\frac{1+\omega}{2}, \omega\right).
\ee
There are two other equivalent ways of rewriting the spectral density 
$F(\beta_1, \, \beta_2, \, \beta_3, \, \alpha_1, \, \alpha_2, \alpha_3)$ 
in terms of spectral variables, corresponding to two other choices of quark-diquark coordinates 
(\ref{Def_qDq_coord}). 

The functional form of the spectral representation (\ref{eq:spectral_represention_wv}) 
suggests trying a factorized Ansatz for 
$F_3(\sigma,\, \rho,\, \omega,\, \nu)$,
in analogy to 
(\ref{Rad_Ansatz}):
\be
F_3(\sigma,\, \rho,\, \omega,\, \nu)= f (\sigma, \rho) h(\sigma,\, \rho,\, \omega,\, \nu),
\ee
where the profile function is normalized as
\be
\int_{-1+| \sigma| }^{1-|  \rho- \frac{\sigma}{2}| -| \rho+ \frac{\sigma}{2}| } d \omega
\int_{-\frac{1}{2}+| \rho- \frac{\sigma}{2}| +\frac{\omega}{2}}^{\frac{1}{2}-| \rho+ \frac{\sigma}{2}| -\frac{ \omega}{2}} d \nu h(\sigma,\, \rho,\, \omega,\, \nu)=1.
\label{Normalization_h1}
\ee
This ensures that in the forward limit 
\be
H(w, v, \xi=0)=f(w,v).
\ee
However, unlike to GPDs, the forward limit of TDAs is unconstrained, and turns to be a subject to modeling assumptions.

\item an alternative form, suggested in Ref.~\cite{Lansberg:2011aa},  instead, allows imposing constraints in the limit $\xi=1$, in which $\pi N$
TDAs reduce to certain combinations of nucleon DAs (\ref{Set_of_DAs}), due to the soft pion theorem~\cite{Pire:2011xv} (see Appendix~\ref{App_Soft_pion} for a brief overview;
the origin of the normalization factor $1/4$ is clarified in the discussion surrounding Eq.~(\ref{Explain14})
):
\begin{equation}
H(x_1, x_2, x_3, \xi=1
) = \frac{1}{4} \times \Phi \left( \frac{x_1}{2}, \frac{x_2}{2}, \frac{x_3}{2} \right) \,.
\label{eq:HtoDAs}
\end{equation}
For this purpose, for the choice 
$w \equiv w_3$ and $v \equiv v_3$,
the spectral representation~\eqref{eq:spectral_represention_x1x2} is rewritten in the following form:
\begin{align}
H(w, v, \xi) & =
\int_{-1}^1 d \kappa \int_{- \frac{1-\kappa}{2}}^{ \frac{1-\kappa}{2}} d\theta
\int_{-1}^1 d \mu \int_{- \frac{1-\mu}{2}}^{ \frac{1-\mu}{2}} d\lambda
\, \delta \left(w- \frac{\kappa-\mu}{2} (1-\xi) - \kappa \xi \right) \nonumber \\
& \times \delta\left(v - \frac{\theta-\lambda}{2} (1-\xi) - \theta \xi \right)  
F_3(\kappa, \theta, \mu, \lambda) \,,
\label{eq:spectral_represention_wv_kappatheta}
\end{align}
where
\begin{align}
\kappa = \alpha_3 + \beta_3 \,, \quad 
2\theta = \alpha_1 + \beta_1 - \alpha_2 - \beta_2 \,, \quad
\mu = \alpha_3 - \beta_3\,, \quad  
2\lambda = \alpha_1 - \beta_1 - \alpha_2 + \beta_2 \,;
\label{Def_kappa_theta_mu_lambda_3}
\end{align}
and 
\be
&&
F_3(\kappa, \theta, \mu, \lambda)\nn \\ && \equiv F\left(\frac{2 \theta-\kappa-2 \lambda+\mu}{4},\frac{-2 \theta-\kappa+2 \lambda+\mu}{4}, \frac{\kappa-\mu}{2},\frac{-2+2 \theta-\kappa+2 \lambda-\mu}{4}, \frac{-2-2 \theta-\kappa-2 \lambda-\mu}{4}, \frac{\kappa+\mu}{2} \right).
\nn \\ &&
\ee
The functional form of Eq.~(\ref{eq:spectral_represention_wv_kappatheta}) is adjusted for the alternative form of the factorized Ansatz:
\be
F_3\left(\kappa , \theta , \mu , \lambda \right)= \phi\left(\kappa , \theta \right) h\left(\mu , \lambda \right),
\label{Factorized_Ansatz_xi1}
\ee
with the profile function $h(\mu,\lambda)$ normalized according to
\be
\int_{-1}^1 d \mu  \int_{- \frac{1-\mu}{2}}^{ \frac{1-\mu}{2}} d\lambda \, h\left(\mu , \lambda \right)=1.
\label{Normalization_h_mu_lambda}
\ee
This ensures that in the limit $\xi=1$
\be
H(w, v, \xi=1)=\phi(w,v).
\ee
This constraint turns to be constructive due to the threshold pion theorem, that fixes a specific TDA at 
$\xi=1$ in terms of the combination of nucleon DAs (\ref{eq:HtoDAs}), see Ref.~\cite{Pire:2011xv},
\be
\phi(w, v) \equiv \frac{1}{4} \times \Phi\left(\frac{1-w +2 v }{4}, \frac{1-w -2 v }{4}, \frac{w +1}{2}\right).
\label{phi_of_w_v}
\ee
\end{enumerate}
 
Similarly to the case of GPDs, within the spectral representation (\ref{eq:spectral_represention_x1x2}), TDAs do not
satisfy the polynomiality condition 
(\ref{Def_Mellin_moments_TDAs}) 
in its full form.
Specifically, by considering the general form factor decomposition of pion-nucleon matrix elements of local operators~(\ref{O_Tw3_local}), one finds that the maximal power of $\xi$ appearing in the $(n_1, n_2, n_3)$-th Mellin moments of the TDAs $V_{1,2}^{\pi N}$, $A_{1,2}^{\pi N}$, and $T_{1,2}^{\pi N}$ is $n_1 + n_2 + n_3 + 1$, whereas the spectral representation (\ref{eq:spectral_represention_x1x2}) only produces terms up to order $n_1 + n_2 + n_3$. 
This discrepancy suggests the necessity of an additional ``$D$-term-like'' contribution
with a pure ERBL-like support, complementing
the spectral representation 
(\ref{eq:spectral_represention_x1x2}).
The most natural source for such contribution is the cross-channel exchange with a nucleon,
see Sec.~5.1 of Ref.~\cite{Pire:2021hbl}.

This led to the development of the phenomenological model of Ref.~\cite{Lansberg:2011aa}, which incorporates a spectral part constructed using the factorized Ansatz (\ref{Factorized_Ansatz_xi1}) 
with input from the soft pion theorem
supplemented by the cross-channel nucleon exchange contribution. The only phenomenological input required by this model is the parametrization of the leading-twist-3 nucleon distribution amplitudes (DAs). In principle, the model should be supplemented by a parametrization of the form factor describing the factorized $u$-dependence of the spectral contribution; however, this element was not included in the original analysis of Ref.~\cite{Lansberg:2011aa}.

This model was used to estimate the backward pion electroproduction cross-section under the kinematical conditions of experiments at JLab. However, it lacks flexibility, as it contains no adjustable parameters that can be fitted to the data. Additionally, the model’s predictions are overly sensitive to the shape of the input nucleon DAs. Solutions based on nucleon DAs close to the asymptotic form tend to underestimate the limited available data points, whereas adopting asymmetric (so-called Chernyak–Zhitnitsky) shapes typically results in an overestimation. By its construction, the model lacks flexibility needed to correct this kind of behavior. Moreover, for $x_B$ values corresponding to the valence region, the contribution of the nucleon exchange part into the cross-section was largely dominating over the contribution of the spectral part. Such behavior may be considered somewhat unnatural, given that the position of the cross-channel nucleon pole at $u = m_N^2$ is far from the threshold of the direct-channel at $u = u_0$.
These evident shortcomings of the model proposed in Ref.~\cite{Lansberg:2011aa} motivate the development of a more refined strategy for modeling of $\pi N$ TDAs, in anticipation of detailed experimental data from JLab.

\subsection{General modeling strategy}

We propose a novel approach for modeling $\pi N$ TDAs that is based on the spectral representation of TDAs. We construct a two-component model for the corresponding 
spectral densities 
\be
F(\beta_1,\beta_2,\beta_3, \alpha_1, \alpha_2,\alpha_3; \, u)=
F^{(0)}(\beta_1,\beta_2,\beta_3, \alpha_1, \alpha_2,\alpha_3; \, u)+(1-\xi)F^{(1)}(\beta_1,\beta_2,\beta_3, \alpha_1, \alpha_2,\alpha_3; \, u).
\label{eq:master_model}
\ee

\bi
\item
The first component, with the spectral density $F^{(0)}$, 
is supposed to be constructed according to lines of Ref.~\cite{Lansberg:2011aa} with help of the
factorized Ansatz of the form 
(\ref{Factorized_Ansatz_xi1}) 
with input at $\xi=1$.
It ensures the normalization of TDAs in the limit $\xi=1$ arising from the soft pion theorem.

\item 
The second component, 
$F^{(1)}$, 
is designed to be flexible, and is primarily intended to be constrained by current and future experimental data. The prefactor $(1 - \xi)$ in Eq.~\eqref{eq:master_model} is introduced by analogy with the two-component GPD parametrization~(\ref{Two_component_GPD}). It ensures that this contribution vanishes in the $\xi = 1$ limit, thereby preserving the normalization imposed by the soft pion theorem. In addition, it restores the complete form of the polynomiality condition for the TDAs $V_{1,2}^{\pi N}$, $A_{1,2}^{\pi N}$, and $T_{1,2}^{\pi N}$, allowing for the maximal power of $\xi$ in the Mellin moments without the need for any supplementary ``$D$-term-like'' contributions.
\ei

In the following, we will review the modeling of spectral density $F^{(0)}$ 
and introduce a  model of spectral density $F^{(1)}$.
In particular, we show how various types of $\pi^+p$ and $\pi^0p$ TDAs can be constructed from a single master TDA $H$, resulting from either the first or second component, using the relations \eqref{eq:spectral_represention_wv} and \eqref{eq:spectral_represention_wv_kappatheta}, respectively. 
This is followed by a discussion on incorporating evolution effects. Finally, we discuss the normalization of our model to existing experimental data and present an analysis of the sensitivity of the modeling to various choices of model 
parameters and on input phenomenological solutions for nucleon DAs. We  conclude this section by presenting predictions for various polarization observables. 

\subsubsection{Modeling spectral distribution $\boldsymbol{F^{(0)} }
$}

For the sake of completeness, we review the construction of the phenomenological 
model for the spectral density $F^{(0)}$ with the input from the soft pion theorem for TDAs at $\xi=1$ $u=m_N^2$.
In terms of the spectral parameters 
(\ref{Def_kappa_theta_mu_lambda_3}), 
corresponding to the choice of quark-diquark coordinates $w=w_3$, $v=v_3$, the factorized Anstaz 
(\ref{Factorized_Ansatz_xi1}) 
looks as
\begin{equation}
F^{(0)}_3(\kappa, \theta, \mu, \lambda;\, u) = 
\phi(\kappa, \theta)\, h(\mu, \lambda)\, G(u).
\end{equation}
The profile function is chosen in a simple form, that ensures vanishing of the spectral density 
at the borders of its domain of definition in variables $\mu$, $\lambda$:
\begin{equation}
h(\mu,\, \lambda) = \frac{15}{16} \, (1+\mu) ((1-\mu)^2-4 \lambda^2) \,
\label{Profile_h_0},
\end{equation}
and is normalized according to (\ref{Normalization_h_mu_lambda}).
The profile function (\ref{Profile_h_0}) is completely symmetric in
pairs of spectral parameters $\beta_{i}$, $\alpha_{i}$:
\be
\left.h\left(\beta_1, \beta_2, \beta_3 ; \alpha_1, \alpha_2, \alpha_3\right)\right|_{\substack{\beta_1+\beta_2+\beta_3=0 \\ \alpha_1+\alpha_2+\alpha_3=-1}}= \frac{15}{4}(1+\alpha_1-\beta_1)(1+\alpha_2-\beta_2)(1+\alpha_3-\beta_3).
\ee

In the limit $\xi = 1$  the resulting TDA reduces to $\phi(w, v)$ (\ref{phi_of_w_v}) expressed in terms of a combination of
nucleon DAs, specified in Appendix~\ref{App_Soft_pion}, multiplied by the form factor $G(u)$ ($G(m_N^2)=1$):
\begin{equation}
H(w, v, \xi=1, u) = \phi(w, v)\, G(u)\,.
\label{eq:xiEq1Limit}
\end{equation}
In this analysis, the dependence on $u$ is described by a modified dipole Ansatz,
\begin{equation}
G(u) = \left( 1 - \frac{u - m_{N}^2}{m_{D}^2}\right)^{-2}\,,
\label{eq:udep_dipole}
\end{equation}
where  $m_{D}^2 = 0.71\,\mathrm{GeV}^2$. 
In the present analysis, $G(u)$ Eq.~(\ref{eq:udep_dipole}) is employed purely as an empirical parametrization -- without attributing intrinsic physical significance -- and is normalized by the condition $G(m_N^{2}) = 1$. The functional form of
$G(u)$ should be revisited once experimental data capable of probing the $u$-dependence of observables become available.

Within this model  $\pi^0 p$ and $\pi^+ p$ TDAs can be expressed in terms of a single master TDA, that is
constrained at the pion production threshold as
\begin{equation}
H^{(0)}(x_1, x_2, x_3, \xi=1, u = m_{N}^2) = \frac{1}{4}  \varphi_N \left( \frac{x_1}{2}, \frac{x_2}{2}, \frac{x_3}{2} \right) \,.
\label{eq:HtoDAsmodel}
\end{equation}
The corresponding set of formulas for TDAs $V_1$, $A_1$ and $T_1$ can be read off Eq.~(\ref{eq:reductionToDAs}):
\be
&&
V_1^{(0) \, \pi^0 p}(x_1, x_2, x_3, \xi, u)     =
-\frac{1}{4} \left(H^{(0)}(x_1, x_2, x_3, \xi, u) + H^{(0)}(x_2, x_1, x_3, \xi, u) \right) \,, \nonumber \\  &&
V_1^{(0) \,\pi^+ p}(x_1, x_2, x_3, \xi, u)   \nonumber \\  && =
-\frac{1}{2 \sqrt{2}} \left(H^{(0)}(x_1, x_2, x_3, \xi, u) + H^{(0)}(x_2, x_1, x_3, \xi, u) + 2H^{(0)}(x_3, x_1, x_2, \xi, u) + 2 H^{(0)}(x_3, x_2, x_1, \xi, u)\right) \,, \nonumber \\ &&
A_1^{(0) \,\pi^0 p}(x_1, x_2, x_3, \xi, u)   =
-\frac{1}{4} \left(-H^{(0)}(x_1, x_2, x_3, \xi, u) + H^{(0)}(x_2, x_1, x_3, \xi, u)\right) \,, \nonumber \\  &&
A_1^{(0) \,\pi^+ p}(x_1, x_2, x_3, \xi, u)   \nonumber \\  && =
-\frac{1}{2 \sqrt{2}} \left(-H^{(0)}(x_1, x_2, x_3, \xi, u) + H^{(0)}(x_2, x_1, x_3, \xi, u) - 2H^{(0)}(x_3, x_1, x_2, \xi, u) + 2H^{(0)}(x_3, x_2, x_1, \xi, u)\right) \,, \nonumber \\ && 
T_1^{(0) \,\pi^0 p}(x_1, x_2, x_3, \xi, u)   =
\frac{3}{4} \left(H^{(0)}(x_1, x_3, x_2, \xi, u) + H^{(0)}(x_2, x_3, x_1, \xi, u)\right) \,, \nonumber \\ &&
T_1^{(0) \,\pi^+ p}(x_1, x_2,x_3, \xi, u)   =
-\frac{1}{2 \sqrt{2}} \left(H^{(0)}(x_1, x_3, x_2, \xi, u) + H^{(0)}(x_2, x_3, x_1, \xi, u)\right) \,, \nonumber \\ 
\label{eq:HtoTDAs_F0}
\ee
TDAs $V_2$, $A_2$ and $T_2$ are fixed from the constraint (\ref{Constraint_TDA12_soft}):
\be
&&
\left\{ V_2^{(0)\, \pi p}, \, A_2^{(0)\, \pi p}, \, T_2^{(0)\, \pi p} \right\} (x_1, x_2, x_3, \xi, u)   =
-\frac{1}{2}\left\{ V_1^{(0)\, \pi p}, \, A_1^{(0)\, \pi p}, \, T_1^{(0)\, \pi p} \right\}(x_1, x_2, x_3, \xi, u);  
\ee
and TDAs $T_3$ and $T_4$ are set to zero in this model:
\be
T_3^{(0) \, \pi p}(x_1, x_2, x_3, \xi, u)   = T_4^{(0) \, \pi p}(x_1, x_2,  x_3, \xi, u)=0 \,,
\ee

\subsubsection{Modeling spectral distribution $\boldsymbol{F^{(1)}}$}

We  now focus on the second component of our model, for which we adapt the factorized Ansatz 
with input at $\xi=0$.
In terms of the spectral parameters (\ref{Spectral_par_3}) it takes the form  
\begin{equation}
    F^{(1)}(\sigma,\rho,\omega,\nu; \, u) = f(\sigma,\rho)\, h(\sigma,\rho,\omega,\nu)\, G(u)\,.
    \label{Ansatz_F1}
\end{equation}
The profile function,
\begin{equation}
h(\sigma,\,\rho,\, \omega,\, \nu) = 
\frac{\Gamma(3b+3)}{2^{5b+2} \Gamma(1+b)^3} 
\frac{\left(1+2 \nu -\omega -2 \left|\rho -\frac{\sigma }{2}\right|\right)^b \left(1-2 \nu -\omega -2 \left|\rho +\frac{\sigma}{2}\right| \right)^b (1 -|\sigma |+\omega)^b}
{\left(1-\frac{1}{2} \left(\left|\rho -\frac{\sigma }{2}\right|+\left|\rho+\frac{\sigma }{2}\right|+|\sigma |\right) \right)^{3b+2}} \,,
\label{eq:profile_function}
\end{equation}
is normalized according to (\ref{Normalization_h1}).
The parameter $b$ in Eq.~\eqref{eq:profile_function} controls the strength of
$\xi$-dependence of the resulting TDA model, similarly to Radyushkin’s double distribution Ansatz~\cite{Radyushkin:1998bz}. As in the GPD case, see for instance Ref.~\cite{Goloskokov:2006hr}, we keep this parameter small and set \mbox{$b=2$}.
The dependence on $u$ variable is again assumed to be of a dipole type, that is, $G(u)$ is given by Eq.~\eqref{eq:udep_dipole}. Overall, this makes this dependence fully factorized, which, contrary to the case of GPDs, is not ruled out by the current understanding of the TDA formalism. The modeling assumption we make should be scrutinized by future experimental data and possible developments of the theory.

In terms of the spectral parameters 
$\beta_{i}$, $\alpha_{i}$ 
the profile function 
(\ref{eq:profile_function}) 
is written as
\be 
 \left.h\left(\beta_1, \beta_2, \beta_3 ; \alpha_1, \alpha_2, \alpha_3\right)\right|_{\substack{\beta_1+\beta_2+\beta_3=0 \\
\alpha_1+\alpha_2+\alpha_3=-1}} 
=\frac{\Gamma(3b+3)}{2^{3b+2} \Gamma(1+b)^3}  \frac{\left(1+\alpha_1-\left|\beta_1\right|\right)^b\left(1+\alpha_2-\left|\beta_2\right|\right)^b\left(1+\alpha_3-\left|\beta_3\right|\right)^b}{\left(1-\frac{1}{2}\left(\left|\beta_1\right|+\left|\beta_2\right|+\left|\beta_3\right|\right)\right)^{3b+2}}.
\ee
It turns to be completely symmetric in pairs of spectral variables, which allows equivalently rewriting the Ansatz 
(\ref{Ansatz_F1}) 
for any choice of quark-diquark coordinates.

The factorized Ansatz 
(\ref{Ansatz_F1}) 
includes an arbitrary function defined on a hexagon expressing
the forward limit of the TDA~(\ref{eq:spectral_represention_wv}):
\begin{equation}
H^{(1)}(w,v,\xi=0,u)=f(w,v)\, G(u)\,.
\label{Forward_limit_H1}
\end{equation}

We construct 
$f(\sigma, \rho)$, 
with help of suitable basis  system of polynomials 
$p_{i}(\sigma, \rho)$ 
orthogonal on the hexagon 
$|\sigma| \le 1 \cap |\rho| \le 1-{|\sigma|}/{2}$
with the weight 
$W(\sigma, \rho)$.
This implies
\begin{equation}
\int_{-1}^{1} d\sigma \int_{-1 + |\sigma|/2}^{1 - |\sigma|/2} d\rho\, W(\sigma, \rho) p_{i}(\sigma, \rho)p_{j}(\sigma, \rho)
=
\begin{cases}
    1\,,& \mathrm{if}~i = j\,,\\
    0\,,& \mathrm{otherwise}\,.
\end{cases}
\end{equation}
The weight function 
$W(\sigma, \rho)$,
depending on a parameter $d$, is chosen to be
\begin{equation}
W(\sigma, \rho) = N \left(1-\sigma ^2\right)^d \left((\rho -1)^2-\frac{\sigma^2}{4}\right)^d \left((\rho +1)^2-\frac{\sigma ^2}{4}  \right)^d \,,
\label{eq:def_weight}
\end{equation}
where the normalization constant  $N$ is fixed by requiring
\begin{equation}
\int_{-1}^{1} d\sigma \int_{-1 + |\sigma|/2}^{1 - |\sigma|/2} d\rho\, W(\sigma, \rho) = 1\,.
\end{equation}
The weight 
(\ref{eq:def_weight}) 
ensures that the resulting forward limit vanishes at the boundaries of the domain of definition, that improves the convergence of the convolution integrals 
(\ref{Amplitude_M}).
Orthogonal polynomials on the hexagon are considered in the literature {\it e.g.} in Ref.~\cite{Hexagon}. They are applied, for instance, in the context of analysis of hexagonal optical elements~\cite{Optics}, and they share some common features with the more familiar Zernike polynomials~\cite{BornWolf} orthogonal on the unit disk.

\begin{figure}[!ht]
\begin{center}
\includegraphics[width=0.6\textwidth]{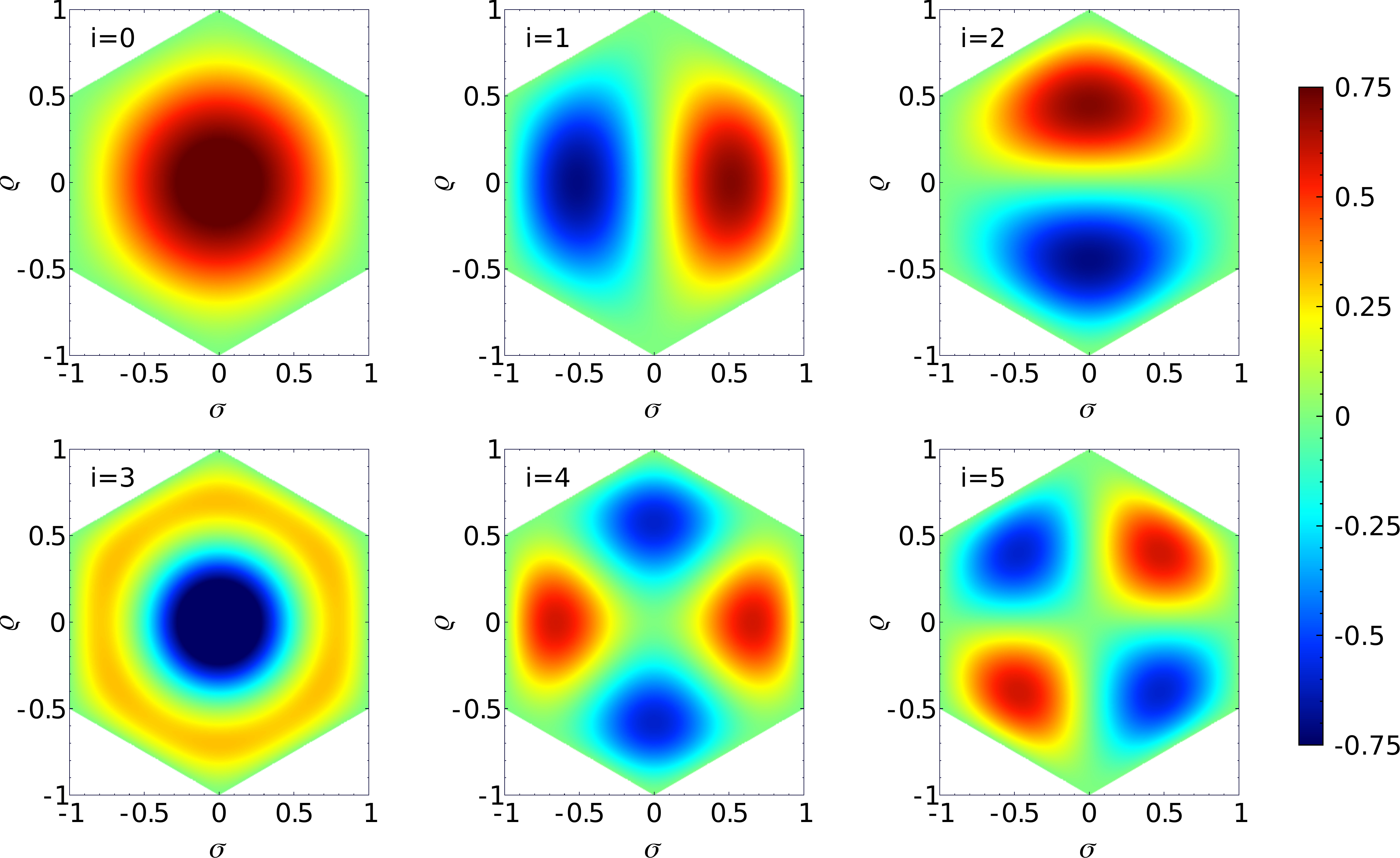}
\end{center}
Fig.\caption{Orthogonal polynomials on the hexagon multiplied by the corresponding weight function, $W(\sigma, \rho)\,p_{i}(\sigma,\rho)$, for $i=0,\ldots,5$.}
\label{fig:fig_Hex}
\end{figure}

In our present analysis, we employ the set of orthogonal polynomials on the hexagon with the weight \eqref{eq:def_weight} for $d=1$. It can be constructed using the standard Gram–Schmidt orthogonalization procedure.  The corresponding normalization coefficients, along with the explicit expressions for the first six polynomials, are the following:
\begin{gather}
N = \frac{40}{47}\,,
\quad
p_{0}(\sigma, \rho) = 1\,, \nonumber \\ 
p_{1}(\sigma, \rho) = \sqrt{\frac{4935}{781}} \sigma\,, 
\quad
p_{2}(\sigma, \rho) = 2\sqrt{\frac{1645}{781}} \rho\,, \nonumber \\ 
p_{3}(\sigma, \rho) = \frac{9870 \left(\rho ^2+\frac{3 \sigma ^2}{4}-\frac{781}{3290}\right)}{\sqrt{3149891}}\,, 
\quad
p_{4}(\sigma, \rho) = 3 \sqrt{\frac{3290}{1313}} \left(\frac{3 \sigma ^2}{4}-\rho ^2\right)\,, 
\quad
p_{5}(\sigma, \rho) = 3 \sqrt{\frac{9870}{1313}} \rho  \sigma\,.
\label{eq:hex_polynomials_dEq1}
\end{gather}
We plot these polynomials as a function of $(\sigma, \,\rho)$ in Fig.~\ref{fig:fig_Hex}. The specific choice of $d$ is another modeling assumption we make. 

Using the presented basis of orthogonal polynomials we construct a flexible parametrization of the forward function:
\begin{align}
f(\sigma, \rho)= W(\sigma, \rho)
\sum_{i=0}^{n(n+3)/2} 
c_{i}\, p_{i} (\sigma, \rho)
\,.
\label{eq:forward_limit_ansatz}
\end{align}

Although $\pi N$ TDAs are subject to isospin and permutation symmetry constraints, allowing for arbitrary forward-limit input functions  for the  TDA families 
$\{V_1^{\pi N},\, A_1^{\pi N},\, T_1^{\pi N}\}$ and $\{V_2^{\pi N},\, A_2^{\pi N},\, T_2^{\pi N}\}$ 
as well as 
$\{ T_3^{\pi N}, \,  T_4^{\pi N}\}$,
introduces excessive model freedom, which cannot be meaningfully constrained 
in the absence of precise experimental data. It this exploratory study we choose to introduce 
just a single input forward function for the master TDA of the isospin-$\frac{1}{2}$ component of the
TDA family 
$\{V_1^{\pi N},\, A_1^{\pi N},\, T_1^{\pi N}\}$. 
Thus in our model all isospin channels are described by the same function and differ just by an isotopic factor. 

To constrain the TDAs of the family 
$\{V_2^{\pi N},\, A_2^{\pi N},\, T_2^{\pi N}\}$ 
we employ the same relation between TDAs as those, which arises within the cross-channel nucleon exchange model 
(\ref{Constraint_TDA12_Nexchange}). Also TDAs $ T_{3,4}^{ \pi N}$ are set to zero. 
 This turns to be a natural 
assumption, since the primary purpose of introducing the component $F^{(1)}$ in 
(\ref{eq:master_model}) 
is to account for the cross-channel nucleon exchange contribution into TDAs through ``smearing'' it from the pure ERBL-like support over the complete support domain of TDAs. 

Thus, in our present model the contribution of the second component of 
(\ref{eq:master_model})  
into the relevant TDA
is expressed in terms of a single master TDA 
$H^{(1)}$ 
with the forward limit 
(\ref{Forward_limit_H1}) with the forward function Eq.~(\ref{eq:forward_limit_ansatz}):
\begin{align}
V_1^{(1) \, \pi^0 p}(x_1, x_2, x_3, \xi, u) & =
\frac{1}{2} \left(H^{(1)}(x_1, x_2, x_3, \xi, u) + H^{(1)}(x_2, x_1, x_3, \xi, u) \right); \nonumber \\ 
V_1^{(1) \, \pi^+ p}(x_1, x_2, x_3, \xi, u) & =
-\sqrt{2}\, V_1^{(1) \, \pi^0 p}(x_1, x_2, x_3, \xi, u); \nonumber\\
A_1^{(1) \, \pi^0 p}(x_1, x_2, x_3, \xi, u) & =
\frac{1}{2} \left(-H^{(1)}( x_1, x_2, x_3, \xi, u) + H^{(1)}(x_2, x_1, x_3, \xi, u)\right); \nonumber \\ 
A_1^{(1) \, \pi^+ p}(x_1, x_2, x_3, \xi, u) & =
-\sqrt{2}\, A_1^{(1) \, \pi^0 p}(x_1, x_2, x_3, \xi, u); \nonumber \\
T_1^{(1) \, \pi^0 p}(x_1, x_2, x_3, \xi, u) & =
\frac{1}{2} \left(H^{(1)}(x_1, x_3, x_2, \xi, u) + H^{(1)}(x_2, x_3, x_1, \xi, u)\right); \nonumber \\ 
T_1^{(1) \, \pi^+ p}(x_1, x_2,x_3, \xi, u) & =
-\sqrt{2}\, T_1^{(1) \, \pi^0 p}(x_1, x_2, x_3, \xi, u)\,, 
\label{eq:HtoTDAs_F1}
\end{align}
and
\be 
 \left\{V_2^{(1) \, \pi p}, A_2^{(1) \, \pi p}, T_2^{(1) \, \pi p}\right\} 
\left(x_1, x_2, x_3, \xi, u\right) 
=\frac{1}{2} \left\{V_1^{(1) \, \pi p}, \, A_1^{(1) \, \pi p}, \, T_1^{(1) \, \pi p}\right\} 
\left(x_1, x_2, x_3, \xi, u\right)\,. 
\label{Constraint_TDA12_component1}
\ee

This parametrization involves an arbitrary number of free parameters, controlled by $n$, which describes the maximum order of the polynomials taken into account in (\ref{eq:forward_limit_ansatz}). The free parameters, $c_{i}$, can be constrained by fitting to experimental data. Since only very limited data is available at this time, we will set $n = 1$, i.e. we will only consider $c_0$, $c_1$ and $c_2$. The other reason for setting $n = 1$ (and, in particular, avoiding $n = 0$) is as follows. Due to certain symmetries with respect to $\sigma$ and $\rho$ variables (visible in Fig.~\ref{fig:fig_Hex}), and the way TDAs are constructed (see Eq.~\eqref{eq:HtoTDAs_F1}), keeping, {\it e.g.} only $c_0$ (with $c_1 = c_2 = 0$) would result in $A_{1,2}^{(1) \, \pi p}(x_1, x_2, x_3) = 0$ and $V_{1,2}^{(1) \, \pi p}(x_1, x_2, x_3) = T_{1,2}^{(1) \, \pi p}(x_1, x_2, x_3)$. Similarly, keeping only $c_1$ (with $c_0 = c_2 = 0$) would result in $A_{1,2}^{\pi p}(x_1, x_2, x_3) = 0$ and $V_{1,2}^{(1) \, \pi p}(x_1, x_2, x_3) = -2 T_{1,2}^{(1) \, \pi p}(x_1, x_2, x_3)$. For only $c_2$ we have $V_{1,2}^{(1), \pi p}(x_1, x_2, x_3) = 0$.  We therefore see, that to have all relevant TDAs non-vanishing and independent we need to consider at least $n=1$.

\subsection{Evolution}

The one-loop renormalization for TDAs, which we focus on, has already been derived in the literature and resembles the evolution equations for nucleon DAs~\cite{Pire:2005ax}. Following the common practice in GPD phenomenology, where instead of using the full GPD evolution equation, only the ``forward'' evolution is used (i.e., that of PDFs being the limiting case of GPDs, see for instance the Goloskokov-Kroll model~\cite{Goloskokov:2006hr}), in our analysis, we only use the ``backward'' evolution, i.e., that of DAs to which TDAs reduce at $\xi = 1$ and $u = m_N^2$. We note that not only are DAs entering the modeling of $F^{(0)}(\kappa, \theta, \mu, \lambda, u)$ evolved, see Eq.~\eqref{eq:HtoDAsmodel}, but also DAs convoluted with 
hard scattering kernels in the amplitudes, see Eq.~\eqref{Structure_of_Amplitude}. On the other hand, $F^{(1)}(\sigma,\rho,\omega,\nu, u)$ will not be affected by the evolution as its contribution vanishes at $\xi=1$. The study of the phenomenological implications of using only the backward evolution would require an actual comparison with the full evolution equations and is beyond the scope of this analysis. However, we observe that using the backward evolution instead of $\mathcal{O}(\alpha_S) = 0$ approach (no evolution) has a positive effect on the agreement of our model with existing data, which will be demonstrated in the following. It suggests, that using the backward evolution will result in better predictions of observables for future experiments.   

The evolution of DAs is done via their moments,
\begin{equation}
\phi^{abc} = \frac{1}{N}\left[\prod_{i=1}^{3}\int_0^1 dy_i \right] \delta(1-y_1-y_2-y_3\,) y_1^a y_2^b y_3^c\,\varphi_N(y_1, y_2, y_3)\,,
\end{equation}
where $N$ is yet another normalization coefficient (not to be confused with that in Eq.~\eqref{eq:def_weight}), ensuring that $\phi^{000} = 1$. In this analysis, we perform the evolution by following the prescription described in Ref.~\cite{Lenz:2009ar}, which involves a specific reconstruction of DAs from the evolved moments, focusing only on $\phi^{100}$, $\phi^{010}$, $\phi^{001}$, $\phi^{200}$, $\phi^{020}$, $\phi^{002}$, $\phi^{110}$, $\phi^{101}$ and $\phi^{011}$. 

\subsection{Normalization to existing data}

After fixing the DA solution that our model reproduces at $\xi=1$ and $u=m_{N}^{2}$, the only free parameters left are $c_0$, $c_1$ and $c_2$ introduced in Eq.~\eqref{eq:forward_limit_ansatz}. As the current data does not allow us to simultaneously constrain all three parameters in a firm way, we are forced to assume a specific ratio between them and fit, {\it e.g.}, $c_0$ only. The ratio we have assumed is $c_0 : c_1 : c_2 = 1 : 1: 1$. The sensitivity of prediction to this choice will be discussed in the following.

\begin{figure}[!ht]
\begin{center}
\includegraphics[width=0.35\textwidth]{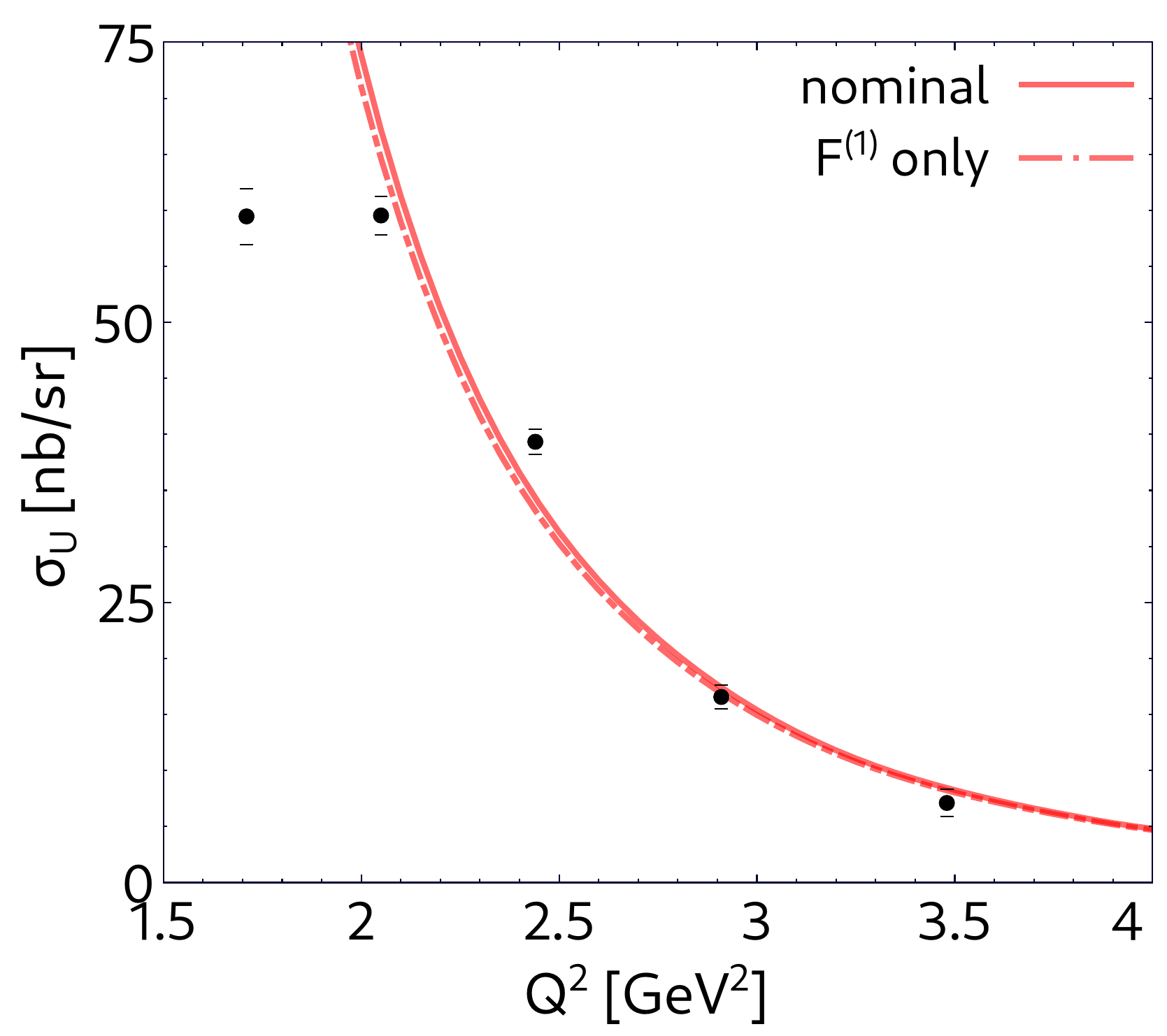}
\quad
\includegraphics[width=0.35\textwidth]{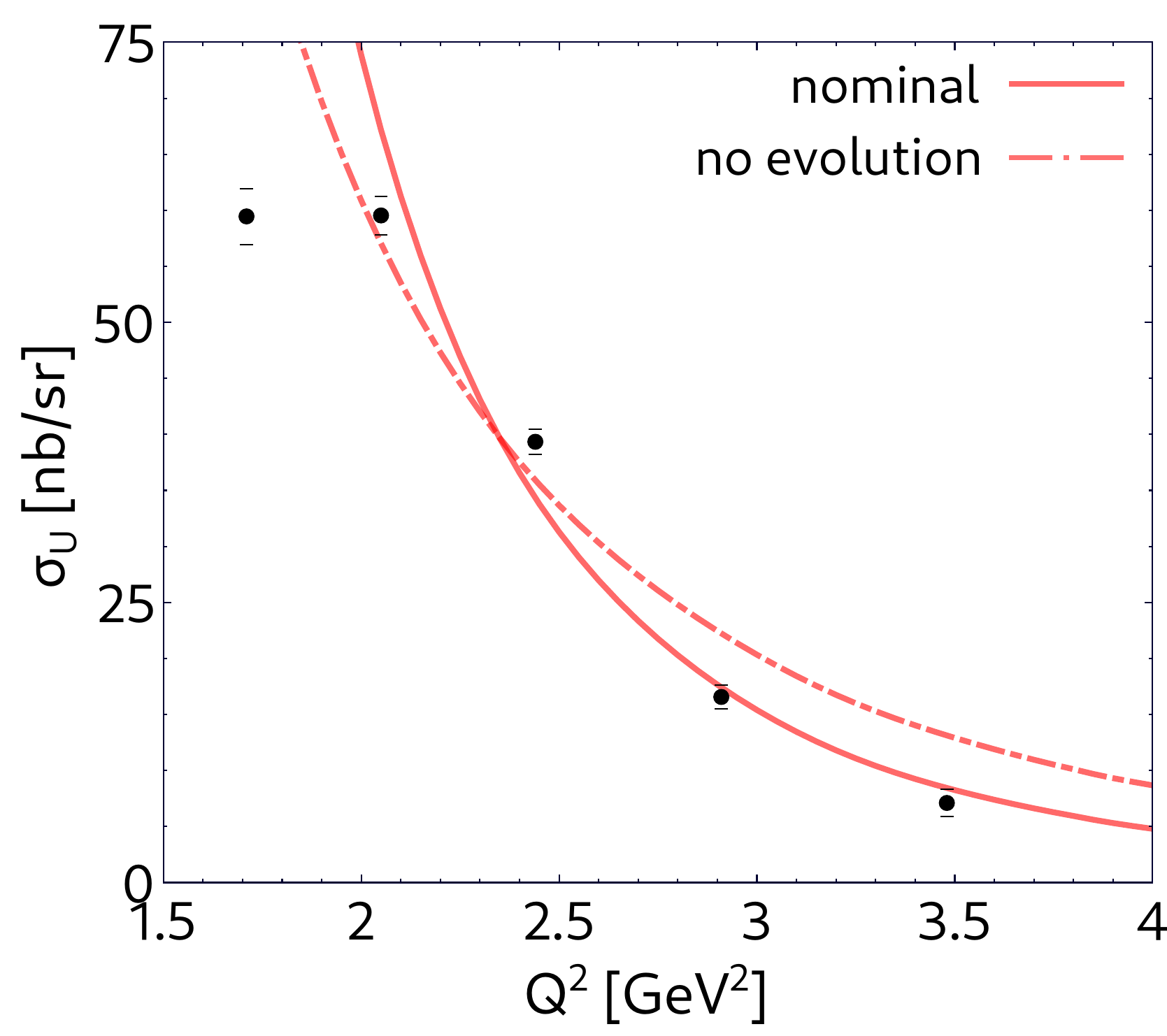}
\end{center}
\caption{CLAS data for the unseparated cross-section $\frac{d \sigma_U}{d \Omega_\pi^*}$ of the $\gamma^* p \to \pi^+ n$ reaction at backward angles~\cite{CLAS:2017rgp}, with $W = 2.2\,\mathrm{GeV}$ and $u = -0.5\,\mathrm{GeV}^2$. The solid curves represent the corresponding results obtained using the default version of our TDA model. The dash-dot curve in the left plot corresponds to a version of our model that includes only the second component, i.e., $F^{(1)}(\sigma,\rho,\omega,\nu, u)$, with $F^{(0)}(\kappa, \theta, \mu, \lambda, u)$ set to zero.  The dotted curve shown in the right plot represents the modeling scenario when all evolution effects are neglected.}
\label{fig:clas_data}
\end{figure}

The value of $c_0$ is determined to replicate the data collected by CLAS~\cite{CLAS:2017rgp} shown in Fig.~\ref{fig:clas_data}. These data are only for the exclusive production of $\pi^+$ mesons. After normalization, our model will reproduce these data, while the resulting predictions for $\pi^0$ mesons will significantly depend on the modeling assumptions we have made, highlighting again the importance of the anticipated measurement of the latter channel. Since CLAS did not provide a separation of the unpolarized cross-section they measured, $\sigma_U$, into transverse and longitudinal components, $\sigma_T$ and $\sigma_L$, respectively, the TDA prediction $\sigma_T \gg \sigma_L$ has not been able to be checked and remains an assumption we make in this analysis.

In the normalization procedure, we exclude the first data point (at $Q^{2} \approx 1.7\,\mathrm{GeV}^2$). This point deviates from the rest of the population in terms of the power behavior in $Q^2$, and its inclusion would therefore lead to a bias. Possible explanations for this point being an outlier include: the significant impact of the $\sigma_L$ contribution, which is neglected in this analysis; the wide kinematics of the CLAS data, which require a multidimensional integration of the differential cross-section rather than probing it at the average values of $Q^2$, $W$, and $u$; and the too small value of $Q^2$, which necessitates the inclusion of higher-twist effects.

The value of $c_0$ fitted to four data points collected by CLAS is $163.75$. The quality of fit is rather poor, as indicated by $\chi^2/4 = 10$. This does not come as a surprise due to the exploratory nature of this analysis. The predictions from the constrained model are shown in Fig.~\ref{fig:clas_data}, and are denoted by solid lines. 

In Fig.~\ref{fig:clas_data}, we also show how minimal the contribution of the $F^{(0)}(\kappa, \theta, \mu, \lambda, u)$ component is to the cross-section at the kinematics covered by the present data. We also use Fig.~\ref{fig:clas_data} to demonstrate the benefits of including evolution effects, even in an incomplete form, namely, the backward evolution. The version of our model constrained by CLAS data without accounting for any evolution effects (with $\alpha_S = 0.3$ and the DA parameterization given at the reference scale $\mu_\mathrm{ref}^2 = 1\,\mathrm{GeV}^2$) is represented by the dotted line. In this case, we obtain significantly worse $\chi^2/4 = 17$. 

\subsection{Sensitivity to modeling options}

\begin{figure}[H]
\begin{center}
\includegraphics[width=0.3\textwidth]{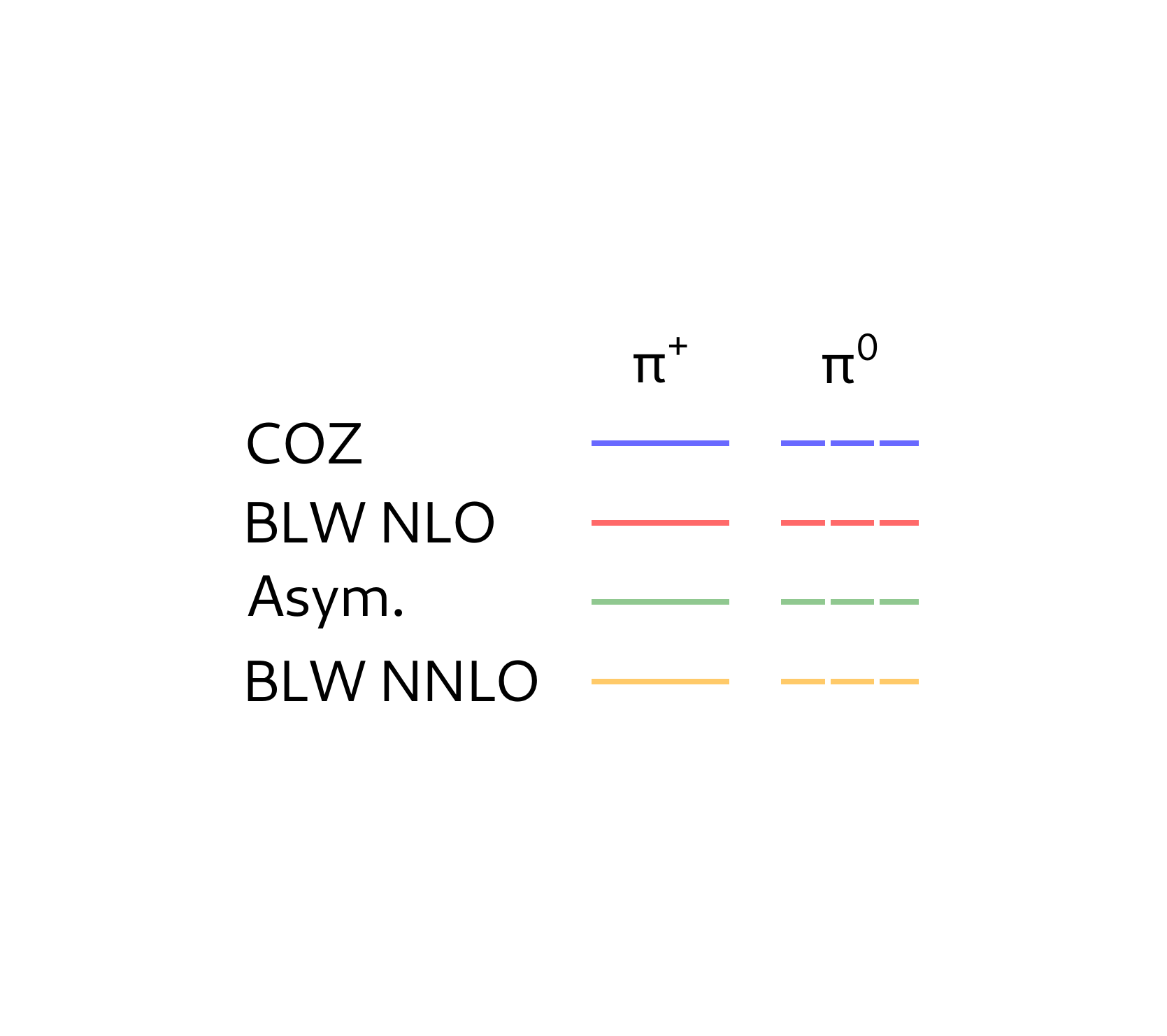}
\includegraphics[width=0.3\textwidth]{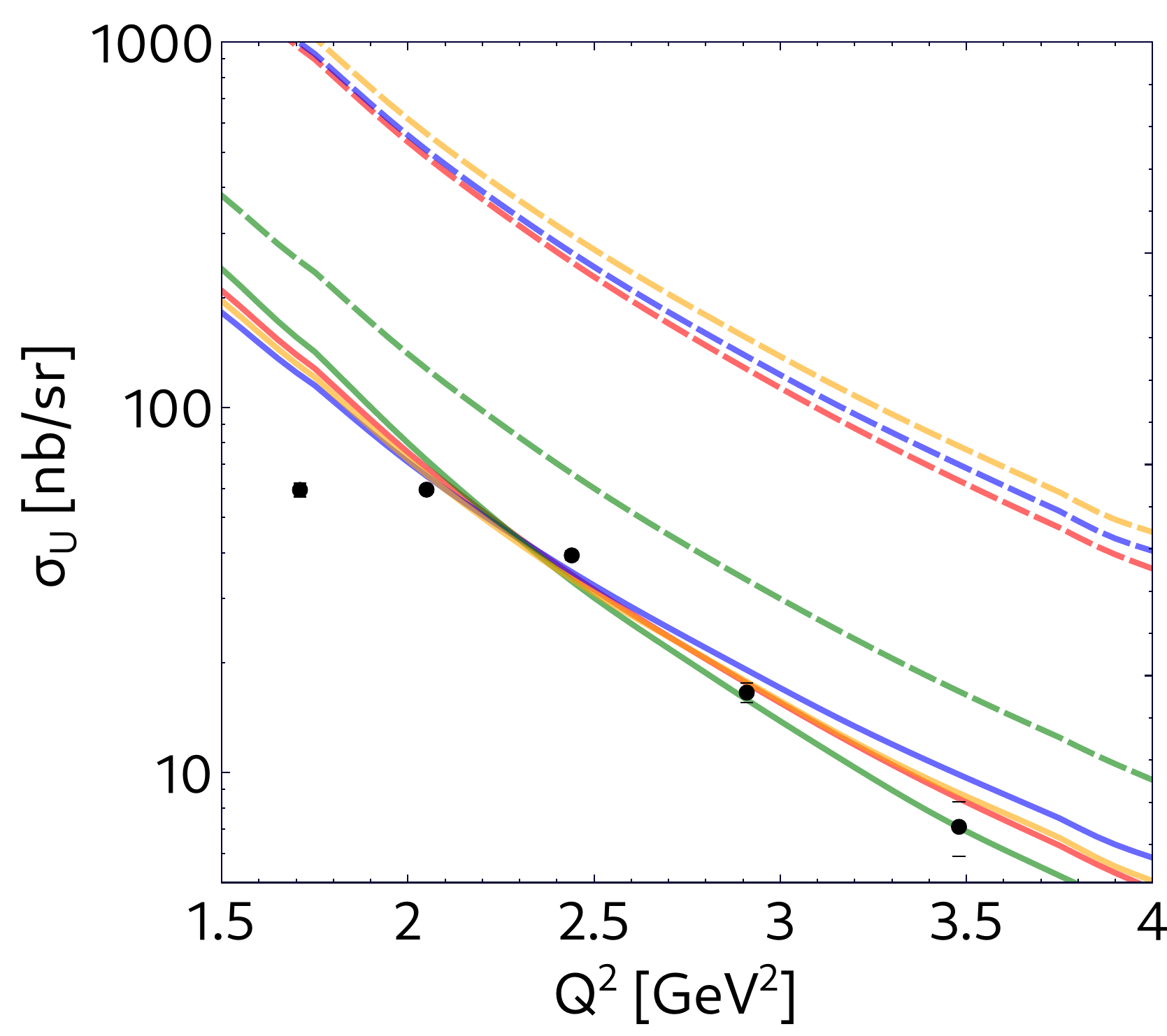}
\includegraphics[width=0.3\textwidth]{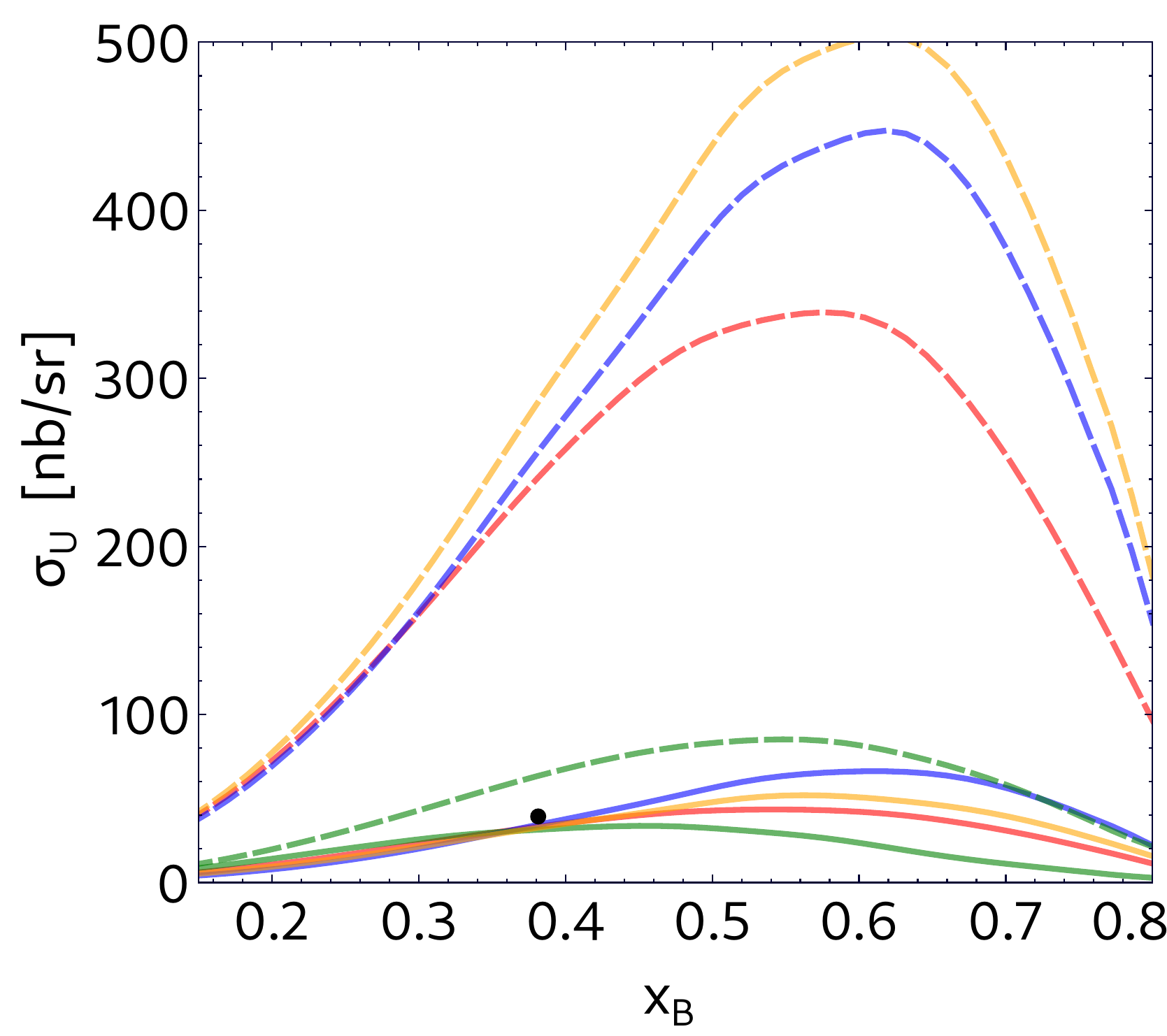}
\\
\includegraphics[width=0.3\textwidth]{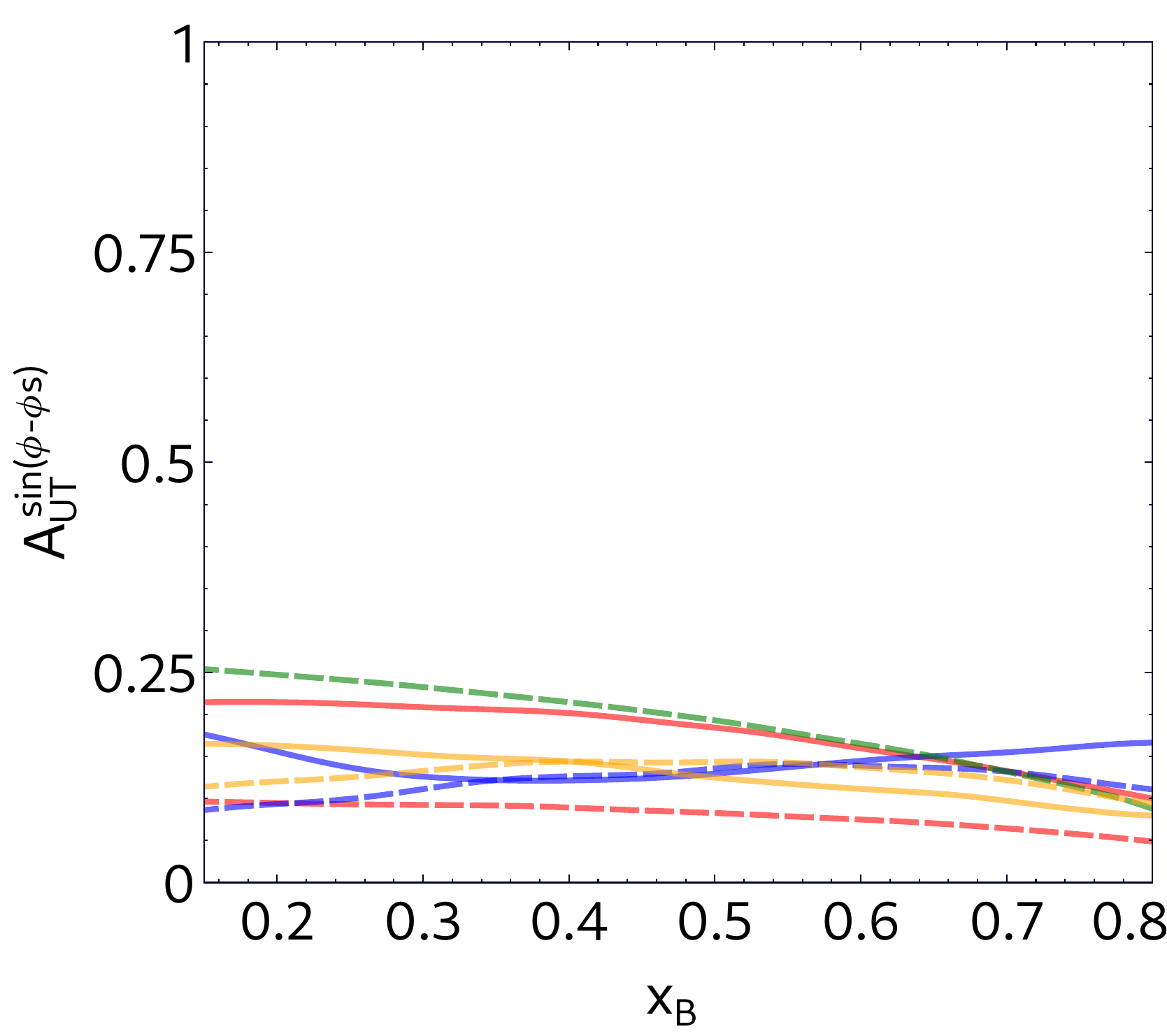}
\includegraphics[width=0.3\textwidth]{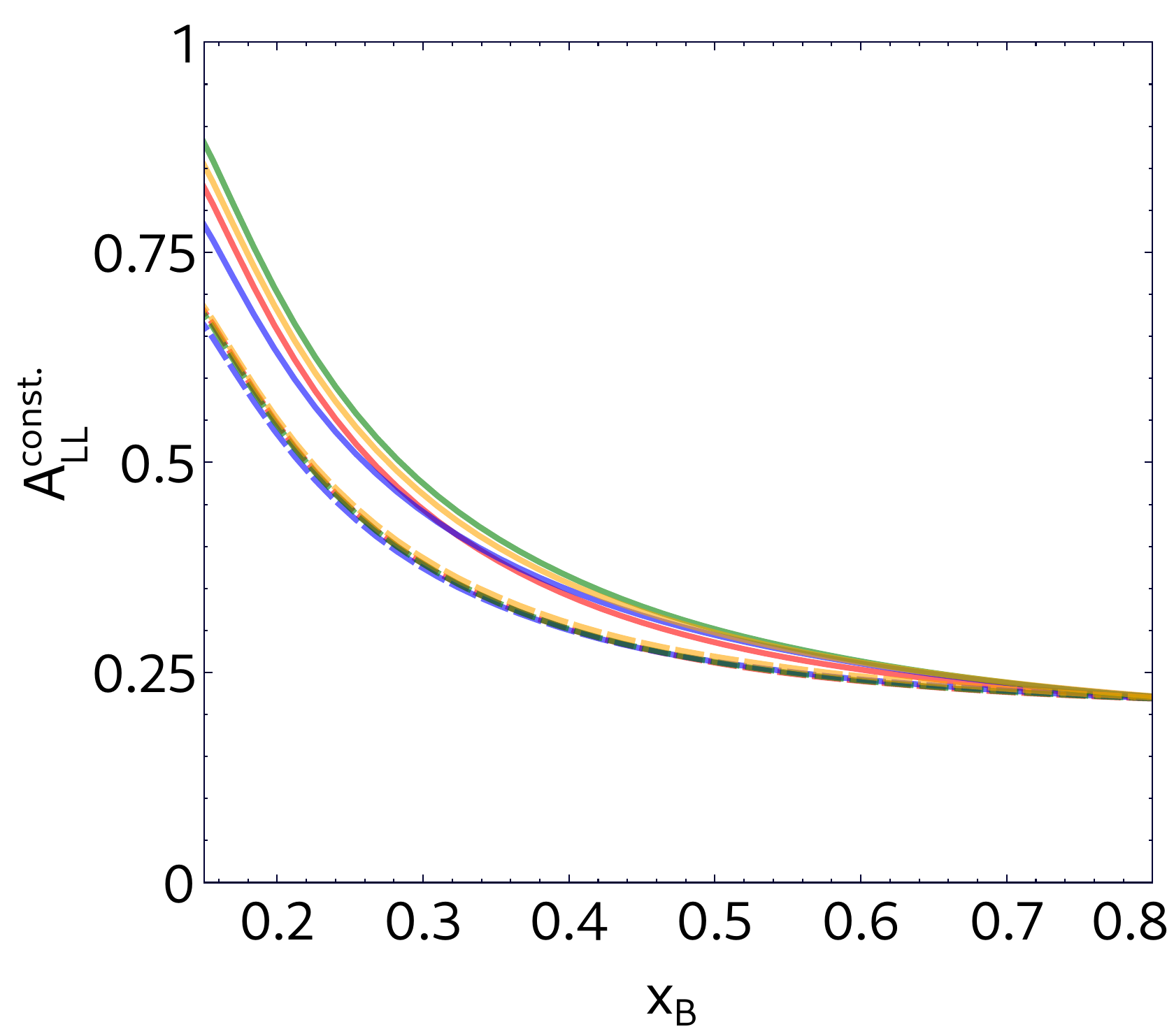}
\includegraphics[width=0.3\textwidth]{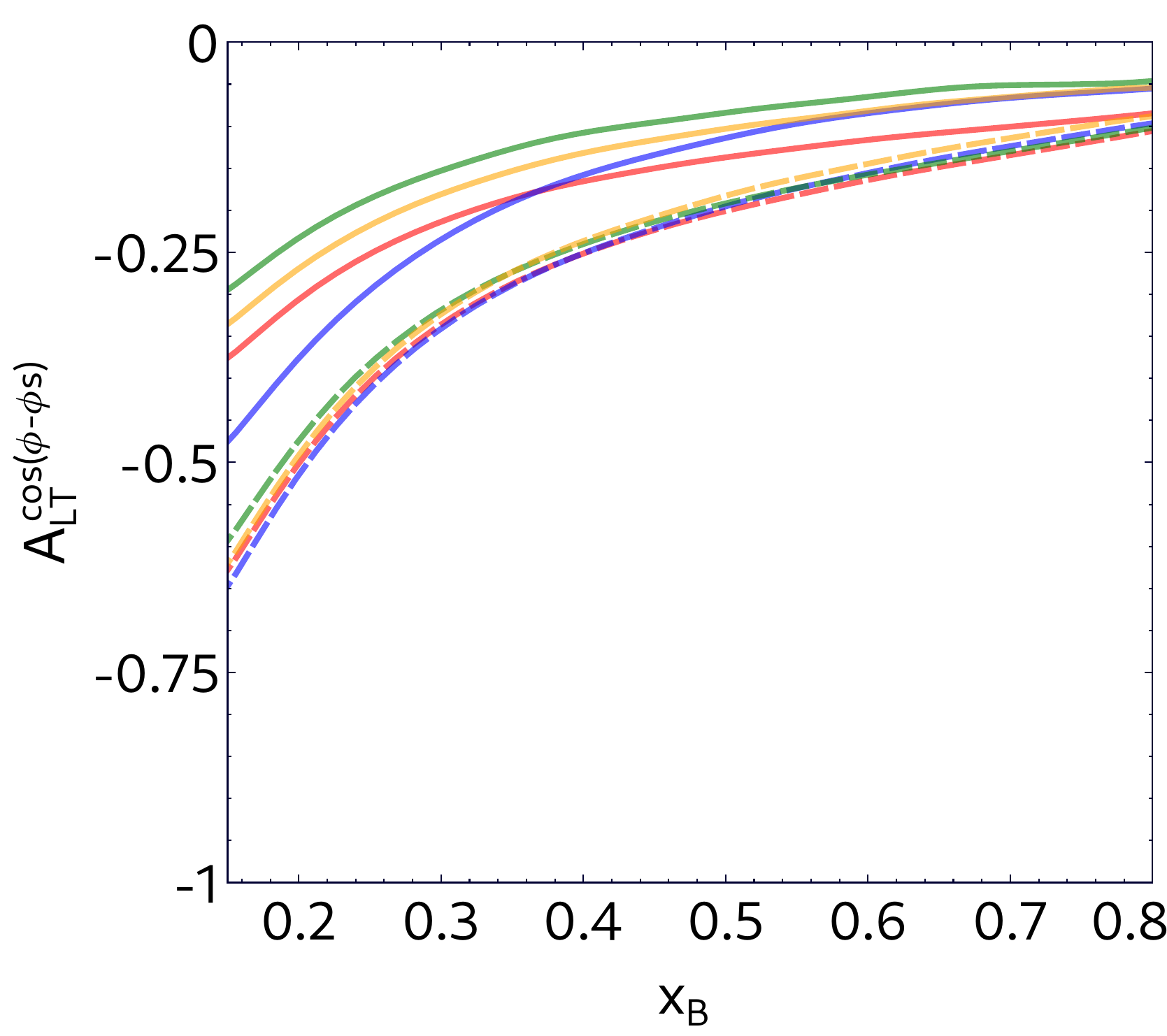}
\caption{Sensitivity to the choice of DA parametrization. Selected observables for the exclusive production of (solid lines) $\pi^+$ and (dashed lines) $\pi^0$ mesons in the backward kinematics are: cross-sections as a function of $Q^2$ for $W = 2.2~\mathrm{GeV}$ and $-u = 0.5~\mathrm{GeV}^2$; cross-sections and asymmetries
as a function of $x_B$ for $Q^2 = 2.44~\mathrm{GeV}^2$ and $-u = 0.5~\mathrm{GeV}^2$. The asymmetries are evaluated for the electron beam energy $E_e = 10.6~\mathrm{GeV}$. The CLAS data~\cite{CLAS:2017rgp} are represented by black markers.}
\label{fig:scan_da}
\end{center}
\end{figure}

\begin{figure}[H]
\begin{center}
\includegraphics[width=0.3\textwidth]{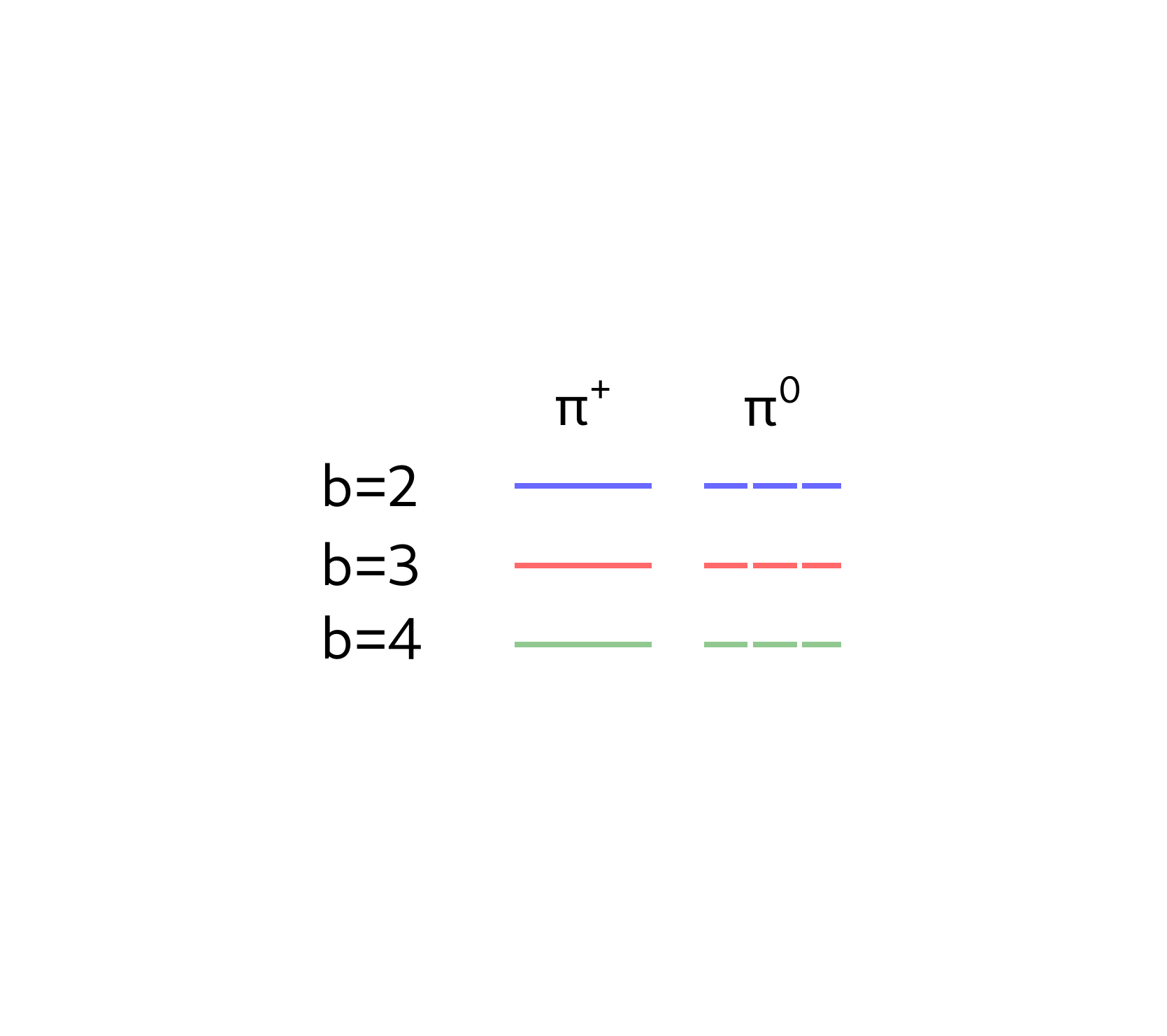}
\includegraphics[width=0.3\textwidth]{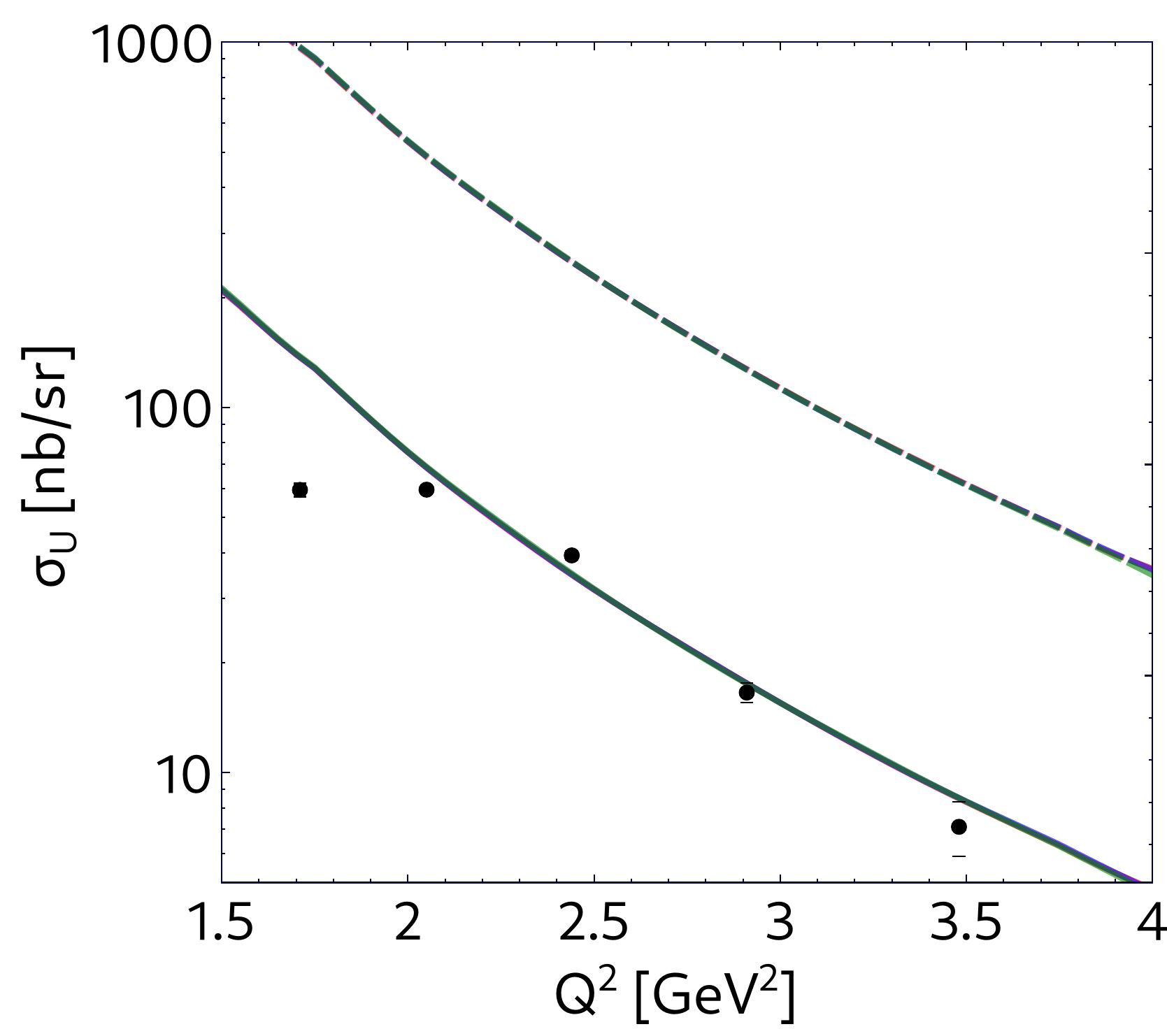}
\includegraphics[width=0.3\textwidth]{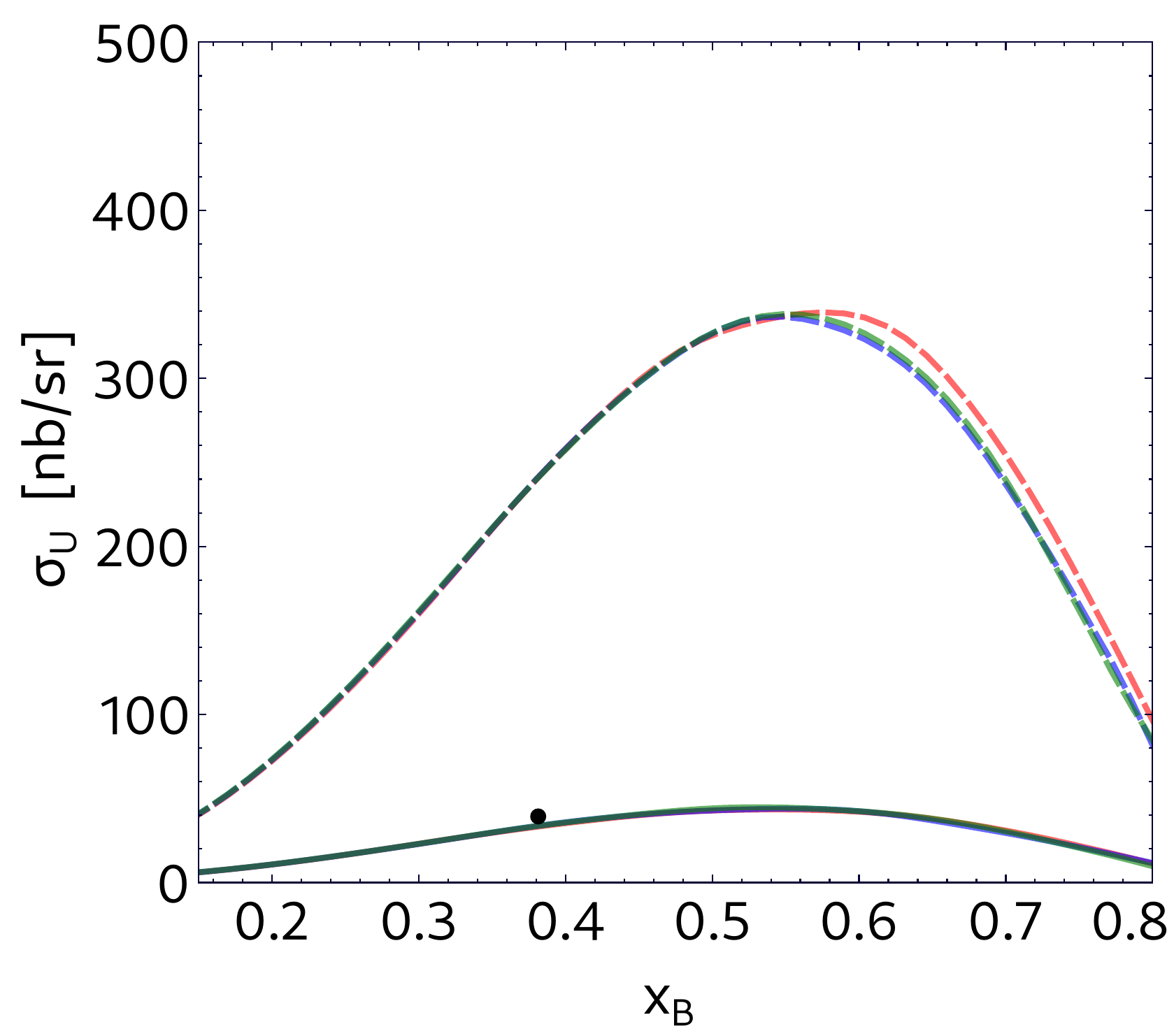}
\\
\includegraphics[width=0.3\textwidth]{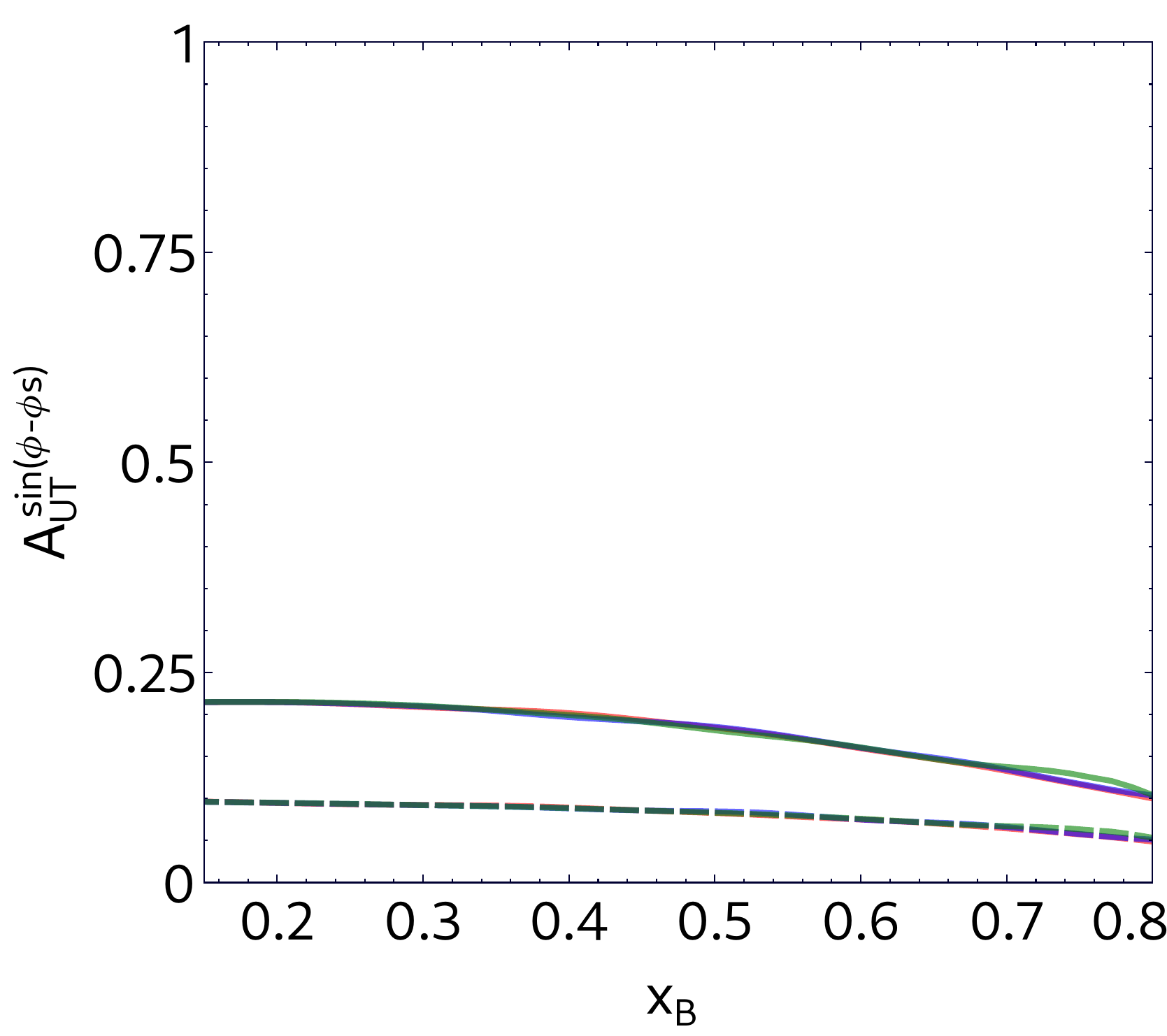}
\includegraphics[width=0.3\textwidth]{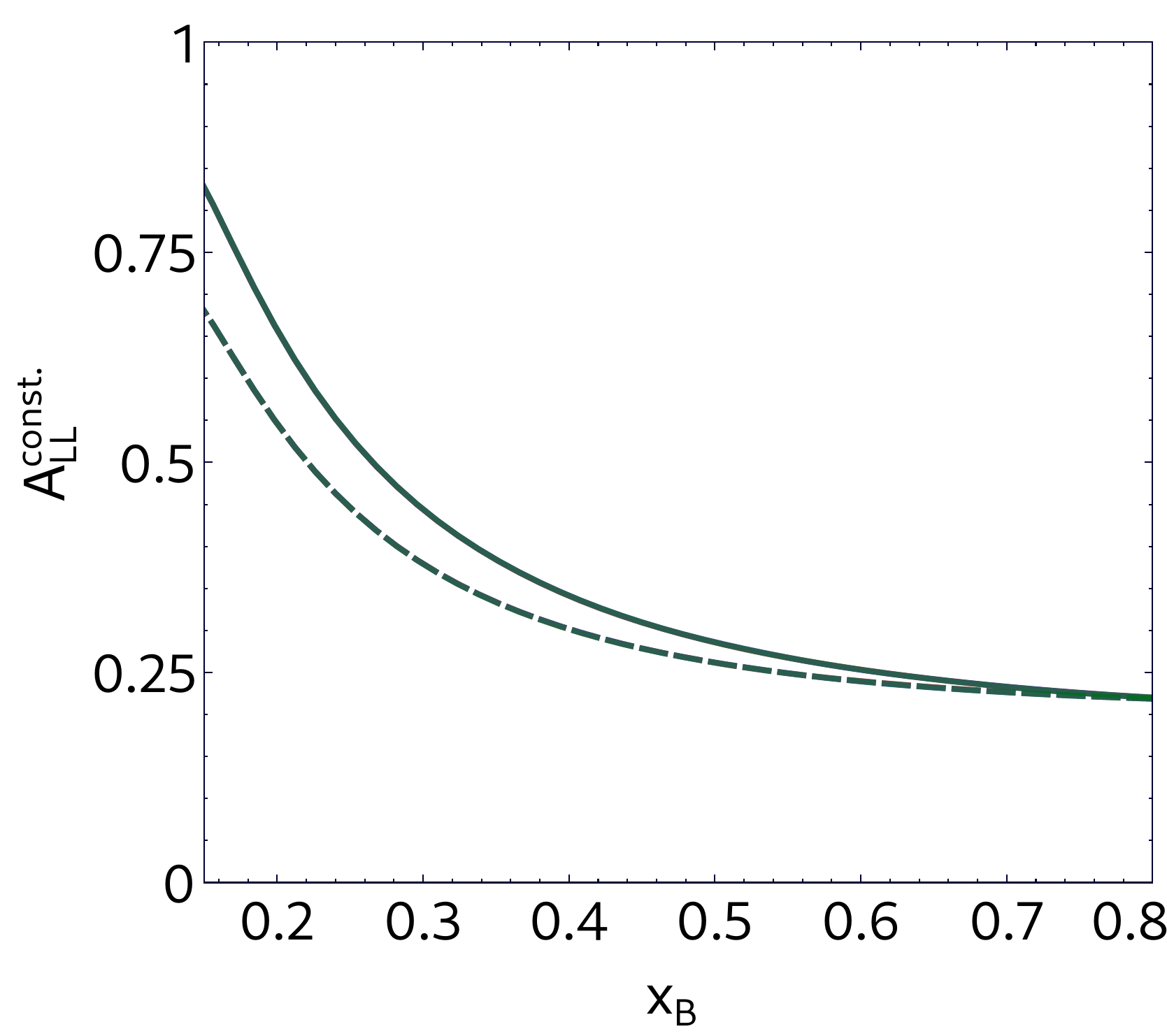}
\includegraphics[width=0.3\textwidth]{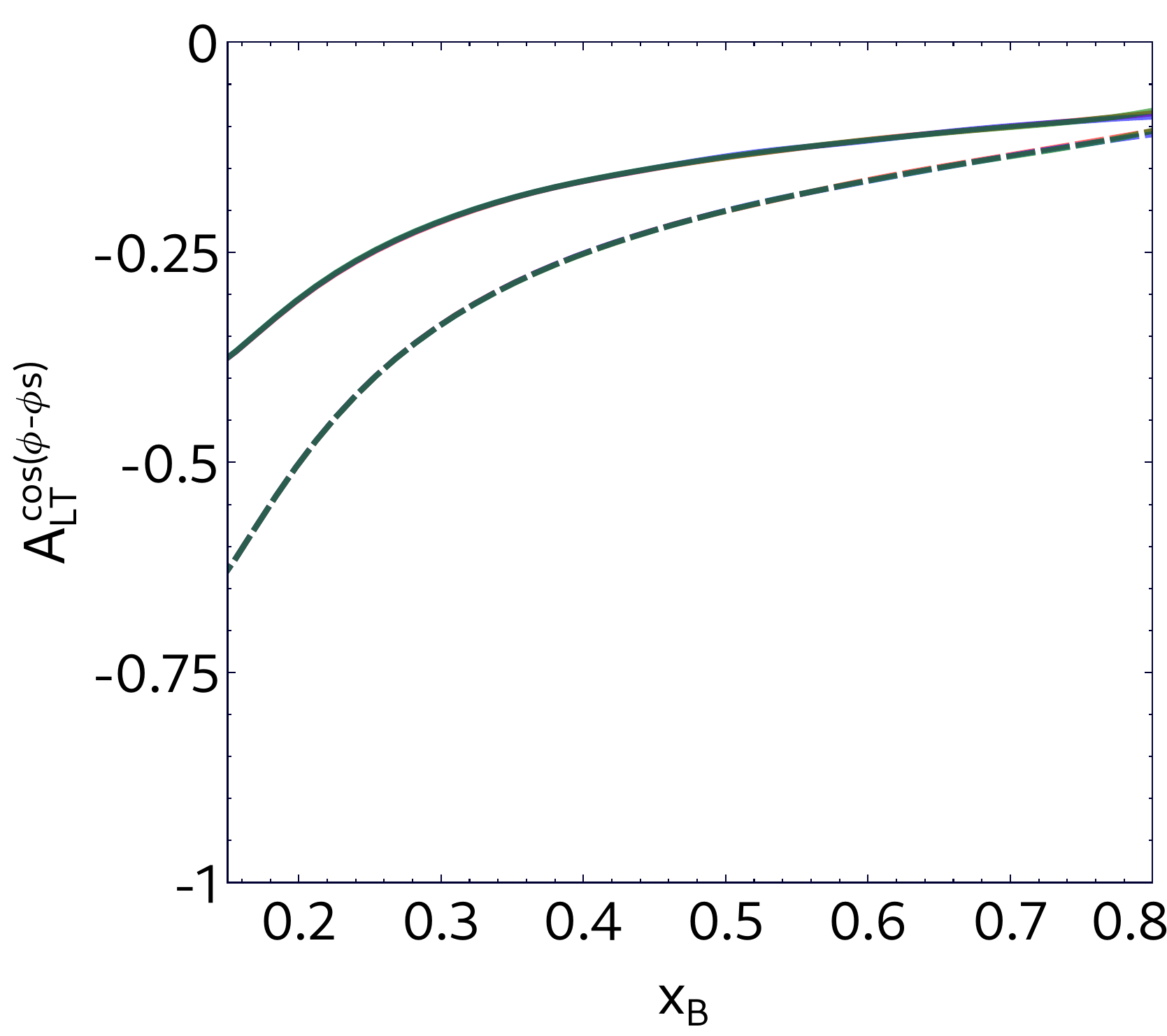}
\caption{Sensitivity to the parameter $b$. For further description see caption of Fig.~\ref{fig:scan_da}.}
\label{fig:scan_b}
\end{center}
\end{figure}

\begin{figure}[!ht]
\begin{center}
\includegraphics[width=0.3\textwidth]{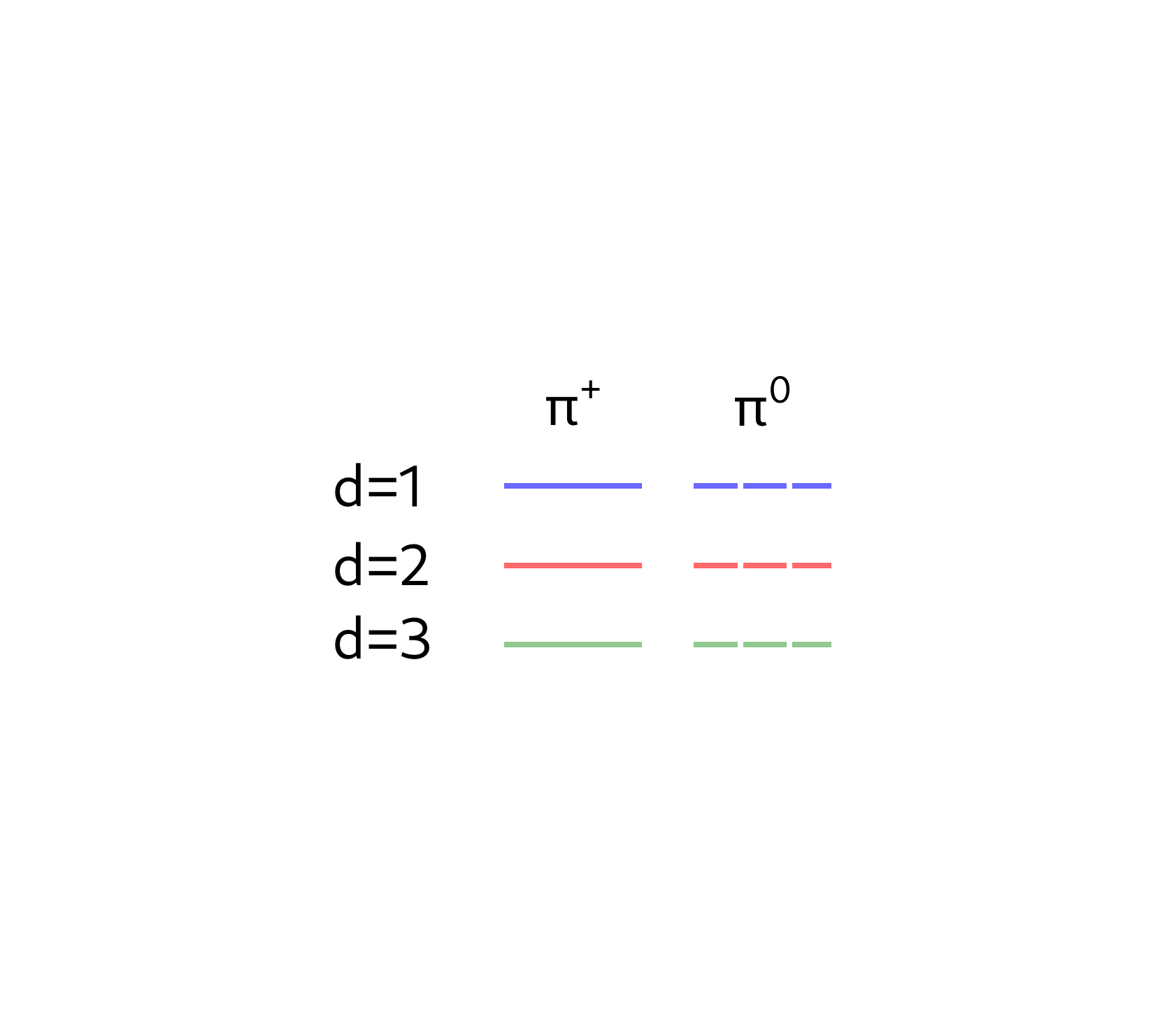}
\includegraphics[width=0.3\textwidth]{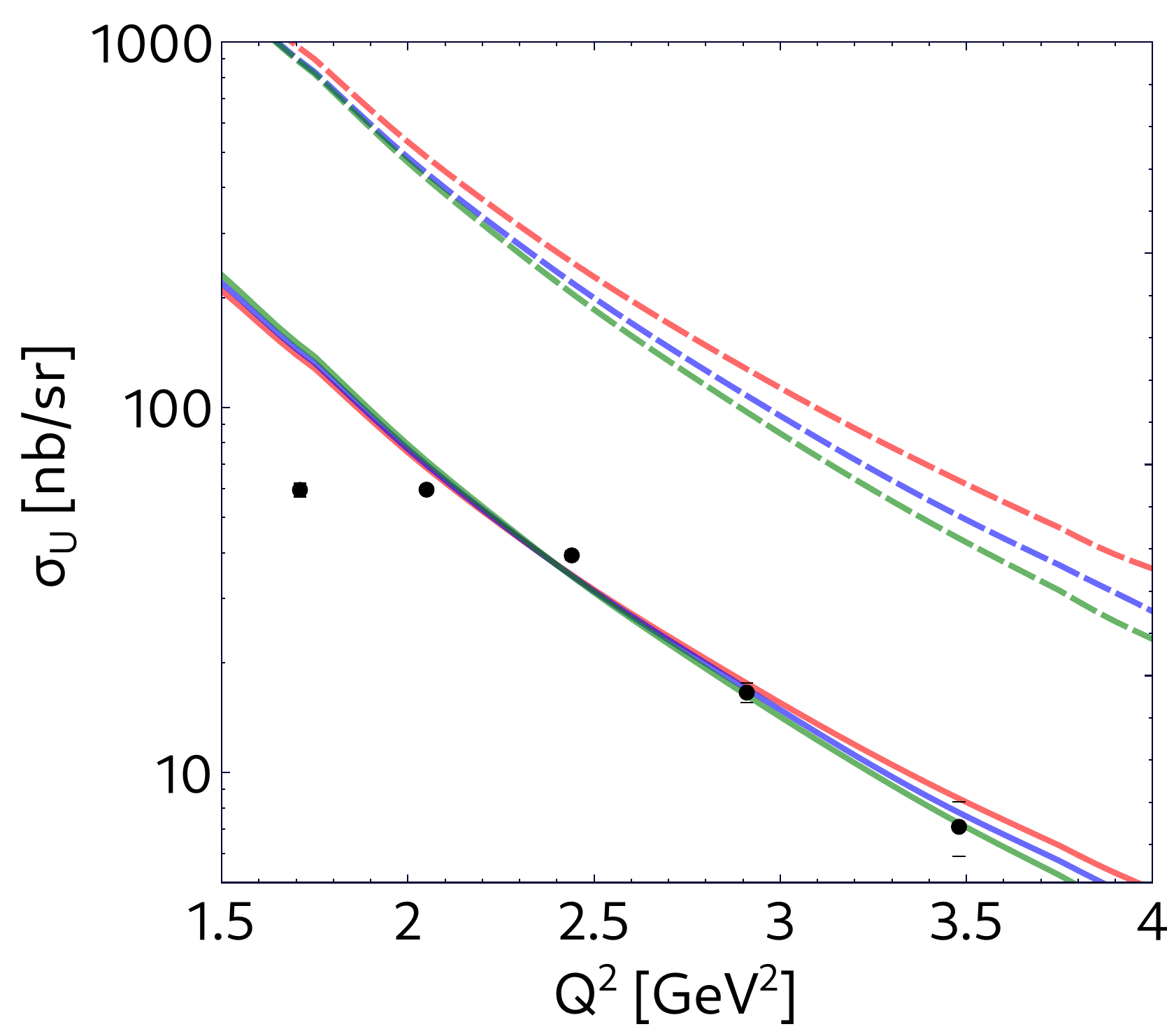}
\includegraphics[width=0.3\textwidth]{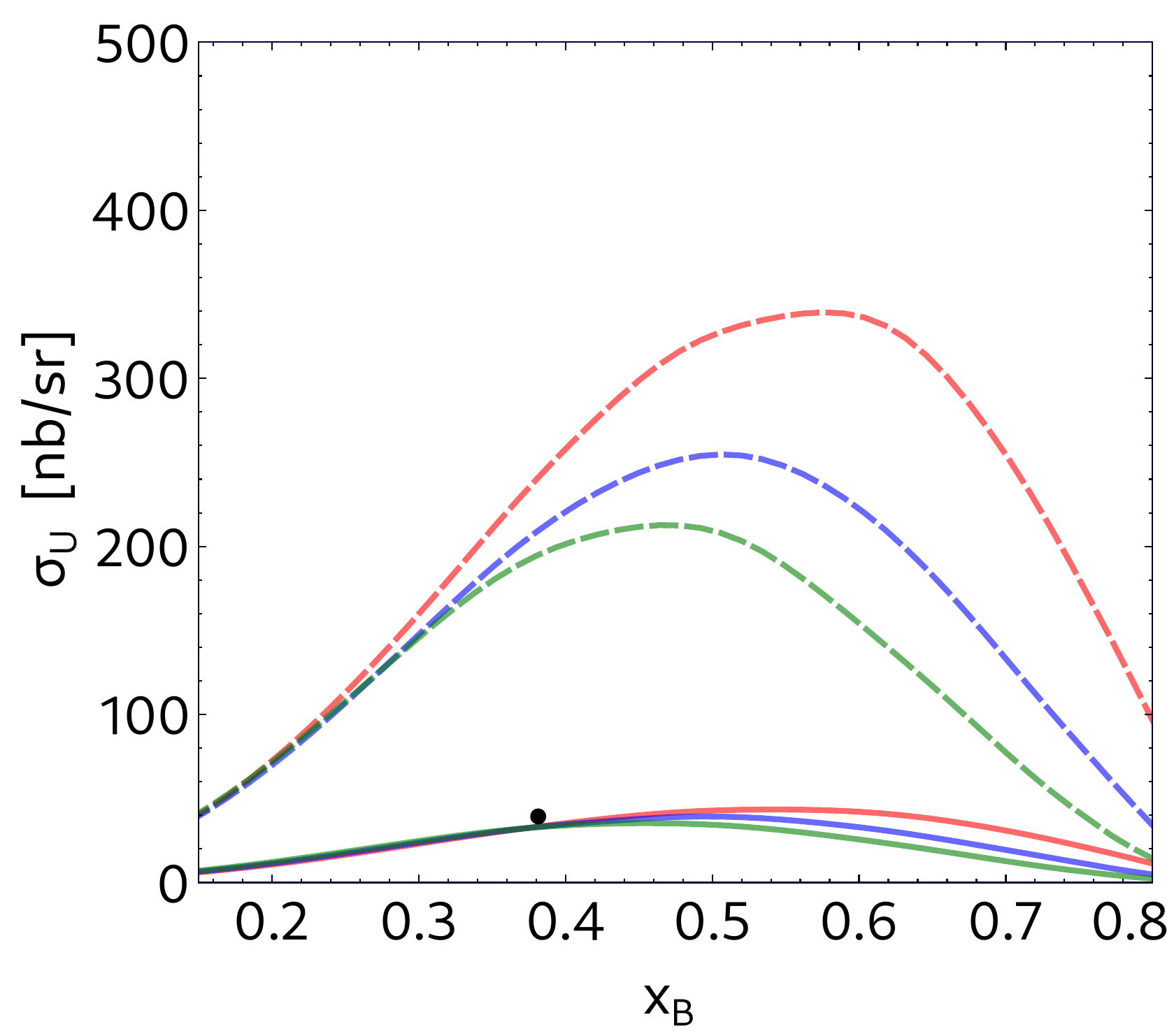}
\\
\includegraphics[width=0.3\textwidth]{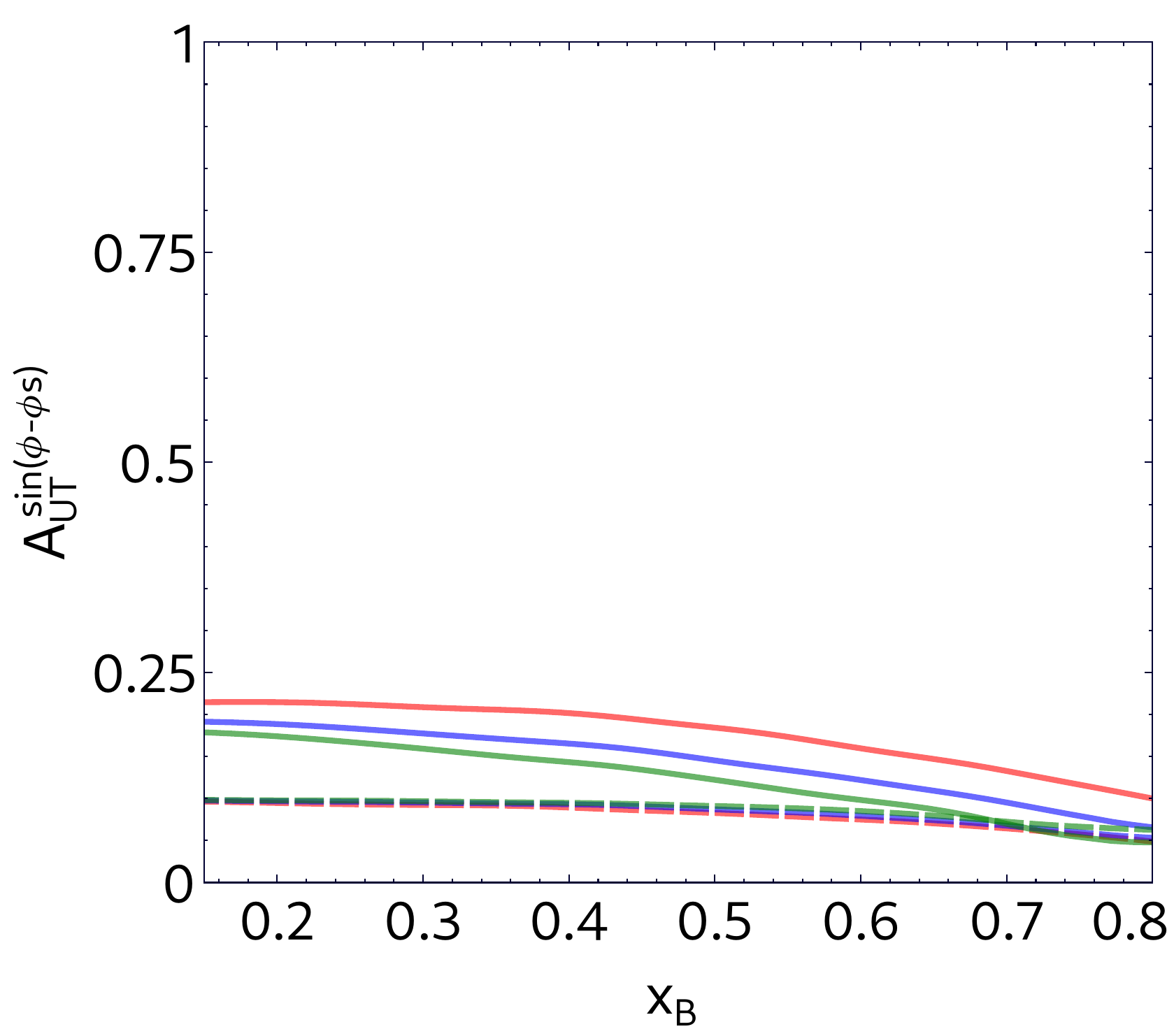}
\includegraphics[width=0.3\textwidth]{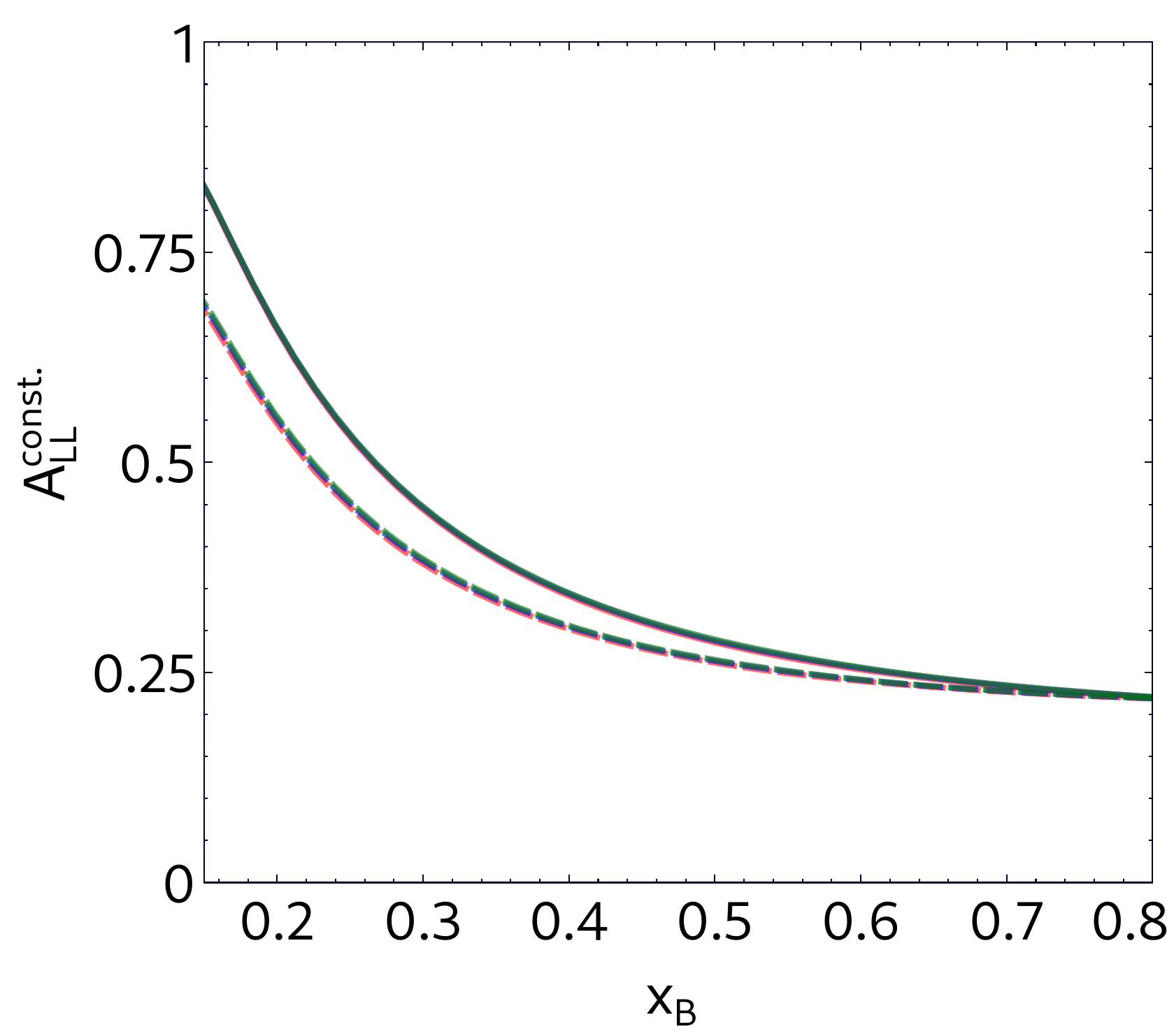}
\includegraphics[width=0.3\textwidth]{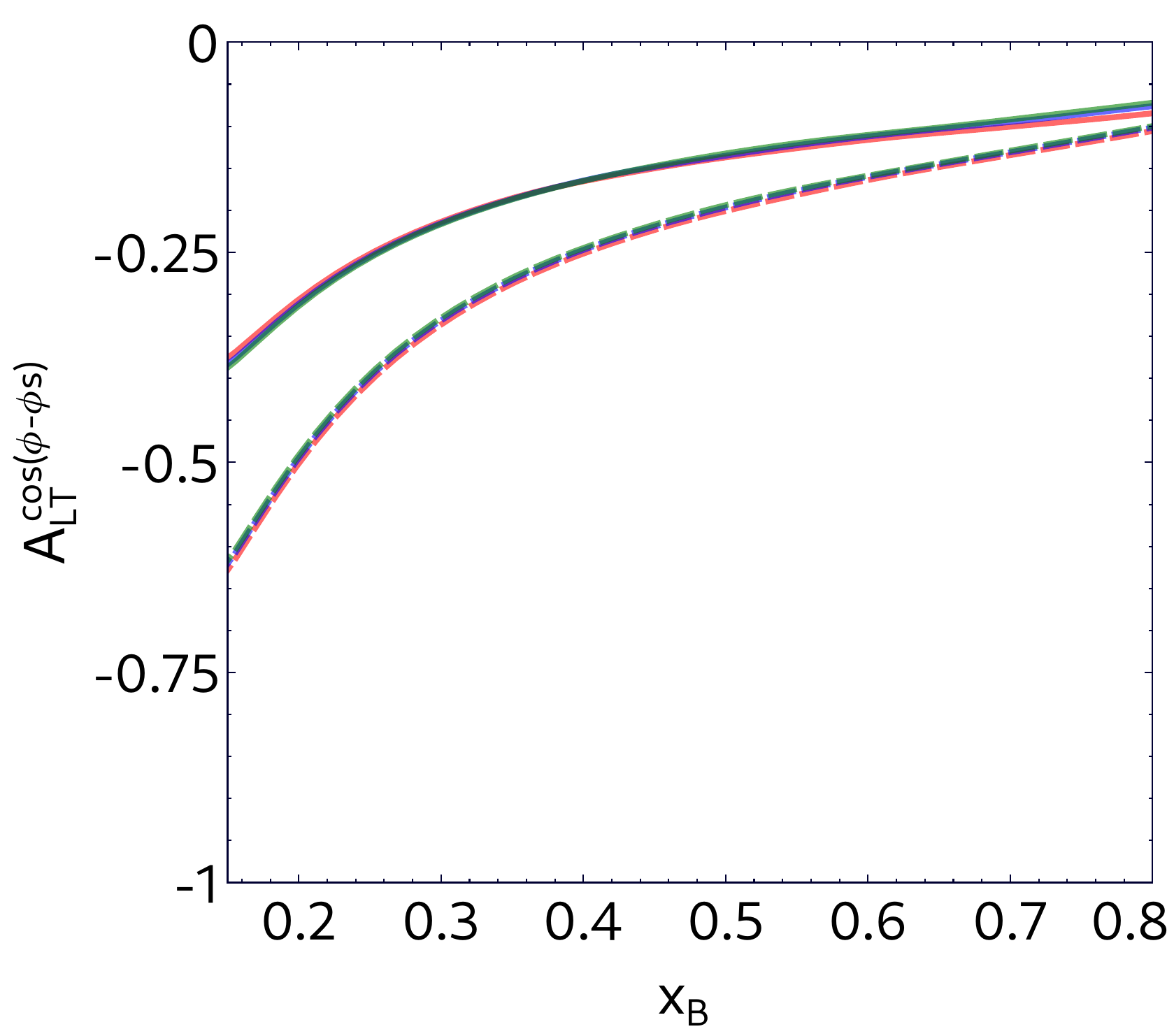}
\caption{Sensitivity to the parameter $d$. For further description see  caption of Fig.~\ref{fig:scan_da}.}
\label{fig:scan_d}
\end{center}
\end{figure}

\begin{figure}[!ht]
\begin{center}
\includegraphics[width=0.3\textwidth]{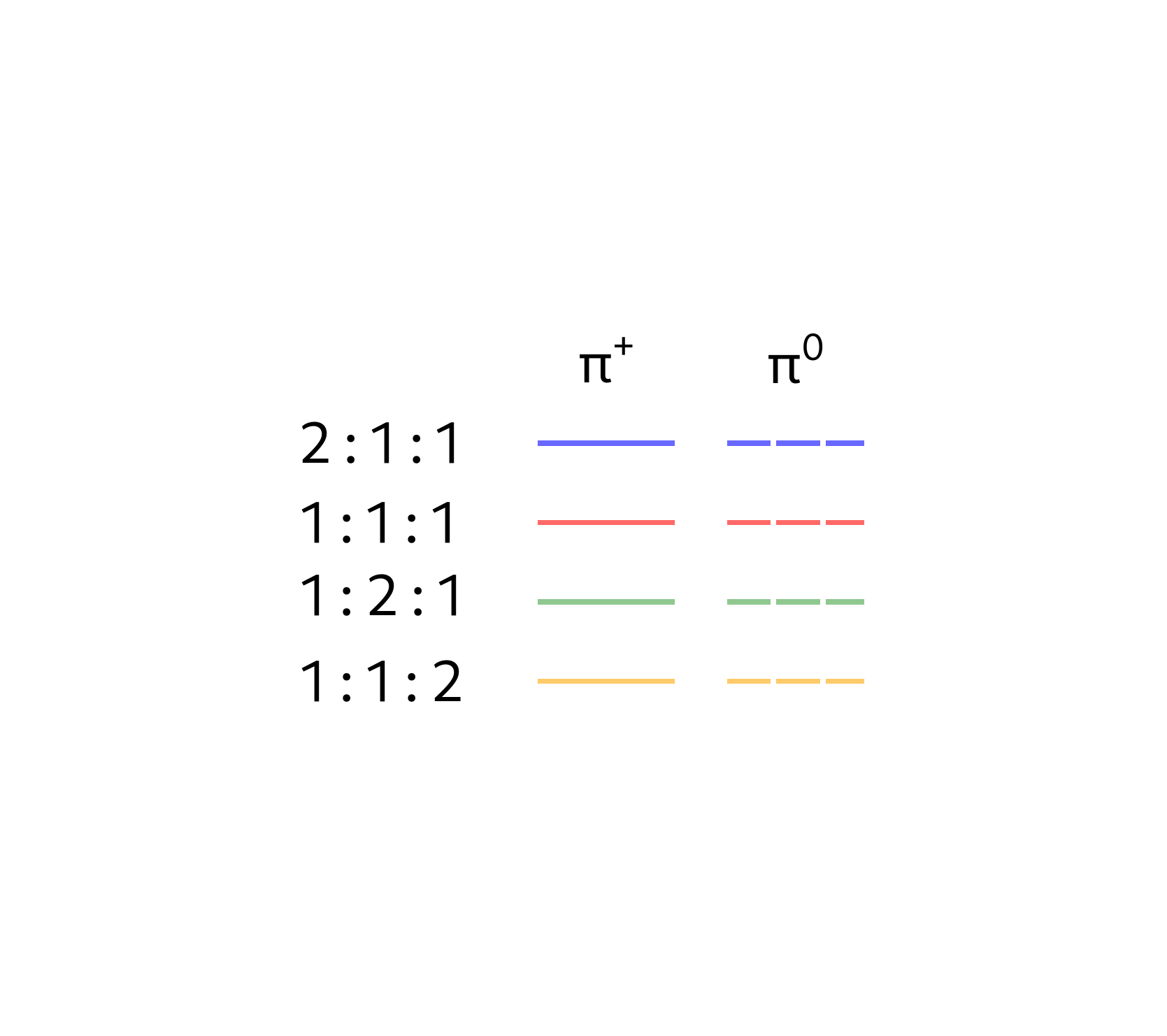}
\includegraphics[width=0.3\textwidth]{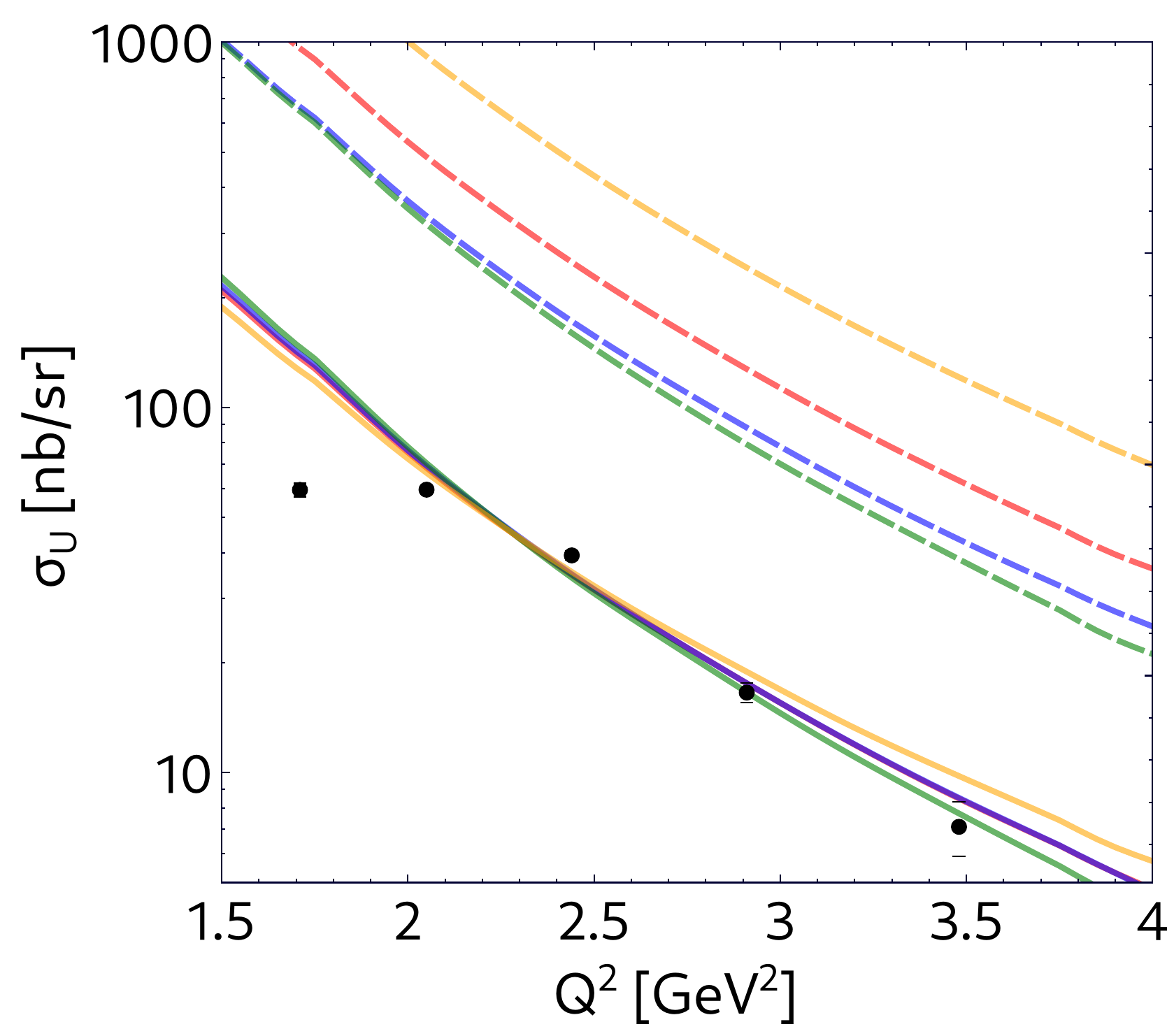}
\includegraphics[width=0.3\textwidth]{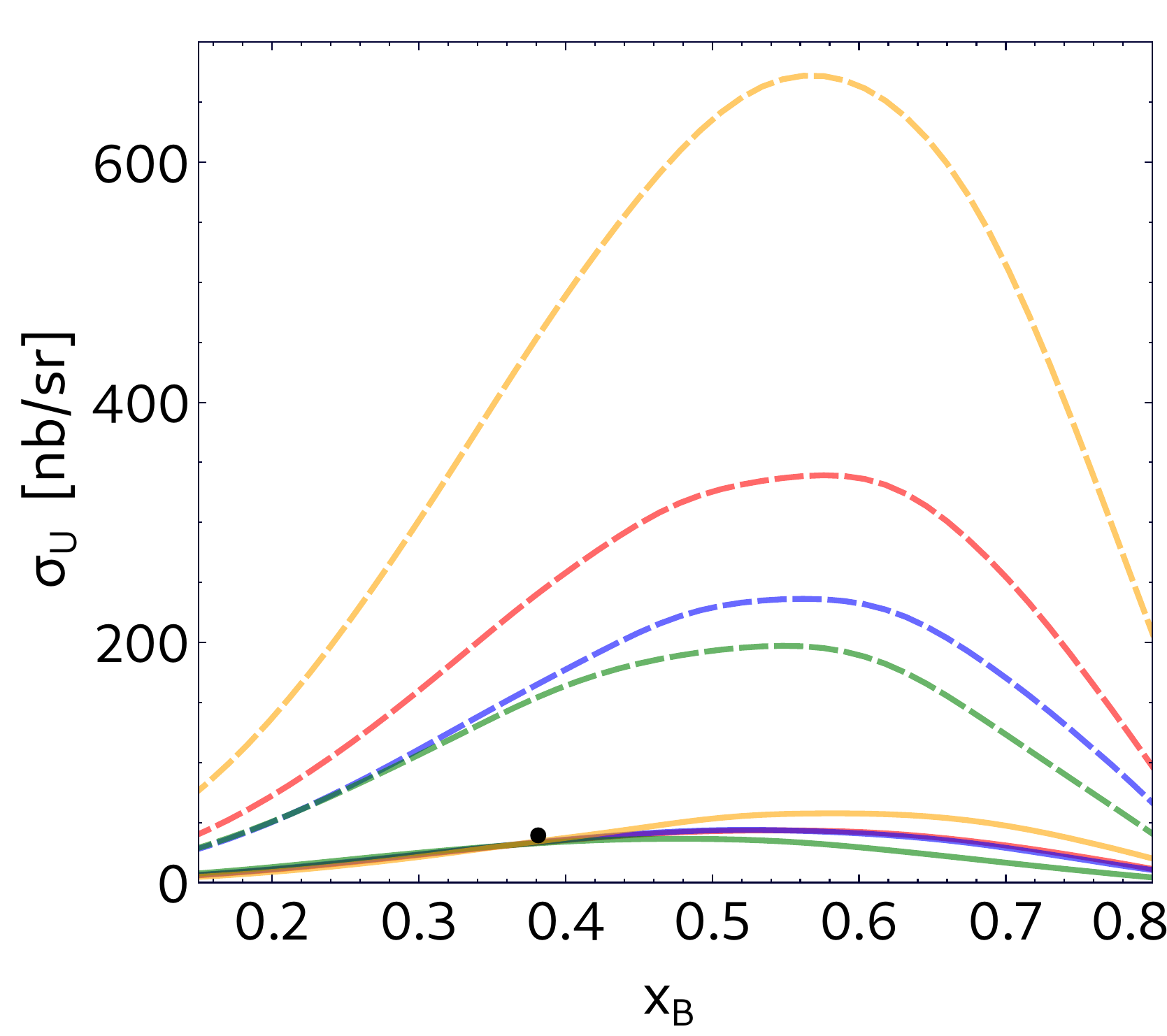} 
\\
\includegraphics[width=0.3\textwidth]{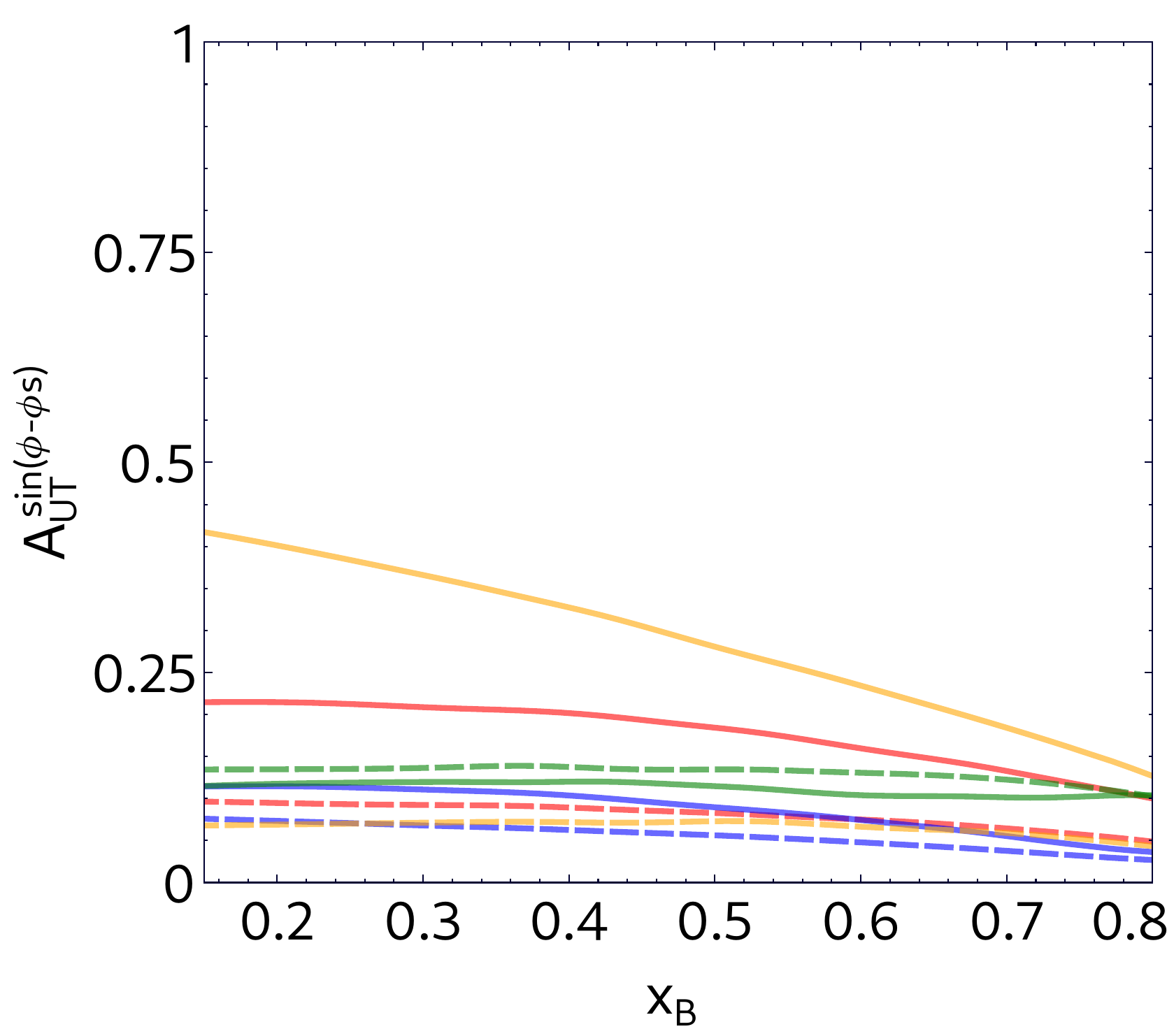}
\includegraphics[width=0.3\textwidth]{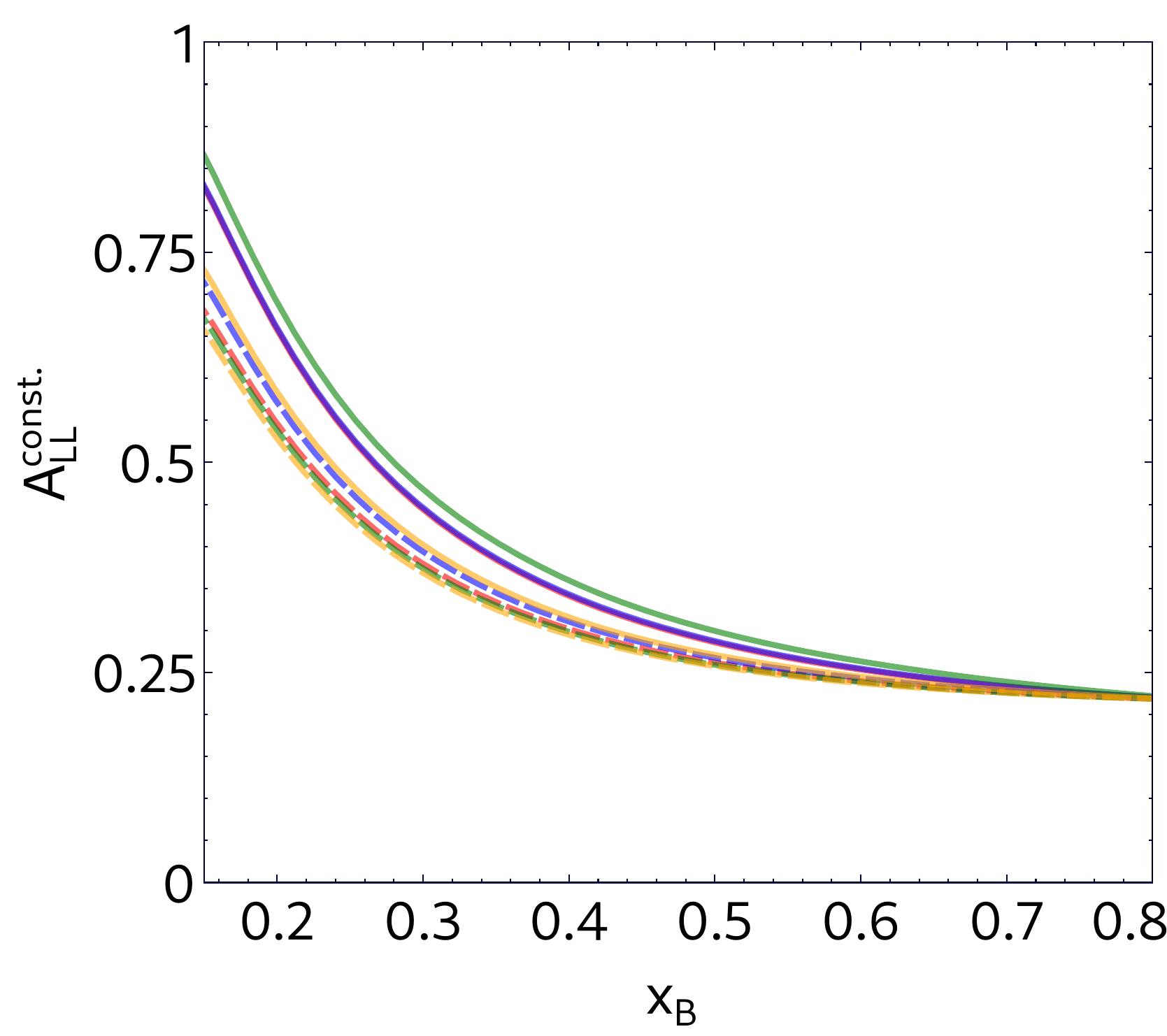}
\includegraphics[width=0.3\textwidth]{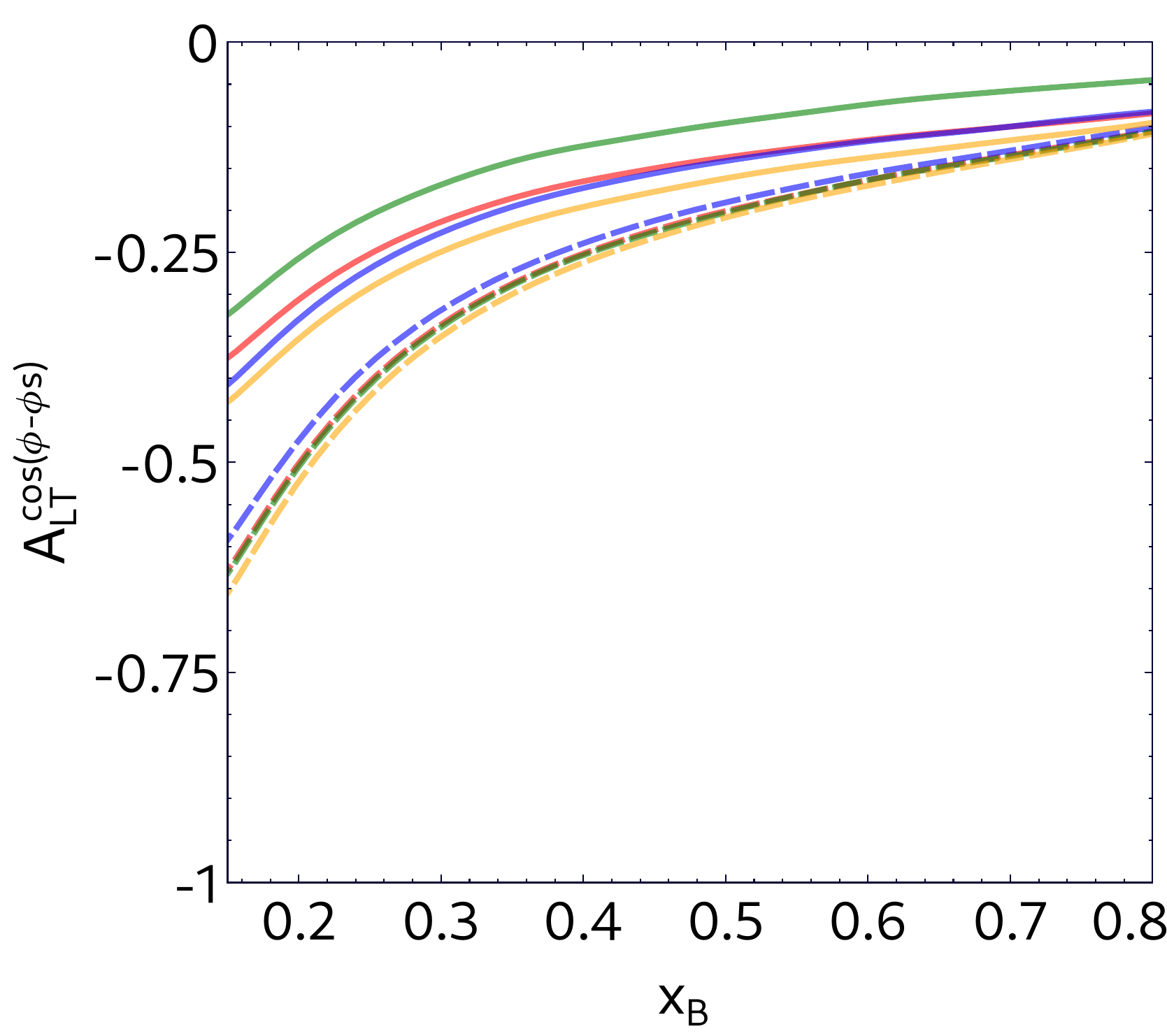}
\caption{Sensitivity to the ratio of the coefficients $c_0:c_1:c_2$ of orthogonal polynomials. For further description see caption of Fig.~\ref{fig:scan_da}.}
\label{fig:scan_pol}
\end{center}
\end{figure}

In the following, we will show the sensitivity of the predictions to various choices made in the modeling procedure. To keep this discussion simple, we will always change only one modeling option while keeping all other parameters at their default values. Various versions of our models will be always independently normalized to the CLAS data as described in the previous paragraph. The versions we consider are indicated in Table~\ref{tab:scan}, along with the values of the fitted $c_0$ parameter and the resulting values of $\chi^2$. The predictions for each version are shown in Figs.~\ref{fig:scan_da} to \ref{fig:scan_pol}, both as a function of $Q^2$ and $x_{B}$ for cross-sections, and as a function of $x_{B}$ for $A_{UT}^{\sin(\phi-\phi_s)}$, $A_{LL}^{\mathrm{const.}}$ and $A_{LT}^{\cos(\phi-\phi_s)}$ asymmetries.

\begin{table}[!h]
\caption{The versions of our model considered in this analysis, along with the value of the fitted $c_0$ coefficient and the resulting value of $\chi^2$. The designated default version is specified in the first line.}
\label{tab:scan}
\begin{tabular}{@{}p{0.05\textwidth}p{0.05\textwidth}p{0.15\textwidth}p{0.15\textwidth}p{0.15\textwidth}p{0.15\textwidth}@{}}
\toprule
$b$ & $d$ & $c_0:c_1:c_2$ & DA & $c_0$ & $\chi^2$  \\ \midrule
$2$ & $1$ & $1:1:1$       & BLW NLO     & $163.75$      & $41$ ~~~(default) \\
$2$ & $1$ & $1:1:1$       & COZ         & $ 43.61$      & $34$ \\
$2$ & $1$ & $1:1:1$       & asymptotic  & $201.98$      & $71$ \\
$2$ & $1$ & $1:1:1$       & BLW NNLO    & $ 95.22$      & $32$ \\[3pt]
$3$ & $1$ & $1:1:1$       & BLW NLO     & $164.13$      & $37$ \\
$4$ & $1$ & $1:1:1$       & BLW NLO     & $164.56$      & $37$ \\[3pt]
$2$ & $2$ & $1:1:1$       & BLW NLO     & $103.96$      & $39$ \\
$2$ & $3$ & $1:1:1$       & BLW NLO     & $ 78.14$      & $57$ \\[3pt]
$2$ & $1$ & $2:1:1$       & BLW NLO     & $ 86.25$      & $38$ \\
$2$ & $1$ & $1:2:1$       & BLW NLO     & $159.23$      & $55$ \\
$2$ & $1$ & $1:1:2$       & BLW NLO     & $133.30$      & $34$ \\
\bottomrule
\end{tabular}
\label{table:scan}
\end{table}

We start by showing the sensitivity of the prediction to the choice of the DA solution, see Fig.~\ref{fig:scan_da}. We remind that our default choice is the Braun-Lenz-Wittmann (BLW) NLO solution from Ref.~\cite{Braun:2006hz}. Other solutions considered are COZ~\cite{Chernyak:1987nv}, the asymptotic solution, i.e.,
\begin{equation}
\varphi_N^{\rm as}(y_1, y_2, y_3) = 120 y_1 y_2 y_3\,,
\end{equation}
and BLW NNLO~\cite{Lenz:2009ar}. We note that the probed DA solution enters both the DA model (via the $F^{(0)}(\kappa, \theta, \mu, \lambda, u)$ component) and, more importantly for the prediction of observables in the presented kinematics, the evaluation of the amplitudes (see Sect.~\ref{Sec_Amp}). From Fig.~\ref{fig:scan_da}, one can see that using different DA solutions results in significantly different predictions for all presented observables. Special attention should be given to the variation of predictions as a function of $Q^2$, although the effect appears moderate on a logarithmic scale. The difference stems from evolution effects. We also note that the asymptotic solution yields significantly different predictions.

In Figs.~\ref{fig:scan_b} and~\ref{fig:scan_d}, we show the sensitivity of the predictions to the choice of $b$ and $d$, respectively. These are the ``shape'' parameters, describing certain aspects of the modelled distributions, specifically, how fast the profile function and the forward limit vanish at the edges of their support domains. The difference in the predictions for observables is small for $b$, but quite significant for $d$ at large $x_B$. We note that we omit $b=1$, as it results in a too mild dependence of the profile function, which, when combined with the forward limit, does not lead to a vanishing TDA at the edges of the support region. In our modeling, $b=2$ is therefore the minimal acceptable value for this parameter.

Finally, in Fig.~\ref{fig:scan_pol}, we show the sensitivity to the ratio of normalization coefficients associated with the forward limit. Varying this ratio results in significantly different predictions for the $\pi^0$ channel. This leads us to conclude that the normalization of our model for the $\pi^0$ channel is not well constrained. In particular, we can vary the ratio between the $c_0$, $c_1$, and $c_2$ coefficients quite freely, resulting in markedly different predictions for the $\pi^0$ channel, while still maintaining reasonable agreement with the currently available experimental data for $\pi^+$ mesons. This highlights the  strong impact of future measurements of exclusive $\pi^0$ production in the backward kinematics on our understanding of the TDA formalism. 

\section{Monte Carlo studies}
\label{sec:mc}

In this section we show exemplary distributions of MC events obtained with the EpIC generator~\cite{Aschenauer:2022aeb} for both ``forward'' and ``backward'' $ep \to ep\pi^0$ processes, which are sensitive to GPDs and TDAs, respectively. The two samples are obtained for an unpolarized beam and fixed target, for the beam energy $E_e = 10.6\,\mathrm{GeV}$, and the following kinematic conditions:  
\begin{itemize}  
\item $0.05 < y < 0.9$,  
\item $2\,\mathrm{GeV}^2 < Q^2 < 10\,\mathrm{GeV}^2$,  
\item $0 < |t-t_0| < 1.5\,\mathrm{GeV}^2$ (for forward events),  
\item $0 < |u-u_0| < 1.5\,\mathrm{GeV}^2$ (for backward events),  
\item $0 < \phi < 2\pi$.  
\end{itemize}  
where $y$ is the inelasticity variable, while $\phi$ is the azimuthal angle between the leptonic and hadronic planes~\cite{Bacchetta:2004jz}. The forward sample has been generated using the Goloskokov-Kroll GPD model~\cite{Goloskokov:2005sd,Goloskokov:2008ib} and the description of the considered process based on the modified-perturbative approach~\cite{Goloskokov:2009ia,Goloskokov:2011rd} developed by the same authors. We used these developments as they are implemented in the PARTONS framework~\cite{Berthou:2015oaw}.  

The presented distributions do not account for any detector effects, such as geometrical acceptance, energy smearing and detector efficiencies. We show them only to demonstrate key features of both processes. Moreover, the distributions of forward and backward events are not normalized to a common luminosity factor. That is, we simply show distributions of $10^5$ events for each case. The reason for such simplified treatment is the unconstrained normalization of our TDA model in the relevant kinematic region, as presented in the ``Sensitivity to modelling options'' part of the previous section.  

\begin{figure}[!ht]
\begin{center}
\includegraphics[width=0.45\textwidth]{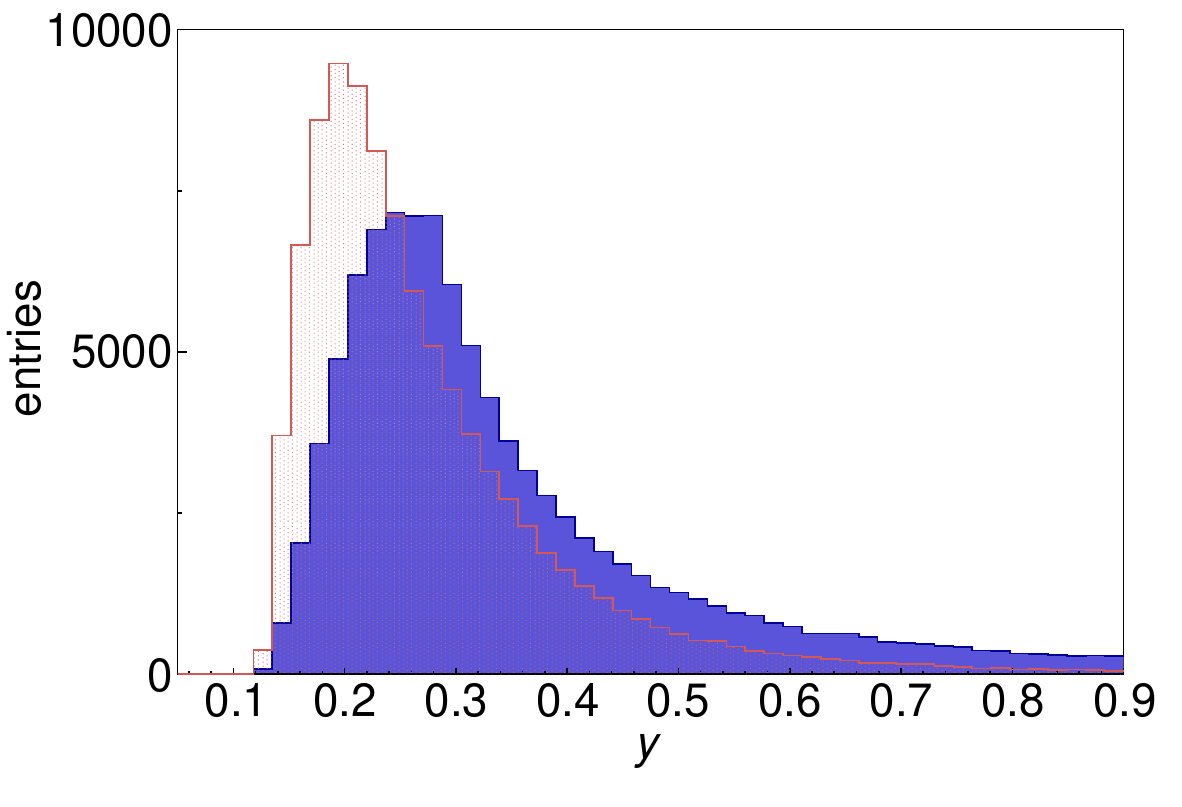}
\includegraphics[width=0.45\textwidth]{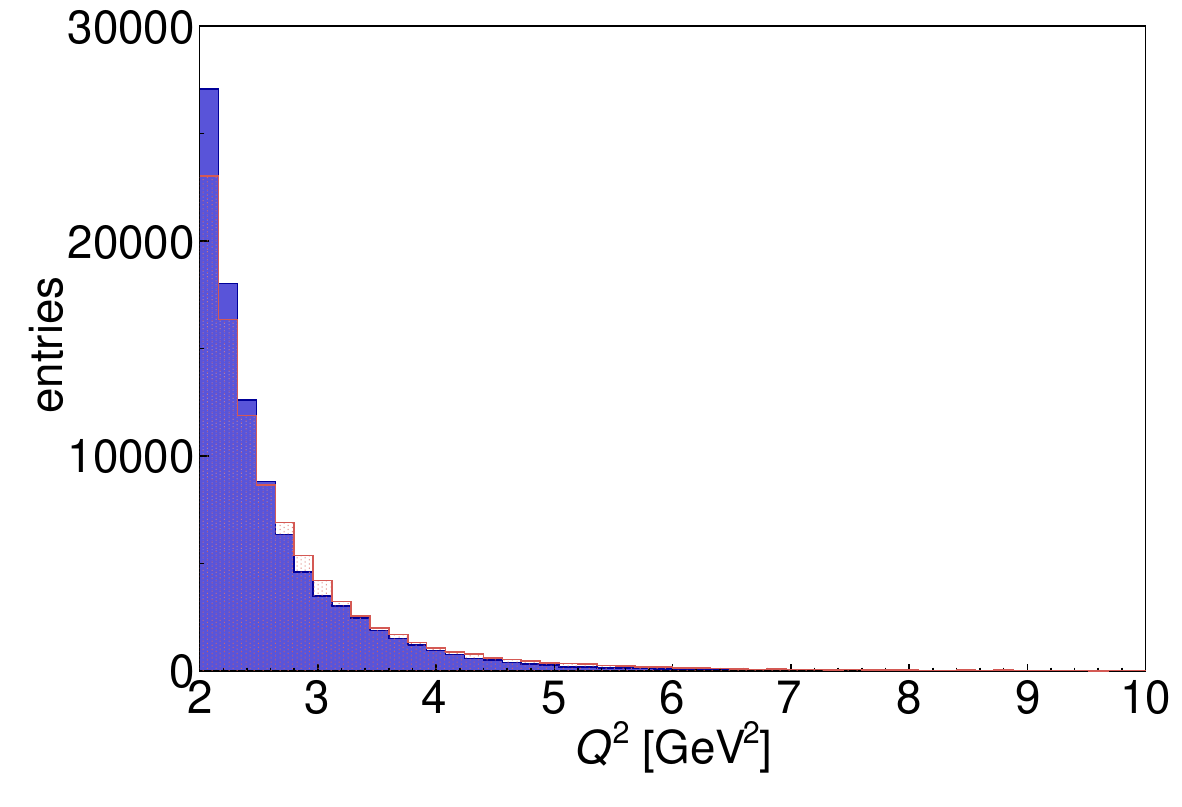}
\includegraphics[width=0.45\textwidth]{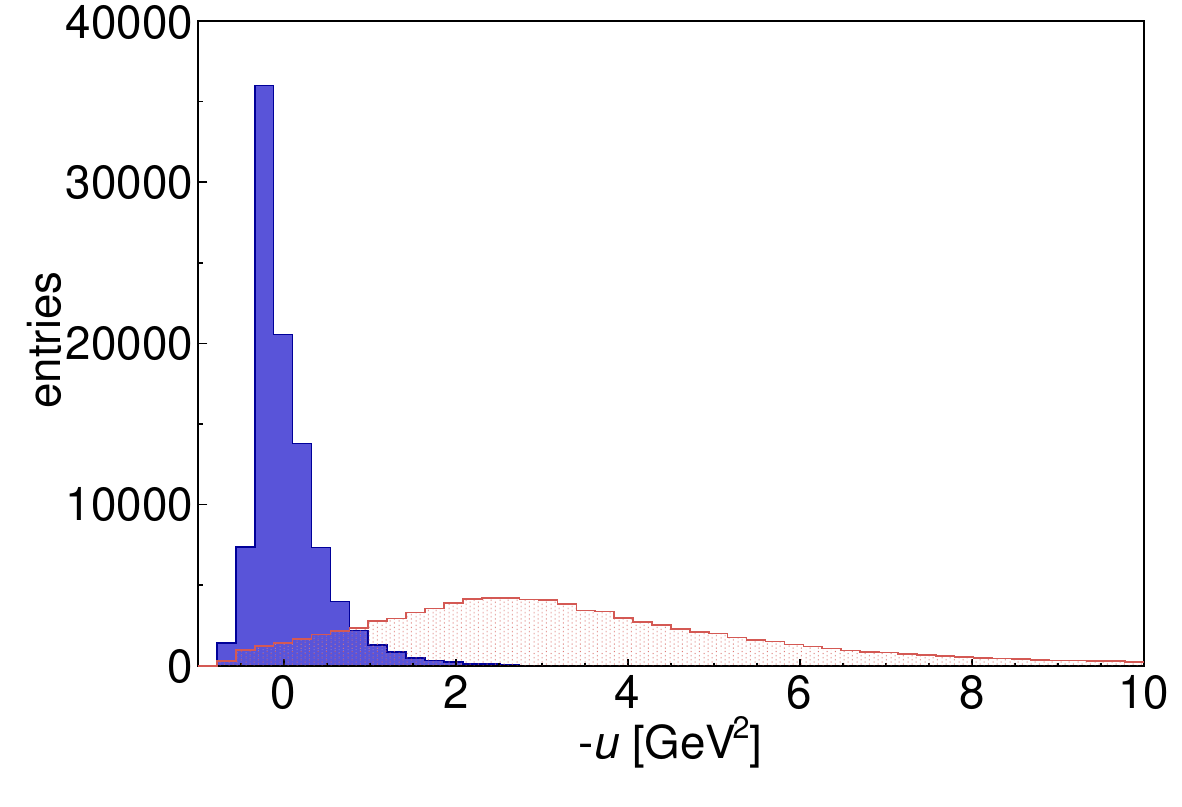}
\includegraphics[width=0.45\textwidth]{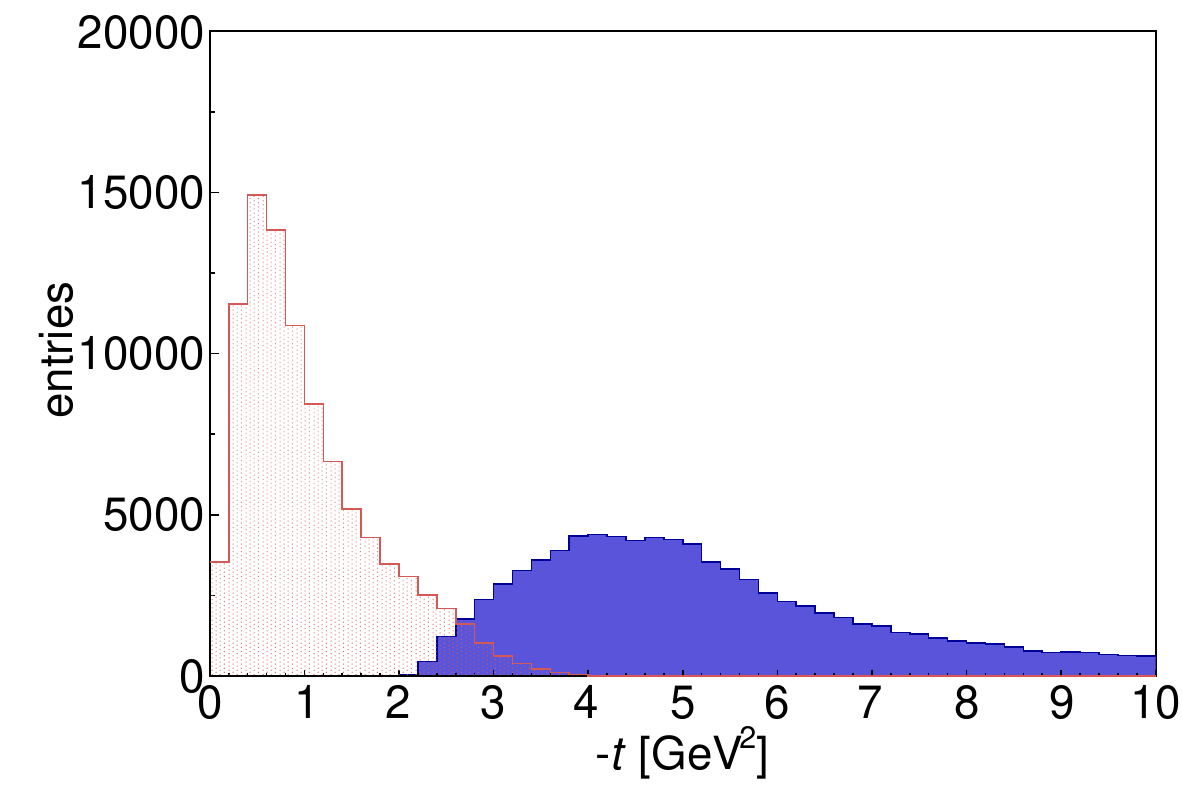}
\end{center}
\caption{Distributions of MC events for forward (red dotted histograms) and backward (blue solid histograms) processes in the variables $y$, $Q^2$, $-u$, and $-t$. Each distribution was obtained from $10^5$ events, without any normalization of the two processes to a common luminosity factor (see the text for details).}
\label{fig:mc}
\end{figure}

The distributions of MC events in the variables $y$, $Q^2$, $-u$, and $-t$ are shown in Fig.~\ref{fig:mc}. Similar distributions are observed for $y$ and $Q^2$, while those for $-u$ and $-t$ differ significantly, with a clear separation between the forward and backward cases. It is important noting that the forward (backward) peak in the $-u$ ($-t$) distribution is significantly smeared, which is a consequence of the presence of $Q^2$ in the relation
\begin{equation}
    u = 2 m_{N}^{2} + m_{\pi}^2 - Q^{2} - s - t \,.
\end{equation}
It is therefore imperative to look for the signature of backward processes by analyzing distributions of events in the $u$ variable, and not $t$ (just as we look for forward processes by examining distributions in the $t$ variable, and not $u$). 

A closer inspection of the histograms reveals that the $Q^2$ dependence of the reaction involving TDAs is steeper than that of the reaction sensitive to GPDs. Such suppression is expected, since the forward process is sensitive to twist-2 GPDs and DAs (even if the modified-perturbative approach \`a la GK involves some of higher-twist terms), whereas the backward process is sensitive to twist-3 TDAs.

An additional feature we deduce from Fig.~\ref{fig:mc} is a milder dependence on $t$ for the forward process, compared to the dependence on $u$ for the backward process. This is a consequence of different choices in the modelling of GPDs and TDAs, and can be tuned once data for the latter become available.  
\section{Summary}
\label{sec:summary}

In this article, we introduce a new approach to the modeling of TDAs for the processes $ep \to en\pi^+$ and $ep \to ep\pi^0$. TDAs are well-defined objects in the QCD collinear factorization framework and are complementary to GPDs. They provide valuable new information about the transition of a hadron into different particles, encoding correlations between partons during such transitions. The phenomenology of TDAs and the experimental programme related to them are, however, still very much in the exploratory phase and underdeveloped, which was a direct motivation for our study. Let us note that dedicated experiments \cite{Li:2020nsk, Huber:2022wns} are proposed or planned at JLab and more data could be gathered from various set-ups in electroproduction but also in photoproduction or with meson or antiproton beams~\cite{PANDA:2016scz,Pire:2022kwu}.

The modeling of TDAs we propose is flexible, which is a significant improvement with respect to the previously considered models, which had a tendency to overshoot existing experimental data without any means of correction. The model we propose correctly reproduces TDAs in the backward limit, while the forward limit is mapped with the help of 2D orthogonal polynomials. The evolution of TDAs is used, however, only in a limited form, i.e. through DAs.

The model is constrained by existing $\pi^+$ data measured by CLAS. The predictions for the $\pi^0$ channel turn out to be unconstrained in terms of normalization. This leads to a twofold conclusion: i) $\pi^+$ and $\pi^0$ channels provide complementary information about TDAs; ii) measurement of the $\pi^0$ channel will be impactful, as it will provide unique information about TDAs.

The study is complemented by an analysis of the sensitivity of predictions to various modeling assumptions. It provides guidance on how certain parameters can be constrained by experimental data, including the choice of DA solution. We particularly stress the importance of polarization observables that nicely complement cross-sections studies and allow to check some basic predictions of our approach together with a first separation of various TDAs. Indeed, the TDA factorization predicts that in addition to the single transverse target spin asymmetry, there are two double spin asymmetries which are non vanishing at leading twist; checking that they do not diminish at higher $Q^2$ values, contrarily to the case for forward meson electroproduction, will be an important test of the whole picture. We moreover proved that these asymmetries were not parametrically small under reasonable model assumptions for the TDAs.

In addition, we provide a tentative MC study. Its purpose is to announce the availability of the MC generator for TDAs, which offers direct support to experimentalists. Furthermore, we highlight the importance of measuring the TDA-related processes in the $u$-channel, i.e. by analyzing distributions of events as a function of the $u$ variable, and not $t$, which is a common misconception.

As a final remark, let us stress that  lattice studies could also shed light on the magnitude and variations of the TDAs, for instance by computing the $u$ dependence of various $x_i$ moments of these functions. Other non-perturbative approaches such as models based on light-cone hadronic wave functions~\cite{Pasquini:2009ki, Pasquini:2024qxn, Castro:2025rpx} should also be pursued.

\begin{acknowledgments}
We thank Wenliang Li and Garth Huber for fruitful discussions. The work of P.~S. and L.~S. was supported by the Grant No.~2024/53/B/ST2/00968 of the National Science Centre.
K.S. was supported by Basic Science Research Program through the National Research Foundation of Korea (NRF) funded by the Ministry of Education RS-2023-00238703 and RS-2018-NR031074.

\end{acknowledgments}

\appendix
\section{Nucleon DAs}

In this Appendix we present our notations for the leading twist-3 nucleon DAs (for a review see {\it e.g.} Ref.~\cite{Braun:1999te}).
In particular, proton DA is defined as \cite{Chernyak:1984bm}
\be
\begin{aligned}
& 4 
(p_N \cdot n)^3 \int\left[\prod_{j=1}^3 \frac{d \lambda_j}{2 \pi}\right] e^{i \sum_{k=1}^3 y_k \lambda_k(p_N \cdot n)} 
\left\langle 0 \right| \widehat{O}_{\rho \tau \chi}^{\,uud}\left(\lambda_1 n, \lambda_2 n, \lambda_3 n\right)\left|N^p\left(p_N, s_N\right)\right\rangle \\
& =\delta\left(y_1+y_2+y_3-1 \right)  f_N \sum_s\left(s^{N}\right)_{\rho \tau, \chi} \Phi_s^{ N}\left(y_1, y_2, y_3; \mu^2\right).
\label{Def_N_DA}
\end{aligned}
\ee
The sum in 
(\ref{Def_N_DA}) 
stands over the set of $3$ leading-twist-$3$
Dirac structures
\be
(s^{ N})_{\rho \tau, \chi}=\left\{ (v^{N} )_{\rho \tau, \chi}, (a^{ N})_{\rho \tau, \chi}, (t^{ N})_{\rho \tau, \chi}\right\}.
\ee
We employ the shortened notations for the set of invariant leading twist-$3$ nucleon  DAs
\be
\Phi^{N}=\left\{V^p, A^p, T^{p}\right\}.
\label{Set_of_DAs}
\ee
Due to isospin and permutation symmetry (see {\it e.g.} ref.~\cite{Pire:2011xv}), the leading twist-$3$ nucleon  DAs 
$V^p, A^p, T^{p}$
possess specific properties under permutation of variables and can be expressed in terms of 
a single function $\varphi_N(y_1,y_2,y_3)$ as
\be
&&
\left\{ V^p, \, A^p\right\}(y_1,y_2,y_3)= \frac{1}{2} \left( \pm \varphi_N(y_1,y_2,y_3)+ \varphi_N(y_2,y_1,y_3) \right); 
\nn \\ &&
T^p(y_1,y_2,y_3)= \frac{1}{2} \left(\varphi_N(y_1,y_3,y_2)+ \varphi_N(y_2,y_1,y_3) \right);
\label{DAs_through_varphi}
\ee

\section{Cross-channel nucleon exchange contribution for $\pi N$ TDA}
\label{App_Nucl_exchange}

\begin{figure}[!ht]
\begin{center}
\includegraphics[width=0.25\textwidth]{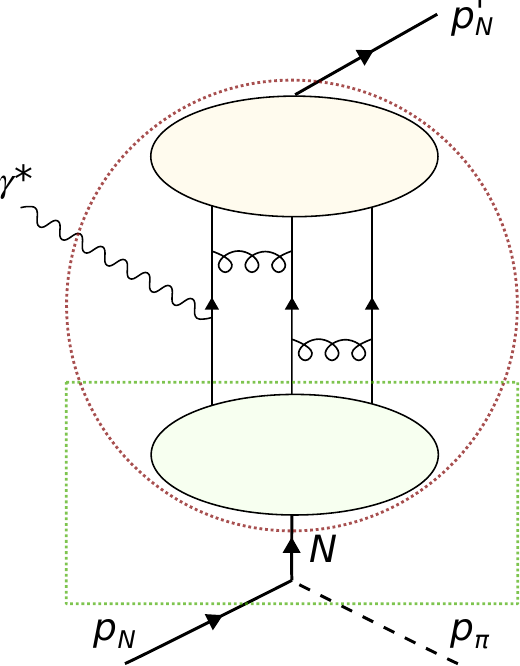}
\end{center}
\caption{Cross-channel
nucleon exchange model for $\pi N$ TDAs; the dotted circle contains a typical LO graph for the nucleon electromagnetic form factor in perturbative QCD;
the rectangle contains the cross-channel nucleon contribution into $\pi N$ TDAs.}
\label{Fig_Npole}
\end{figure}

The cross-channel nucleon exchange model for $\pi N$ TDAs provides a most obvious way 
to estimate them. It originates from a simple cross-channel exchange diagram, Fig.~\ref{Fig_Npole}, involving LO
graphs for nucleon electromagnetic form factor in perturbative QCD and the standard $\pi \bar{N}N$ interaction vertex. The phenomenological status
of this TDA model is similar to that of cross-channel meson exchange models for GPDs broadly employed in the early days of the  GPD phenomenology, see {\it e.g.} Refs.~\cite{Mankiewicz:1997uy, Frankfurt:1998et}.

The cross-channel nucleon exchange contributes only into the isospin $I=\half$ $\pi N$ TDAs expressing $\pi N$ TDAs in terms of nucleon DAs.
This mechanism makes contributions into $\pi N$ TDAs $V_{1,2}^{\pi N_{I=1/2}}$, $A_{1,2}^{\pi N_{I=1/2}}$, $T_{1,2}^{\pi N_{I=1/2}}$, while $T_{3,4}^{\pi N}=0$ within this model.
Moreover, it imposes a constraint between TDAs:
\be 
\left.\left\{V_2, A_2, T_2\right\}^{\pi N_{I=1 / 2}}
\left(x_1, x_2, x_3, \xi, u\right)\right|_{N(940)}
=\frac{1}{2} \left.\left\{V_1, A_1, T_1\right\}^{ \pi N_{I=1 / 2}}
\left(x_1, x_2, x_3, \xi, u \right)\right|_{N(940)}.
\label{Constraint_TDA12_Nexchange}
\ee
Accounting for the isospin and permutation symmetry properties (see Ref.~\cite{Pire:2011xv}), it is possible to express all $\pi N$ TDAs within the cross-channel nucleon exchange model in terms of a single function 
\be
H^{\pi N_{I=1 / 2}}(x_1,x_2,x_3,\xi,u) \Big|_{N(940)}= \prod_{k=1}^3 \theta\left(0 \leq x_k \leq 2 \xi\right)
\left(g_{\pi N N}\right) \frac{m_N f_\pi}{u-m_N^2} 2 \xi \frac{1}{(2 \xi)^2} \varphi_N\left(\frac{x_1}{2 \xi}, \frac{x_2}{2 \xi}, \frac{x_3}{2 \xi}\right).
\label{H_from_N_exchange}
\ee
Here $\varphi_N(y_1,y_2,y_3)$ is the leading twist-$3$ nucleon DA, {\it cf.} Eq.~(\ref{DAs_through_varphi}) and $g_{\pi NN} \simeq 13$ is the dimensionless pion-nucleon coupling constant. 
Note that (\ref{H_from_N_exchange}) is a pure ``$D$- term-like'' contribution. It is non-zero only in the ERBL-like region, and its $(n_1, n_2, n_3)$-th
Mellin moments give rise to monomials of $\xi$ of the maximal possible power $n_1+n_2+n_3+1$ for 
TDAs  $V_{1,2}^{\pi N_{I=1/2}}$, $A_{1,2}^{\pi N_{I=1/2}}$, $T_{1,2}^{\pi N_{I=1/2}}$.

The TDAs $V_{1}^{\pi N_{I=1/2}}$, $A_{1}^{\pi N_{I=1/2}}$, $T_{1}^{\pi N_{I=1/2}}$ 
are expressed from (\ref{H_from_N_exchange})
with help of the usual $I=1/2$ relation, analogous to Eq.~(\ref{DAs_through_varphi}) for nucleon DAs:
\be
&&
\left\{ V_1^{ \pi N_{I=1 / 2}}, \, A_1^{ \pi N_{I=1 / 2}} \right\}(x_1,x_2,x_3,\xi,u)= 
\frac{1}{2} \left(\pm H^{\pi N_{I=1 / 2}}(x_1,x_2,x_3,\xi,u) + H^{\pi N_{I=1 / 2}}(x_2,x_1,x_3,\xi,u) \right); \ \ 
  \nn \\ &&
T_1^{ \pi N_{I=1 / 2}}(x_1,x_2,x_3,\xi,u)=
\frac{1}{2} \left( H^{\pi N_{I=1 / 2}}(x_1,x_3,x_2,\xi,u) + H^{\pi N_{I=1 / 2}}(x_2,x_3,x_1,\xi,u) \right);
\label{TDAs_through_varphi}
\ee
and the TDAs $V_{2}^{\pi N_{I=1/2}}$, $A_{2}^{\pi N_{I=1/2}}$, $T_{2}^{\pi N_{I=1/2}}$ are expressed
through the cross-channel nucleon exchange model constraint (\ref{Constraint_TDA12_Nexchange}).

Expressing TDAs of physical channels from (\ref{H_from_N_exchange}) requires multiplication by the isospin symmetry factors:
\be
&&
H^{\pi^0 p_{I=1/2}}=H^{\pi^0 n_{I=1/2}}=H^{\pi N_{I=1 / 2}}; \nn \\ &&
H^{\pi^+ p_{I=1/2}}=-H^{\pi^- n_{I=1/2}}= -\sqrt{2} H^{\pi N_{I=1 / 2}}.
\ee

\section{Constraints for $\pi N$ TDAs from the soft-pion theorem}
\label{App_Soft_pion}

According to the partial conservation of the axial current (PCAC) hypothesis (see, {\it e.g.}, \cite{Alfaro_red_book}), the following soft-pion theorem \cite{Pobylitsa:2001cz,Braun:2006td} holds for the matrix element of the three-quark light-cone operator (\ref{Def_3q_operator}), which defines the $\pi N$ generalized distribution amplitudes (GDAs) -- the 
cross-conjugate counterparts of the $\pi N$ TDAs:
\be
\left.\langle 0| \widehat{O}_{\rho \tau \chi}^{\alpha \beta \gamma}(\lambda_1n,\lambda_2n,\lambda_3n)\left|\pi_a N_\iota\right\rangle\right|_{\substack{\text { soft } \\ \text { pion }}}=-\frac{i}{f_\pi}\langle 0|\left[\widehat{Q}_5^a, \widehat{O}_{\rho \tau \chi}^{\alpha \beta \gamma}(\lambda_1n,\lambda_2n,\lambda_3n)\right]\left|N_\iota\right\rangle.
\label{soft_pion_th_GDA}
\ee
Eq. (\ref{soft_pion_th_GDA}) is written with explicit SU$(2)$ isotopic indices (see Ref.~\cite{Pire:2011xv} for a detailed discussion of the isospin symmetry properties of TDAs):
$\alpha, \, \beta, \, \gamma=1,2$ stand for quark isotopic indices;
$\iota=1,2$ is the nucleon isotopic index;
$a=1,2,3$ is the pion isotopic index.

The commutator of the chiral charge operator $\widehat{Q}_5^a$
a with quark field operators is given by
\be
\left[\widehat{Q}_a^5, \Psi_\eta^\alpha\right]=-\frac{1}{2}\left(\sigma_a\right)_\delta^\alpha \gamma_{\eta \tau}^5 \Psi_\tau^\delta,
\ee
where $\sigma^a$ stands for the Pauli matrices. This relates the
matrix element (\ref{soft_pion_th_GDA}) to the leading twist nucleon DA (\ref{Def_N_DA}).

The soft-pion theorem~(\ref{soft_pion_th_GDA}) is expected to hold for the $\pi N$ GDA at a scale $Q^2 \gg \Lambda_{\mathrm{QCD}}^3 / m_\pi$, at the pion production threshold of the cross-channel, which corresponds to $\Delta^2 = (m_N + m_\pi)^2$ and $\xi = \frac{m_N + m_\pi}{m_N - m_\pi}$. In general, this kinematical point differs from the threshold of the direct-channel reaction, where $\Delta^2 = (m_N - m_\pi)^2$ and $\xi = \frac{m_N - m_\pi}{m_N + m_\pi}$,
see  Fig. 22 of \cite{Pire:2021hbl}.
To apply the physical normalization provided by the soft-pion theorem for $\pi N$ TDAs -- while avoiding the subtle issue of analytic continuation in $\Delta^2$ between the cross-channel and direct-channel -- we have to take the chiral limit, setting the pion mass $m_\pi = 0$. In this limit, the physical domains of the cross-channel and direct-channel reactions share a common threshold point at $\xi = 1$ and $\Delta^2 = m_N^2$.

Under these assumptions the soft pion theorem (\ref{soft_pion_th_GDA})  yield the following normalization
for TDAs describing the $p \to \pi^0$ and $p \to \pi^+$ transitions for
$\xi=1$ and $u=M_{N}^{2}$ in terms of the leading twist nucleon DA $\varphi_N(y_1,y_2,y_3)$  
\cite{Pire:2011xv}:
\be
&&
V_1^{\pi^0 p}(x_1, x_2, x_3, \xi=1, u = m_{N}^2)   = -\frac{1}{4} \times \frac{1}{4}\left(\varphi_N\left(\frac{x_1}{2}, \frac{x_2}{2}, \frac{x_3}{2}\right) + \varphi_N\left(\frac{x_2}{2}, \frac{x_1}{2}, \frac{x_3}{2}\right)\right); \nonumber \\ && 
V_1^{\pi^+ p}(x_1, x_2, x_3, \xi=1, u=m_N^2)    \nonumber \\ && 
=-\frac{1}{4} \times \frac{1}{2 \sqrt{2}} \left(\varphi_N\left(\frac{x_1}{2}, \frac{x_2}{2}, \frac{x_3}{2}\right) + \varphi_N\left(\frac{x_2}{2}, \frac{x_1}{2}, \frac{x_3}{2}\right) + 2\varphi_N\left(\frac{x_3}{2}, \frac{x_1}{2}, \frac{x_2}{2}\right) + 2\varphi_N\left(\frac{x_3}{2}, \frac{x_2}{2}, \frac{x_1}{2}\right)\right); \nonumber \\ &&
A_1^{\pi^0 p}(x_1, x_2, x_3, \xi=1, u = m_{N}^2)   = -\frac{1}{4} \times\frac{1}{4}\left(-\varphi_N\left(\frac{x_1}{2}, \frac{x_2}{2}, \frac{x_3}{2}\right) + \varphi_N\left(\frac{x_2}{2}, \frac{x_1}{2}, \frac{x_3}{2}\right)\right) ; \nonumber \\ &&
A_1^{\pi^+ p}(x_1, x_2, x_3, \xi=1, u=m_N^2)   \nonumber \\ && =
- \frac{1}{4} \times \frac{1}{2 \sqrt{2}} \left(-\varphi_N\left(\frac{x_1}{2}, \frac{x_2}{2}, \frac{x_3}{2}\right) + \varphi_N\left(\frac{x_2}{2}, \frac{x_1}{2}, \frac{x_3}{2}\right) - 2\varphi_N\left(\frac{x_3}{2}, \frac{x_1}{2}, \frac{x_2}{2}\right) + 2\varphi_N\left(\frac{x_3}{2}, \frac{x_2}{2}, \frac{x_1}{2}\right)\right); \nonumber \\ &&
T_1^{\pi^0 p}(x_1, x_2, x_3, \xi=1, u = m_{N}^2)   = \frac{1}{4} \times \frac{3}{4} \left(\varphi_N\left(\frac{x_1}{2}, \frac{x_3}{2}, \frac{x_2}{2}\right) + \varphi_N\left(\frac{x_2}{2}, \frac{x_3}{2}, \frac{x_1}{2}\right)\right) ;\nonumber \\ && 
T_1^{\pi^+ p}(x_1, x_2, x_3, \xi=1, u=m_N^2)   =
-\frac{1}{4} \times \frac{1}{2 \sqrt{2}} \left(\varphi_N\left(\frac{x_1}{2}, \frac{x_3}{2}, \frac{x_2}{2}\right) + \varphi_N\left(\frac{x_2}{2}, \frac{x_3}{2}, \frac{x_1}{2}\right)\right). 
\label{eq:reductionToDAs}
\ee
Note that  TDAs are defined as functions of $x_{1,2,3}$ satisfying $x_1+x_2+x_3=2\xi$, while nucleon DAs are  of $y_{1,2,3}$, satisfying $y_1+y_2+y_3=1$.
Therefore, to keep the overall normalization of the Mellin moments it is convenient to single out the overall normalization factor $\frac{1}{4}$ in r.h.s. of Eqs.~(\ref{eq:reductionToDAs}).
Then {\it e.g.} for the asymptotic form of nucleon DA, $\varphi_N^{\rm as} \left(y_1,y_2,y_3\right) \equiv 120 y_1 y_2 y_3$,
\be
&&
 \int_0^2 dx_1 \int_0^2 dx_2 \int_0^2 dx_3 \delta(x_1+x_2+x_3-2) \frac{1}{4} \varphi_N^{\rm as} \left(\frac{x_1}{2}, \frac{x_3}{2}, \frac{x_2}{2}\right) \nn \\ &&
 =\int_{-1}^1 dw \int_{\frac{-1+w}{2}}^{\frac{1-w}{2}}\frac{1}{4} \varphi_N^{\rm as} \left(
 \frac{1-w+2 v}{4}, \frac{1-w-2 v}{4}, \frac{1+w}{4}
 \right)=  1.
 \label{Explain14}
\ee

The TDAs $V_2^{\pi p}$, $A_2^{\pi p}$, $T_2^{\pi p}$ at the pion threshold 
are expressed through the corresponding TDAs $V_2^{\pi p}$, $A_2^{\pi p}$, $T_2^{\pi p}$
(\ref{eq:reductionToDAs}) 
as
\be 
 \left\{V_2, A_2, T_2\right\}^{\pi p}
\left(x_1, x_2, x_3, \xi=1, u=m_N^2\right) 
=\frac{1}{2}  \left\{V_1, A_1, T_1\right\}^{ \pi p}
\left(x_1, x_2, x_3, \xi=1, u=m_N^2 \right). 
\label{Constraint_TDA12_soft}
\ee
Note that the sign in the r.h.s. of Eq.~(\ref{Constraint_TDA12_soft}) is opposite to that in the cross-channel nucleon exchange model (\ref{Constraint_TDA12_Nexchange}).
Finally, the TDAs $T_3^{\pi p}$ and $T_4^{\pi p}$ vanish at the pion threshold:
\be
&&
T_3^{\pi p}(x_1, x_2, x_3, \xi=1, u = m_{N}^2)   = T_4^{\pi p}(x_1, x_2, x_3, \xi=1, u = m_{N}^2) = 0.
\ee

\bibliography{main.bbl}
\end{document}